\documentclass[11pt, american,english,british, a4paper]{article}
\pdfoutput=1
\usepackage{jheppub}
\usepackage[T1]{fontenc}
\usepackage[latin9]{inputenc}
\usepackage{xcolor}
\usepackage{float}
\usepackage{amsmath}
\usepackage{amssymb}
\usepackage{stackrel}
\usepackage{graphicx}
\usepackage{wasysym}
\usepackage{esint}
\makeatletter

\providecommand{\tabularnewline}{\\}

\usepackage{babel}

\@ifundefined{showcaptionsetup}{}{%
 \PassOptionsToPackage{caption=false}{subfig}}
\usepackage{subfig}
\makeatother

\usepackage{babel}

\global\long\def\ud{\mathrm{d}}

\global\long\def\fii{\varphi}

\global\long\def\lam{\lambda}

\global\long\def\th{\vartheta}
\global\long\def\l{\langle}
\global\long\def\r{\rangle}
\global\long\def\oo{\mathcal{O}}

\global\long\def\eps{\varepsilon}

\title{Hydrodynamics of massless integrable RG flows and a non-equilibrium
c-theorem }

\author[a,b]{D. X. Horváth}

\affiliation[a]{MTA-BME \textquotedbl{}Momentum\textquotedbl{} Statistical
Field Theory Research Group,\\
1111 Budapest, Budafoki út 8, Hungary}

\affiliation[b]{Department of Theoretical Physics,\\
Budapest University of Technology and Economics,\\
1111 Budapest, Budafoki út 8, Hungary}

\emailAdd{esoxluciuslinne@gmail.com}
 
\abstract{
We study Euler scale hydrodynamics of massless integrable quantum
field theories interpolating between two non-trivial renormalisation
group fixed points after inhomogeneous quantum quenches. Using a partitioning
protocol with left and right initial thermal states and the recently
developed framework of generalised hydrodynamics, we focus on current
and density profiles for the energy and momentum as a function of
$\xi=x/t$, where both $x$ and $t$ are sent to infinity. Studying
the first few members of the $A_{n}$ and $D_{n}$ massless flows
we carry out a systematic treatment of these series and generalise
our results to other unitary massless models.

In our analysis we find that the profiles exhibit extended plateaux
and that non-trivial bounds exist for the energy and momentum densities
and currents in the non-equilibrium stationary state, i.e. when $\xi=0$.
To quantify the magnitude of currents and densities, dynamical central
charges are defined and it is shown that the dynamical central charge
for the energy current satisfies a certain monotonicity property.
We discuss the connection of the Landauer-Büttiker formalism of transport
with our results and show that this picture can account for some of
the bounds for the currents and for the monotonicity of the dynamical
central charge. These properties are shown to be present not only
in massless flows but also in the massive sinh-Gordon model suggesting
their general validity and the correctness of the Landauer-Büttiker
interpretation of transport in integrable field theories. Our results
thus imply the existence of a non-equilibrium c-theorem as well, at
least in integrable models. Finally we also study the interesting
low energy behaviour of the $A_{2}$ model that corresponds to the
massless flow from the tricritical to the critical Ising field theory.
}

\begin{document}
\maketitle

\section{Introduction}

Understanding the out-of-equilibrium dynamics of isolated quantum
many-body systems and giving a rigorous foundation of quantum statistical
mechanics are one of the most challenging problems in contemporary
physics. Thanks to the recent advances in laboratory techniques \cite{NewtonCradle,ExperimentalNoThermalization1,ExperimentalNoThermalization3,GGEExperimental,ColdAtomSchm1,ColdAtomSchm2,Nagerl,Fukuhara,Kaufman}
the experimental realisability of closed quantum systems provides
a direct insight into quantum \textcolor{black}{statistical physics.
As a result of intensive experimental and theoretical investigations,}
significant progress has been made in the study of non-equilibrium
behaviour and recent investigations have led to a series of interesting
discoveries. Perhaps the most unusual behaviour is related to integrable
quantum systems and the experimental observation of the lack of thermalisation
therein \cite{NewtonCradle,ExperimentalNoThermalization1,ExperimentalNoThermalization3,ExperimentalNoThermalization2}.
Because of their unusual properties, their experimental relevance
and analytic tractability, non-equilibrium physics in integrable models
continues to attract a lot of attention.

A paradigmatic setting for non-equilibrium dynamics is provided by
so-called quantum quenches corresponding to a sudden change in the
parameters of a closed quantum system \cite{CardyCalabrese,CardyCalabrese2}.
One of the simplest example is that of a homogeneous system where
time evolution after a quantum quench is expected to lead to relaxation
to thermal equilibrium. However, in the case of integrable system
thermalisation is absent and the steady state was proposed to be described
by the generalised Gibbs ensemble (GGE) \cite{GGEProposal} which
is supported by experimental and theoretical studies \cite{GGEExperimental,GETH,ExcitedBetheStates,CauxNoGGE,PozsgayNoGGE,Pozsi2,Goldstein,EsslerMussardo,ProsenCGGE,MarciNoGGE,IlievskiStringChargeDuality,PozsgaytGGE,IlievskiCauxQPPicture,IntegrableQuench,IlievskiQuasiLocal}.
Nevertheless, important issues such as a theoretical description of
the eventual time evolution, as well as the complete set of relevant
conserved quantities necessary for the construction of the steady
state ensemble are still open in general.

Inhomogeneous quenches, in general, pose a more difficult problem
than homogeneous ones with fewer exact or approximate results \cite{2005PlatiniIsingChain,DeLucaVitiBernardDoyonIsing,ColluraKarevski,JeromeFreeFermion,DeLuca1,FagottiSpinChains,MarciInHomQQ,PerfettoGambassi,MarciSemiClInHom}.
One promising approach exploits hydrodynamical description assuming
the separation of spatial and temporal scales and a local (GGE) equilibrium,
and is supported both by numerical simulations and experimental observations
\cite{TakatoGHDCheckedWithAbacus,VasseurKarraschMooreHydroCheckDMRG,GHDNewtonCraddleFirst,GHDNewtonCraddle,DubailInHomCFT}.
Based on the previously mentioned assumptions and on the functional
completeness of conserved quantities a hydrodynamical description
of integrable systems named Generalised Hydrodynamics (GHD) was developed
in \cite{GHDFundation,GHDFundationB,GHDFoundationC}. In its simplest
form the GHD describes the exact average densities and currents associated
with conserved quantities at the Euler scale, that is when $x,t\rightarrow\infty$
such that their ratio is fixed. The appearance of non-trivial physics
in this limit is in accordance with the ballistic spreading of quasi-particles
in integrable models and in many relevant situations the GHD predictions
become valid after a relatively short transient time interval \cite{TakatoGHDCheckedWithAbacus,VasseurKarraschMooreHydroCheckDMRG,GHDNewtonCraddleFirst,GHDNewtonCraddle}.
The GHD approach has been applied to various systems including spin
chains and the Hubbard model \cite{GHDFundationB,GHDPiroli,GHDDeLuca,GHDHubbard,GHDDomainWall,GHDAlba,GHDLowTXXZ,GHDKarrasch,GHDBastianello},
classical gases and fields \cite{GHDDoyonSpohn,GHDCao,GHDClassFT,GHDClassToda}
and quantum gases and fields \cite{VasseurKarraschMooreHydroCheckDMRG,GHDFundation,GHDMarci,GHDSineG}.
Interesting view points on the GHD approach are given in \cite{GHDGeometric,DoyonSolitonGas,MarciFleeGas}.
Besides the Euler scale description of current and density averages
in various integrable models important new directions are understanding
the fluctuations of the hydrodynamic quantities \cite{GHDFluctuations1,GHDFluctuations2}
and incorporating diffusive effects which requires going beyond the
Euler scale in the GHD description \cite{GHDDiff1,GHDDiff2,GHDDiff3}. 

The aim of the present paper is to introduce a renormalisation group
perspective and connect the GHD description of the non-equilibrium
dynamics of integrable quantum field theories to their formulation
as relevant perturbations of conformal field theories (CFT) \cite{BPZ}.
In the works \cite{NonEqCFTSpyros,NonEqCFT1,NonEqCFT2,NonEqCFT3,TTBarCFT,NonEqCFT5,NonEqCFTReview}
a rather extensive description of transport properties of out-of-equilibrium
CFTs was given. The non-equilibrium setting was provided by the (bi-)
partitioning protocol, in which two semi-infinite and independent
systems described by the same Hamiltonian are prepared in different
initial states, typically in two thermal states with different temperatures.
The left and right semi-infinite systems are joined together at a
given time and the subsequent time-evolution is governed by the full
and homogeneous Hamiltonian acting on the whole space. In the resulting
dynamics for CFTs, two shock waves originating from the boundary point
move with the speed of light to the left and right directions and
in the expanding region between the waves a non-equilibrium steady-state
(NESS) emerges. The NESS is obtained in the limit $t\rightarrow\infty$
for finite $x$ and therefore corresponds to $\xi=0$, supporting
non-trivial flows of currents while the asymptotic left ($\xi<0$)
and right ($\xi>0$) regions act as effective heat reservoirs. In
\cite{NonEqCFT1,NonEqCFT2,NonEqCFT3,TTBarCFT,NonEqCFT5} the properties
of the NESS, current and density averages, fluctuation spectrums for
large deviations and correlation functions in the NESS were studied.
In particular for CFTs the average energy density $q_{e}$ and energy
current $j_{e}$ are given by 

\begin{equation}
\begin{split}q_{e} & =\frac{c\pi}{12}\left(T_{l}^{2}+T_{r}^{2}\right)\\
j_{e} & =\frac{c\pi}{12}\left(T_{l}^{2}-T_{r}^{2}\right)\,,
\end{split}
\label{eq:EqCFT1}
\end{equation}
where $T_{l}$ and $T_{r}$ are the initial left and right temperatures
and $c$ is the central charge of the CFT. Due to relativistic invariance
$j_{e}$ equals the momentum density $q_{p}$ and due to conformal
invariance $q_{e}$ equals the momentum current $j_{p}$ (and also
the pressure).

These results are also valid for systems at or sufficiently close
to their critical points but a certain neighbourhood of the critical
point or more precisely a certain type of irrelevant perturbation
of the fixed point CFT was also studied in \cite{TTBarCFT}. In general,
the vicinity of critical points are described by (relevant) perturbations
of the critical theory resulting in either a massive quantum field
theory (QFT) or a crossover to an infrared (long-distance) fixed point.
The high energy behaviour and consequently the high temperature transport
properties of such a system are captured by the fixed point CFT, but
the intermediate and low energy physics is generally not known. If
the perturbed CFT is integrable, which is quite often the case of
interest, the GHD approach can be used to explore the transport properties
at all scales of the field theory.

Some relevant perturbations of CFTs give rise to massless QFTs, which
host massless excitations. For these interpolating renormalisation
group flows describing crossover behaviour the high- and low-energy
properties are dictated by CFTs. The massless models are not conformal
invariant theories and transition from the ultra-violet (UV) CFT to
the infra-red (IR) occurs at the characteristic energy scale $M$
called the crossover scale. 

Integrable massless flows have a long history and many of them have
a detailed description based on the standard techniques of integrability
such as S-matrix bootstrap, thermodynamic Bethe Ansatz (TBA) and form
factor bootstrap. In this paper we primarily focus on the so-called
$A_{n}$ and $D_{n}$ massless flows. The $A_{n}$ models interpolate
between multi-critical Ising field theories \cite{AnZamo123,AnZamo2,AnZamo3}
corresponding to the conformal minimal models $\mathcal{M}_{n+2}\rightarrow\mathcal{M}_{n+1}$
whereas the $D_{n}$ flows describe the crossover between the $\mathbb{Z}_{n}$
parafermion model and $\mathcal{M}_{n+1}$ \cite{DnParafermions}.
The S-matrices of these models and the corresponding TBA equations
are well-known \cite{AnZamo123,AnZamo2,AnZamo3,DnParafermions,DynkinTBAs}
together with certain form factors \cite{MasslessFlow,RefRoaming,RoamingFF}
and correlation functions \cite{MasslessFlow} for the $A_{2}$ case,
and it is worth noting the interesting connection of these flows to
roaming trajectories and staircase models \cite{RefZamo,RGFlow2,RGFlow3,RGFlow4,RGFlow5}.

As these theories interpolate between a UV and an IR conformal field
theory, it is natural to investigate the crossover in terms of transport
behaviour. This is what the present work aims to accomplish by applying
the GHD approach to the partitioning protocol and focusing on the
interpolation between the behaviour of the fixed point CFTs describing
the endpoints of the RG flow.

The paper is organised as follows. In section \ref{sec:ReprStateTBAGHD}
the description of homogeneous and inhomogeneous macro-states in terms
of the TBA type equations is discussed for integrable models and the
main ideas of the GHD approach are summarised. In section \ref{sec:Thermodynamics}
some elements of the thermodynamics of the $A_{n}$ and $D_{n}$ flows
are discussed that are useful to understand the inhomogeneous problem.
Section \ref{sec:A2Hydro} is devoted to hydrodynamics of the $A_{2}$
model. Besides reviewing its interesting low temperature limit, we
identify features that are later shown to be generally present in
massless models and introduce the notion of dynamical central charges.
In section \ref{sec:AnDnHydro} the discussion is generalised to higher
members of the $A_{n}$ and $D_{n}$ hierarchy of massless flows.
In section \ref{sec:LBF} we establish a connection with the Landauer-Büttiker
(LB) theory of transport and discuss its implications for the dynamical
central charges in RG flows. In this section the transport of massive
sinh-Gordon integrable QFT is also studied and compared to the LB
results, and we show that all the results suggest an out-of-equilibrium
version of the c-theorem \cite{OriginalCTheorem,cTheorem}, at least
for relativistic integrable QFTs. Finally we conclude in section \ref{sec:Conclusions}.
Some details are relegated to appendices: in appendix \ref{subsec:Appendix-A-MasslessFF},
the low-temperature expansion of the TBA and GHD equations are performed
for the $A_{2}$ model, while appendix B discusses the hydrodynamics
of the $W_{5}^{3}\rightarrow W_{4}^{3}$ massless flow.

\section{Macro-states in integrable models and the GHD\label{sec:ReprStateTBAGHD}}

In this section we briefly review how macro-states such as homogeneous
thermal or inhomogeneous locally quasi-stationary states (LQSS) can
be characterised in integrable models and discuss the main ideas behind
the GHD approach\foreignlanguage{english}{. For most integrable models,
the spectrum of energy levels can be described in terms of stable
quasi-particles excitations. Averages of local operators in macro
states can be constructed by representing the corresponding density
matrices in terms of continuous densities associated with the distribution
of these stable quasi-particles. In particular the so-called root
density $\rho(\th)$ gives the number of quasiparticles in the range
$[\th,\th+\Delta\th]$, which is $L\rho(\th)\Delta\th$, if $L$ is
the system size and the rapidity $\th$ parametrizes the energy and
momentum of the quasi-particles. In the case of additional quantum
numbers or different species of quasi-particles, multiple root densities
labeled by the appropriate quantum numbers are required. }

Similarly to the root densities one can introduce the hole densities
$\rho_{h}(\th)$ as well, which are associated with unoccupied one-particle
energy levels. The root and hole densities are not independent quantities
due to the interaction between the quasiparticles. Extending our description
to models with $n$ quasiparticle species and with diagonal scattering,
the root and hole densities $\rho_{j}$ and $\rho_{j,h}$ corresponding
to the different species satisfy the Bethe Ansatz equations

\begin{equation}
\rho_{j}(\th)+\rho_{j,h}(\th)=\frac{1}{2\pi}p'_{j}(\th)+\sum_{k=1}^{n}\left(\varphi_{jk}\star\rho_{k}\right)(\th)\,.\label{RootAndHoleDensities}
\end{equation}
In \eqref{RootAndHoleDensities}, $p'_{j}(\th)$ denotes the derivative
of the one-particle momentum of the $j$th species with respect to
the rapidity $\th$, $\fii_{jk}$ is specified by the two-particle
scattering matrix $S_{jk}$ as

\begin{equation}
\fii_{jk}(\th)=-i\frac{\ud}{\ud\th}\ln\left(-S_{jk}(\th)\right)\,,\label{Fi}
\end{equation}
and the symbol $\star$ denotes convolution:

\begin{equation}
\left(f\star g\right)(\th)=\int_{-\infty}^{\infty}\frac{\ud\th'}{2\pi}f(\th-\th')g(\th')\,.
\end{equation}
It is convenient to introduce the filling functions $n_{j}(\th)$
and the pseudo-energies $\eps_{j}(\th)$ via

\begin{equation}
n_{j}(\th)=\frac{\rho_{j}(\th)}{\rho_{j}(\th)+\rho_{j,h}(\th)}=\frac{1}{e^{\eps_{j}(\th)}+1}\,,\label{eq:FillingFunction}
\end{equation}
with $j=1,...,n$ and also the dressing operation for a set of functions
$f_{j}(\th)$ with $j=1,...,n$ by the solution of the integral equation

\begin{equation}
f_{j}^{\text{dr}}(\th)=f_{j}(\th)+\sum_{k=1}^{n}\left(\fii_{jk}\star n_{k}f_{k}^{\text{dr}}\right)(\th)\,.\label{eq:Dressing}
\end{equation}
Given the root and hole densities or the filling functions, the densities
of various extensive quantities can be easily calculated in the macro-state.
For conserved charges $Q_{i}$ with one-particle eigenvalue $h_{j}^{(i)}(\th)$
with respect to the $j$th particle species, that is 

\[
Q_{i}|\th\r_{j}=h_{j}^{(i)}(\th)|\th\r_{j}\,,
\]
the corresponding density $q_{i}$ can be expressed equivalently as

\begin{equation}
\begin{split}q_{i}= & \sum_{j=1}^{n}\int_{-\infty}^{\infty}\frac{\ud\th}{2\pi}\,h_{j}^{(i)}(\th)\rho_{j}(\th)\,,\\
= & \sum_{j=1}^{n}\int_{-\infty}^{\infty}\frac{\ud\th}{2\pi}\,\left(p_{j}'(\th)\right)^{\text{dr}}n_{j}(\th)h_{j}^{(i)}(\th)\\
= & \sum_{j=1}^{n}\int_{-\infty}^{\infty}\frac{\ud\th}{2\pi}\,p_{j}'(\th)n_{j}(\th)\left(h_{j}^{(i)}(\th)\right)^{\text{dr}}\;.
\end{split}
\label{qAv}
\end{equation}
Similarly to the quantum mechanical operator ${\tt q}_{i}$, which
is the density of the charge $Q_{i}$, the average of the corresponding
current ${\tt j}_{i}$ in the macro-state can be expressed in terms
of the filling functions or root densities. Before quoting the expressions
we first introduce the effective velocity $v_{j}^{\text{eff}}(\th)$
defined as

\begin{equation}
v_{j}^{\text{eff}}(\th)=\frac{\left(e_{j}'(\th)\right)^{\text{dr}}}{\left(p_{j}'(\th)\right)^{\text{dr}}}\,.\label{vEff}
\end{equation}
While the velocity of a single particle can be defined as 

\[
\frac{e_{j}'(\th)}{p_{j}'(\th)}\,,
\]
the dressing in \eqref{vEff} accounts for the effect of other quasi-particles
in the macro-state specified by the root densities. Due to scattering
processes, the single-particle velocity is modified according to \eqref{vEff}. 

With the effective velocity, the averages $j_{i}$ of ${\tt j}_{i}$
can be written as

\begin{equation}
\begin{split}j_{i}= & \sum_{j=1}^{n}\int_{-\infty}^{\infty}\frac{\ud\th}{2\pi}\,v_{j}^{\text{eff}}(\th)h_{j}^{(i)}(\th)\rho_{j}(\th)\\
= & \sum_{j=1}^{n}\int_{-\infty}^{\infty}\frac{\ud\th}{2\pi}\,\left(e_{j}'(\th)\right)^{\text{dr}}n_{j}(\th)h_{j}^{(i)}(\th)\\
= & \sum_{j=1}^{n}\int_{-\infty}^{\infty}\frac{\ud\th}{2\pi}\,e_{j}'(\th)n_{j}(\th)\left(h_{j}^{(i)}(\th)\right)^{\text{dr}}\;.
\end{split}
\label{jAv}
\end{equation}
The expressions were first proposed in \cite{GHDFundation} and later
verified for relativistic QFT in \cite{TakatoProofVEff}. These equations
are in accordance with the ballistic transport of conserved quantities
in integrable models. 

Finally, we mention that the entropy density $s$ of the macro-state
can be written as

\begin{equation}
s=\sum_{j=1}^{n}\int_{-\infty}^{\infty}\ud\th\left[\rho_{j,t}(\th)\ln\rho_{j,t}(\th)-\rho_{j}(\th)\ln\rho_{j}(\th)-\rho_{j,h}(\th)\ln\rho_{j,h}(\th)\right]\,.
\end{equation}

\subsection{Thermodynamic Bethe Ansatz description for thermal and GGE states}

The root densities or the filling functions of homogeneous and global
thermal and GGE states can be obtained by solving the Thermodynamic
Bethe Ansatz (TBA) equations

\begin{equation}
\eps_{j}(\th)=w_{j}(\th)-\left(\sum_{k=1}^{n}\fii_{jk}\star\ln\left(1+e^{-\eps_{k}}\right)\right)(\th)\,,\label{TBA}
\end{equation}
where the driving term reads

\begin{equation}
w_{j}(\th)=\sum_{i=1}^{\infty}\beta_{i}h_{j}^{(i)}(\th)\,,\label{DrivingTermTBAGGE}
\end{equation}
if the state to describe is a GGE state with density matrix

\begin{equation}
\rho_{GGE}=\frac{1}{Z}e^{-\sum\beta_{i}Q_{i}}\,,\label{RhoGGE}
\end{equation}
with generalised chemical potentials $\beta_{i}$ associated with
each conserved charge. For the particular case of thermal states,
$w_{j}$ is merely

\begin{equation}
w_{j}(\th)=\frac{1}{T}e_{j}(\th)\,,\label{DrivingTermTBAThermal}
\end{equation}
where $T$ is the temperature. The corresponding free-energy density
or generalised free-energy density $f=F/L$ with $F=\sum\beta_{i}\l Q_{i}\r-S$
can be calculated by

\begin{equation}
f=\sum_{k=1}^{n}\int_{-\infty}^{\infty}\ud\th\,p'_{k}(\th)\ln\left(1+e^{-\eps_{k}(\th)}\right)\,.
\end{equation}
From the thermal free energy density the effective central charge 

\begin{equation}
\tilde{c}(T)=\frac{1}{T}\frac{3}{\pi^{2}}\sum_{k=1}^{n}\int_{-\infty}^{\infty}\ud\th\,p'_{k}(\th)\ln\left(1+e^{-\eps_{k}(\th)}\right)\label{eq:EffectiveC}
\end{equation}
can be obtained, which plays an important role in our subsequent considerations.
Due to the c-theorem \cite{cTheorem}, $\tilde{c}(T)$ increases monotonously
with the temperature\footnote{In its most commonly formulated version, the effective central charge
$\tilde{c}$ entering the c-theorem is a function of the distance
$R$ instead of the temperature $T$. } and signals the amount of the effective degrees of freedom in the
field theory. In the $T\rightarrow\infty$ limit its value is determined
by the UV limiting conformal field theory, in particular 

\begin{equation}
\lim_{T\rightarrow\infty}\tilde{c}(T)=c_{UV}\,,
\end{equation}
where $c_{UV}$ is the central charge of the UV conformal field theory
if it is unitary. In the $T\rightarrow0$ limit, $\tilde{c}$ is zero
in massive models but equals the central charge of the IR conformal
field theory for massless flows

\begin{equation}
\begin{split}\lim_{T\rightarrow0}\tilde{c}(T)= & \begin{cases}
0 & \text{massive case}\\
c_{IR} & \text{massless case .}
\end{cases}\end{split}
\end{equation}

\subsection{GHD and the partitioning protocol\label{subsec:BriefGHD}}

The main purpose of our paper is to study transport properties of
massless integrable models in inhomogeneous initial states. To treat
inhomogeneous situations, it is convenient to apply a hydrodynamic
approach relying on the separation of space and time scales and on
the assumption of local equilibration. The large-scale behaviour of
inhomogeneous systems can be described by a space-time dependent GGE

\begin{equation}
\rho_{GGE}=\frac{1}{Z}e^{-\sum\int\ud x\beta_{i}(x,t){\tt q}_{i}(x)}\,,\label{RhoGGEXT}
\end{equation}
and consequently the large-scale expectation values of local operators
can be obtained as

\begin{equation}
\l\oo(x,t)\r=\l\oo\r_{x,t}\,,\label{LocalRelaxation1}
\end{equation}
where 

\begin{equation}
\l\oo\r_{x,t}=\frac{1}{Z}\text{Tr}\,\oo(0,0)e^{-\sum\beta_{i}(x,t)Q_{i}}\,.\label{LocalRelaxation2}
\end{equation}
Similarly to the homogeneous case, ($x,t$)-dependent root densities
$\rho_{j}(\th,x,t)$ and filling functions $n_{j}(\th,x,t)$ can be
introduced to describe the LQSS. Exploiting eqs. \eqref{LocalRelaxation1}
and \eqref{LocalRelaxation2}, the continuity equation of the conserved
quantities $\partial_{t}{\tt q}_{i}+\partial_{x}{\tt j}_{i}=0$ transforms
into 

\begin{equation}
\partial_{t}q_{i}(x,t)+\partial_{x}j_{i}(x,t)=0\,,\label{GHDContEq}
\end{equation}
where

\begin{equation}
\begin{split}q_{i}(x,t)= & \l{\tt q}_{i}\r_{x,t}\\
j_{i}(x,t)= & \l{\tt j}_{i}\r_{x,t}\,.
\end{split}
\end{equation}
In integrable models, assuming a sufficient functional completeness
of the conserved charges, the continuity equation for the LQSS averages
\eqref{GHDContEq} can be recast in many different forms including
the space-time dependent root densities or filling functions. For
our purposes, the most direct rewriting reads \cite{GHDFundation}

\begin{equation}
\partial_{t}n_{j}(\th,x,t)-v_{j}^{\text{eff}}[\th,\{n_{j}(\th,x,t)\}]\partial_{x}n_{j}(\th,x,t)=0\,,\label{GHDBasicEq1}
\end{equation}
where $v_{j}^{\text{eff}}$ is defined by eq. \eqref{vEff} and the
argument in the bracket stresses that $v_{j}^{\text{eff}}$ is a complicated
functional of the set of filling functions $n_{j}$ which now depend
on space and time besides the rapidity. The equation \eqref{GHDBasicEq1}
is in complete agreement with the ballistic spreading of quasi-particles
in integrable models. The effect of the interactions is incorporated
in the effective velocity of the quasi-particles. For the case of
the partitioning protocol corresponding to an initial density matrix
\begin{equation}
\rho_{0}\propto\rho_{l}\otimes\rho_{r}\,,\label{Rho0RiemannProblem-1}
\end{equation}
which is different on the left and right halves of the system, eq.
\eqref{GHDBasicEq1} can be solved in a particularly simple way. For
the particular case of thermal states considered in this paper, the
initial density matrix reads

\begin{equation}
\rho_{0}=\frac{1}{Z}e^{-\beta_{l}H_{l}}\otimes e^{-\beta_{r}H_{r}}\,,\label{Rho0RiemannProblem}
\end{equation}
where the Hamiltonians $H_{l}$ and $H_{r}$ act in the left and right
half-spaces. The two halves are joined together at time $t=0$ and
subsequent time evolution is governed by the homogeneous Hamiltonian
acting on the whole space. To be precise, some boundary conditions
have to be prescribed for $H_{r}$ and $H_{l}$ at the position $x=0$
before $t=0$; nevertheless, it is expected that their effect becomes
negligible at the Euler scale. Consequently, the initial condition
for $n_{j}(\th,x,0)$ can be written as

\begin{equation}
n_{j}(\th,x,0)=\Theta_{H}(x)n_{j}^{(r)}(\th)+\Theta_{H}(-x)n_{j}^{(l)}(\th)\,,\label{InitCondn_j}
\end{equation}
where $\Theta_{H}$ is the Heaviside function and $n_{j}^{(r)}$ and
$n_{j}^{(l)}$ are the filling functions corresponding to the right
and left density matrices describing homogeneous thermal (or GGE in
general) states with temperatures $T_{r}$ and $T_{l}$. To obtain
a solution of \eqref{GHDBasicEq1} with an initial condition for $n_{j}$
compatible with \eqref{InitCondn_j} it is exploited that both the
differential equations \eqref{GHDBasicEq1} and the initial condition
\eqref{InitCondn_j} are invariant under the the reparametrisation
$x,t\rightarrow\lam x,\lam t$. As a consequence the solution of \eqref{GHDBasicEq1}
depends only on the ratio $x/t$ which we denote by $\xi$ and call
a ray. The corresponding ray-dependent continuity equations read
\begin{equation}
-\xi\partial_{\xi}q_{i}(\xi)+\partial_{\xi}j_{i}(\xi)=0\,,\label{GHDContEqXi}
\end{equation}
and

\begin{equation}
\left(\xi-v_{j}^{\text{eff}}[\{n_{j}(\th,\xi)\}]\right)\partial_{\xi}n_{j}(\th,\xi)=0\,,\label{GHDBasicEq1Xi}
\end{equation}
and the implicit solution of the latter is given by

\begin{equation}
n_{j}(\th,\xi)=\Theta_{H}(v_{j}^{\text{eff}}(\th,\xi)-\xi)n_{j}^{(l)}(\th,\xi)+\Theta_{H}(\xi-v_{j}^{\text{eff}}(\th,\xi))n_{j}^{(r)}(\th,\xi)\,.\label{GHDBasicEq1XiSolution}
\end{equation}
Its interpretation in terms of the ballistic spreading of quasi-particles
is natural; quasi-particles that contribute at the ray $\xi$ come
from the left if their effective velocity is larger than $\xi$ and
they come from the right side if their effective velocity is slower
than $\xi$; the effective velocity depends on all the other particles
due to the elastic scattering between them. The self-consistent numerical
solution of the above equation is usually straightforward to obtain
by iteration. In particular when $v_{j}^{\text{eff}}(\th,\xi)$ is
a monotonously increasing function of $\th$ for all $\xi$, the solution
can be rewritten as

\begin{equation}
n_{j}(\th,\xi)=\Theta_{H}(\th-\th_{j})n_{j}^{(l)}(\th,\xi)+\Theta_{H}(\th_{j}-\th)n_{j}^{(r)}(\th,\xi)\,,\label{GHDBasicEq1XiSolution2}
\end{equation}
where $\th_{j}$ is determined by the implicit equations

\begin{equation}
v_{j}^{\text{eff}}(\th_{j})=\xi\label{vEffXi}
\end{equation}
for all $j$ and $v_{j}^{\text{eff}}$, which is a functional of all
$n_{j}(\th,\xi)$, is determined by \eqref{vEff}. As discussed in
section \ref{sec:Thermodynamics}, the effective velocities of magnonic
particles can be monotonously decreasing or non-monotonous functions
of the rapidity too. In such a case \eqref{GHDBasicEq1XiSolution2}
may include more terms, and more than a singe $\th_{j}$ is necessary
to use to describe the jumps in the filling functions. The values
of the $\th_{j}^{(m)}$ parameters are still to be determined by \eqref{vEff}
for all $m$ in a self-consistent manner. 

In massless models the effective velocities of the quasi-particles
(either the later introduced right- and left-moving particles or the
magnons are regarded) usually do not cover the full $[-1,1]$ interval
but only its subset $[v_{j}^{min},v_{j}^{max}]$. In such a case the
solution of \eqref{GHDBasicEq1Xi} for a ray $\xi\in[-1,v_{j}^{min}]$
is $n_{j}^{(l)}(\th)$ since there are only faster than $\xi$ particles,
which must come from the left and similarly for a ray $\xi\in[v_{j}^{max},1]$
it is $n_{j}^{(r)}(\th)$ (where $n_{j}^{(l)}(\th)$ and $n_{j}^{(r)}(\th)$
are the filling functions of the homogeneous left/right thermal states
or GGE). This situation is similar to what was discussed in \cite{PiroliLL},
which focused on the transport of non-linear Luttinger liquids.

Once $n_{j}(\th,\xi)$ are determined, the ray-dependent averages
of the densities and currents of conserved charges can be straightforwardly
calculated using \eqref{eq:Dressing}, \eqref{qAv} and \eqref{jAv}.

\section{Thermodynamics of the $A_{n}$ and $D_{n}$ massless flows\label{sec:Thermodynamics}}

\begin{figure}[H]
\begin{centering}
\includegraphics[width=0.29\textwidth]{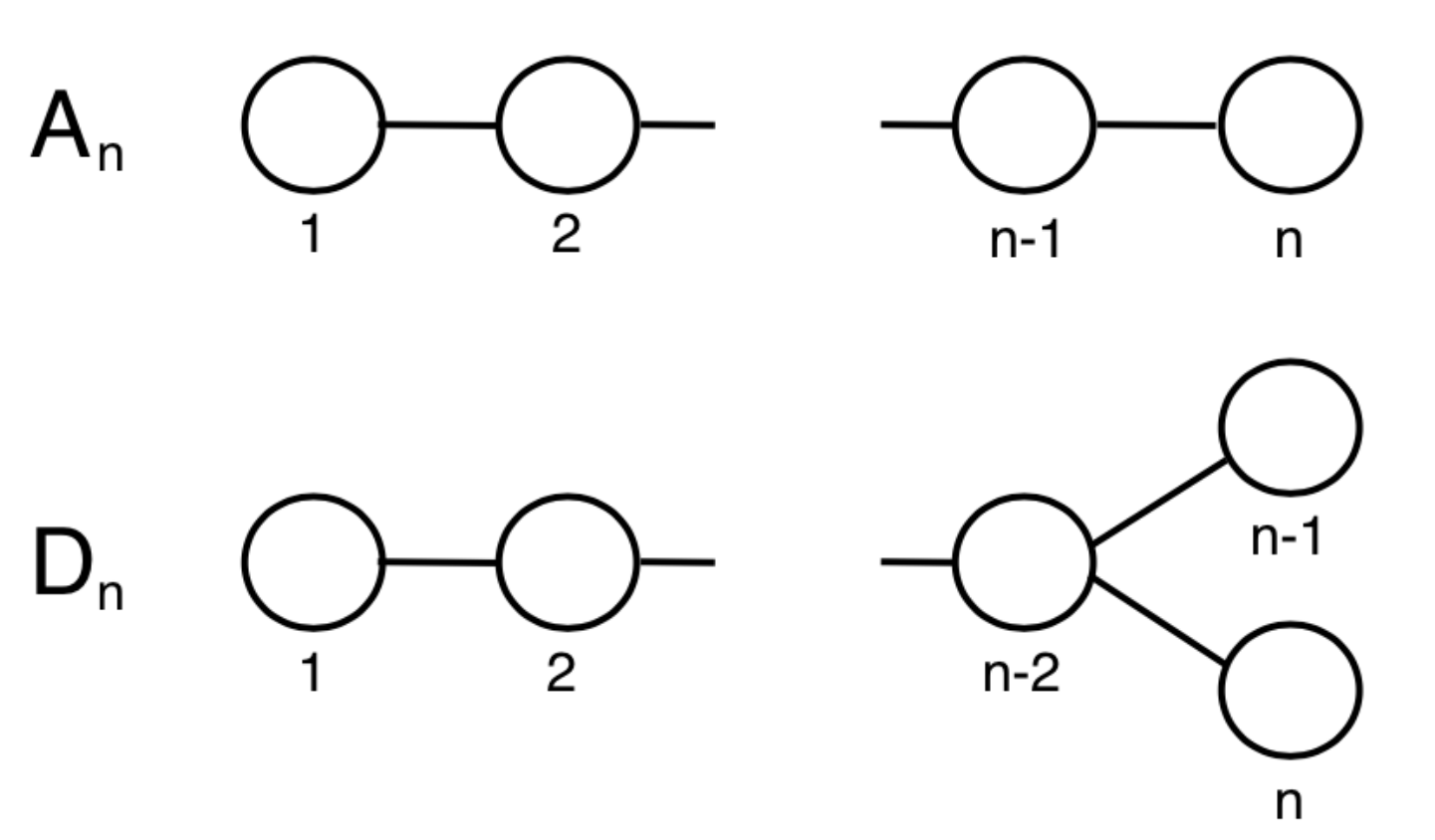}
\par\end{centering}
\caption{\label{figDynkin}{\footnotesize{}Dynkin diagrams of the $A_{n}$
and $D_{n}$ Lie algebras}}
\end{figure}

In this section we review on the finite temperature description of
integrable massless flows using the TBA. The finite temperature filling
functions are essential inputs for the hydrodynamics of the partitioning
protocol, moreover many peculiar features of the emerging hydrodynamics
can be understood by analysing the homogeneous, finite temperature
case. The TBA equations for massless flows associated with the $A_{n}$
and $D_{n}$ series can be written \cite{DynkinTBAs} as

\begin{equation}
\eps_{j}(\th)=w_{j}(\th)-\left(\sum_{k=1}^{n}\fii\star\ln\left(1+e^{-\eps_{k}}\right)G_{jk}\right)(\th)\,,\label{TBAAnDn}
\end{equation}
where 

\begin{equation}
\varphi(\th)=\frac{1}{\cosh\th}\,,
\end{equation}
$G_{jk}$ is the adjacency matrix of the $A_{n}$ and $D_{n}$ Dynkin
diagrams (c.f. figure \ref{figDynkin}) and the source terms read

\begin{equation}
\begin{split}w_{j}(\th)= & \frac{M}{2T}\left(e^{\th}\delta_{j,1}+e^{-\th}\delta_{j,n}\right)\text{for }A_{n},\:n\geq2\\
w_{j}(\th)= & \frac{M}{2T}\left(e^{\th}\delta_{j,n-1}+e^{-\th}\delta_{j,n}\right)\text{for }D_{n},\:n\geq3.
\end{split}
\label{DrivingTermTBAAnDn}
\end{equation}

\selectlanguage{english}%

The nodes in the diagrams correspond to different particle species.
The scattering between these species is non-trivial only for the neighbouring
nodes. According to eq. \eqref{DrivingTermTBAAnDn} only two nodes
in the Dynkin diagrams couple to non-vanishing driving terms. The
source terms $\frac{M}{2T}e^{\pm\th}$ correspond to right- and left-moving
i.e. (RM) and (LM) particles, whose one-particle energy and momentum
are $\frac{M}{2}e^{\pm\th}$ and $\pm\frac{M}{2}e^{\pm\th}$ respectively.
The energy scale $M$ is the crossover scale, which separates the
low and high energy regimes dominated by the UV and IR limiting CFTs.
The other nodes in the Dynkin diagrams correspond to magnons, which
describe internal degrees of freedom of the quasi-particle excitations.
It is important to note that although the magnonic excitations themselves
may be regarded as quasi-particles, they have zero one-particle eigenvalues
with respect to the conserved charge operators, hence quantities such
as the energy or momentum are carried by only the RM and LM species.

The eventual identification of the models described by the TBA equations
is a non-trivial task.\foreignlanguage{british}{ The $A_{n}$ flows
interpolate between multi-critical Ising field theories \cite{AnZamo123,AnZamo2,AnZamo3}
according to $\mathcal{M}_{n+2}\rightarrow\mathcal{M}_{n+1}$ with
\begin{equation}
\begin{split}c_{UV} & =1-\frac{6}{\left(n+2\right)\left(n+3\right)}\\
c_{IR} & =1-\frac{6}{\left(n+1\right)\left(n+2\right)}\,.
\end{split}
\label{eq:CIRUVAn}
\end{equation}
These integrable RG trajectories are the $\phi_{1,3}$ perturbations
with scaling dimension $\Delta=1-\frac{2}{n+3}$ of the UV conformal
theory $\mathcal{M}_{n+2}$ \cite{AnZamo123,AnZamo2,AnZamo3}. For
the $A_{n}$ massless models, both $c_{UV}$ and $c_{IR}$ tends to
1 as $n\rightarrow\infty$ and the difference between the UV and IR
central charges vanishes in this limit. }

\selectlanguage{british}%
The $D_{n}$ flows describe the crossover between the $\mathbb{Z}_{n}$
parafermion model and $\mathcal{M}_{n+1}$ \cite{DnParafermions}
with

\begin{equation}
\begin{split}c_{UV} & =\frac{2\left(n-1\right)}{n+2}\\
c_{IR} & =1-\frac{6}{\left(n+1\right)\left(n+2\right)}
\end{split}
\label{eq:CIRUVDn}
\end{equation}
and are obtained by adding the perturbing operator $\psi_{1}(z)\bar{\psi}_{1}(\bar{z})+\psi_{1}^{\dagger}(z)\bar{\psi}_{1}^{\dagger}(\bar{z})$
with dimension $1-\frac{1}{n}$ \cite{DynkinTBAs} to the UV limiting
CFT. Here in the $n\rightarrow\infty$ limit $c_{UV}=2\neq c_{IR}=1$.

\begin{figure}[H]
\begin{centering}
\subfloat[\foreignlanguage{british}{$n_{RM}(\protect\th)$ and $n_{LM}(\protect\th)$ }]{\begin{centering}
\includegraphics[width=0.45\columnwidth]{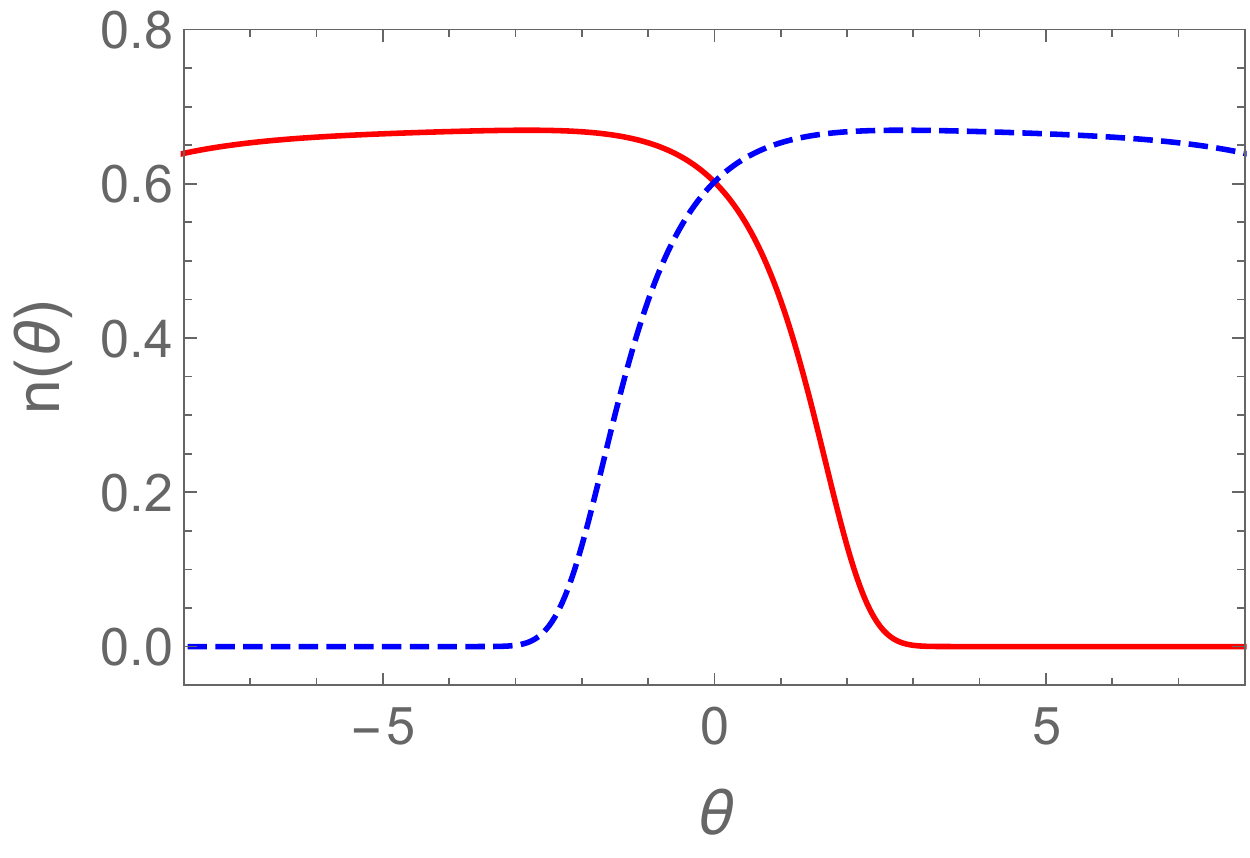} 
\par\end{centering}
}\subfloat[$v_{RM}^{\text{eff}}(\protect\th)$ and $v_{LM}^{\text{eff}}(\protect\th)$]{\begin{centering}
\includegraphics[width=0.45\columnwidth]{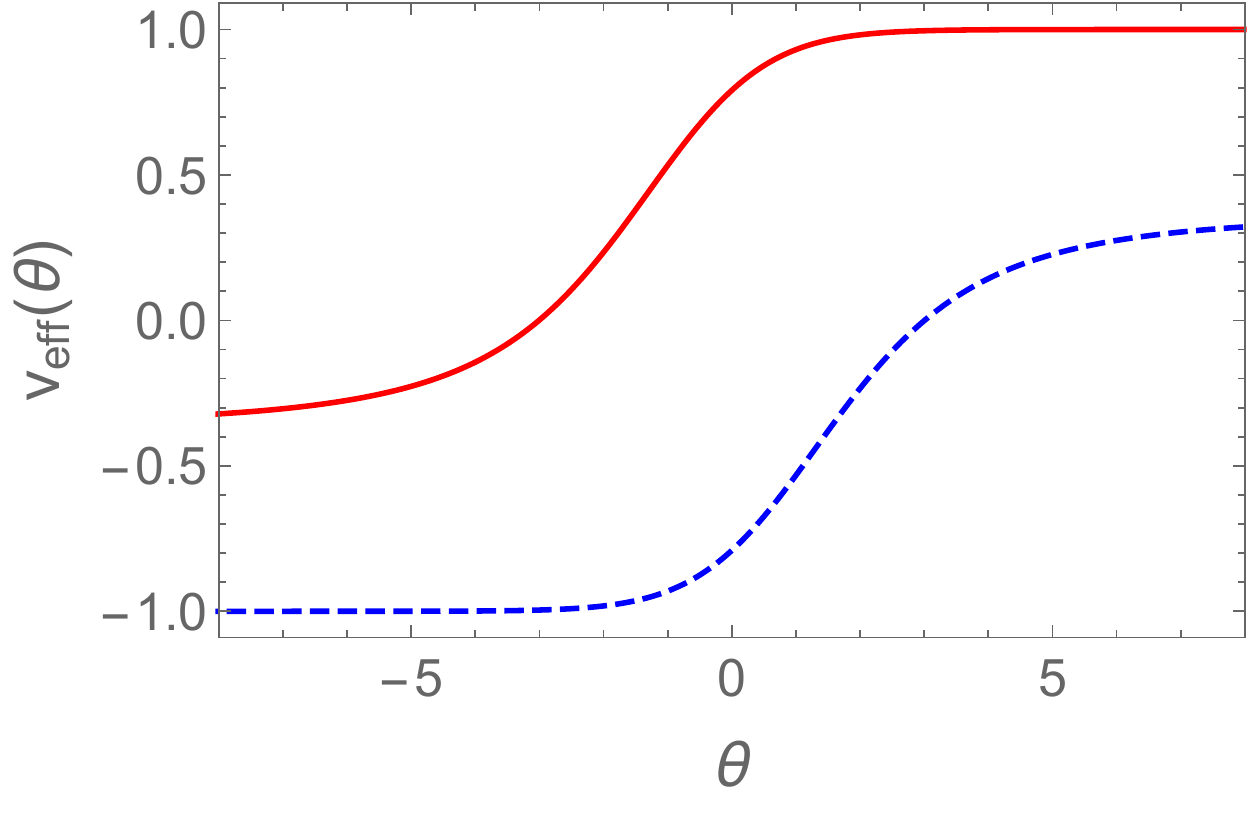} 
\par\end{centering}
}
\par\end{centering}
\caption{{\footnotesize{}\label{fig:PottsnAndvEff}(a) Filling function of
right-moving (red continuous curve) and left-moving particles (blue
dashed curve) and (b) the effective velocities of the right-moving
(red continuous curve) and left-moving (blue dashed curve) particles
in the $D_{4}$ flow in a thermal state with $T/M=1.4$.}}
\end{figure}

\selectlanguage{english}%
For thermal states the filling function of the RM and LM particles
are of a very peculiar form in all massless integrable models: they
are kinks related by $n_{RM}(\th)=n_{LM}(-\th)$. Focusing on $n_{RM}$
for $\th\rightarrow-\infty$ its value is a constant and for $\th\rightarrow\infty$
it goes to zero as $\exp\left(-M(\exp\th)/(2T)\right)$. This behaviour
is illustrated on figure  \ref{fig:PottsnAndvEff} for the case of
the $D_{4}$ massless model.

\selectlanguage{american}%

\begin{figure}[H]
\begin{centering}
\subfloat[$v_{M1}^{\text{eff}}(\protect\th)$ and $v_{M2}^{\text{eff}}(\protect\th)$]{\begin{centering}
\includegraphics[width=0.45\columnwidth]{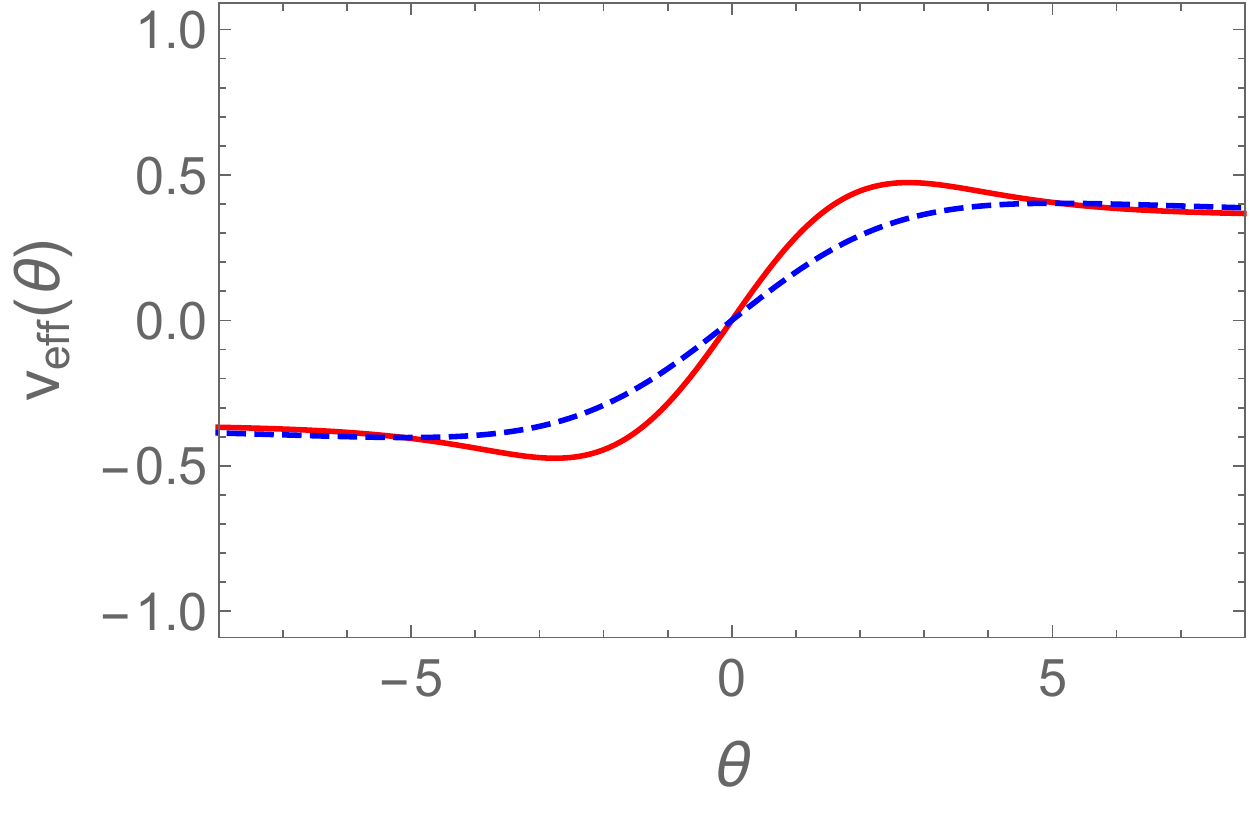} 
\par\end{centering}
}\subfloat[$\tilde{c}(T)$ as a function of $-\log T/M$]{\begin{centering}
\includegraphics[width=0.45\columnwidth]{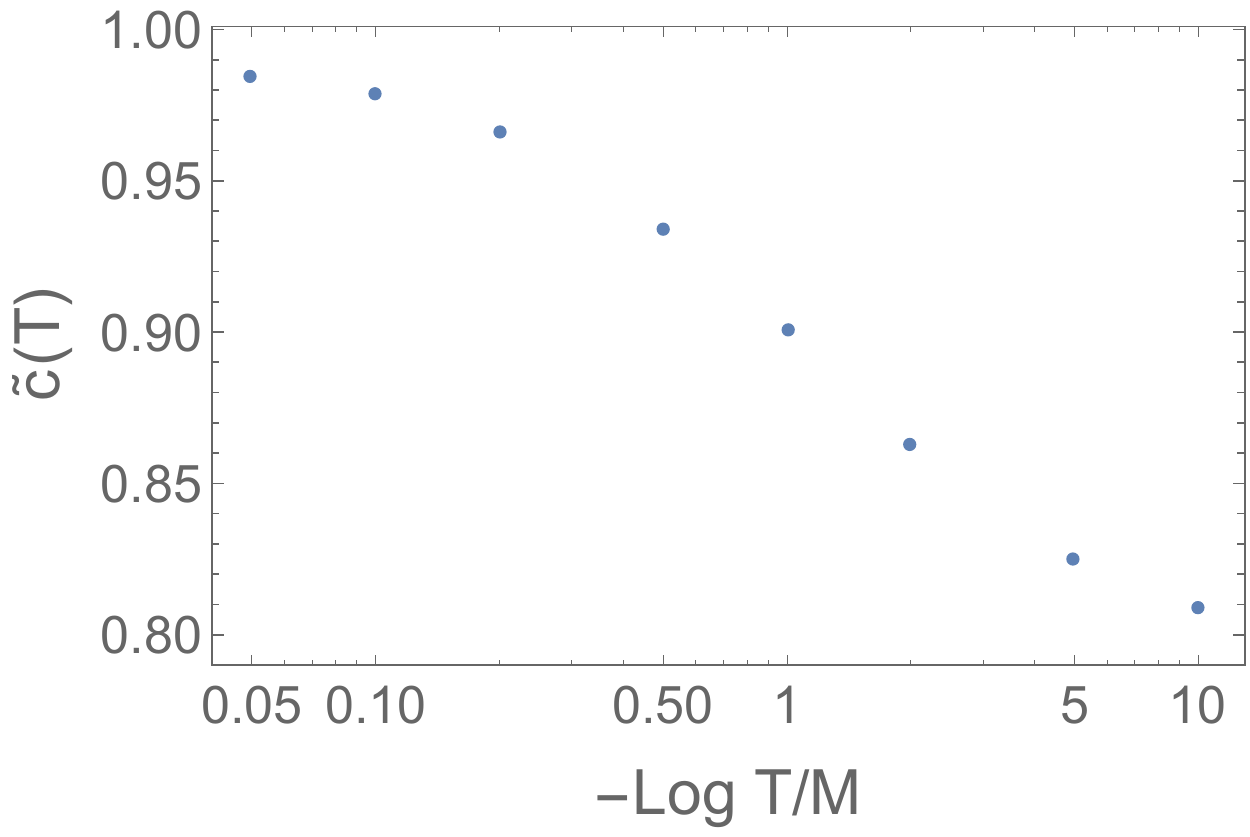} 
\par\end{centering}
}
\par\end{centering}
\caption{{\footnotesize{}\label{fig:PottsvEffMagnonsAndCTheorem}(a) the effective
velocities of the first (red continuous curve) and second (blue dashed
curve) magnons in the $D_{4}$ flow in a thermal state with $T/M=1.4$.
The first and the second magnons correspond to the first and second
nodes in the $D_{4}$ Dynkin diagram. (b) The c-theorem for the $D_{4}$
massless model. The $\widetilde{c}$ function interpolates between
the UV and IR central charges whose values are 1 and 0.8.}}
\end{figure}

Although the bare velocities of the RM and LM particles are $\pm1$
in units of the speed of light, in highly-excited states such as thermal
states the effective velocities are different from $\pm1$ due to
interactions which results in non-trivial kinetics in the GHD setting.
The TBA description of the $D_{4}$ flow also includes two magnonic
particles, whose effective velocities can even be non-monotonous functions
of the rapidity as demonstrated by figure \ref{fig:PottsvEffMagnonsAndCTheorem}.

Finally, to conclude this section subfigure (b) in figure \ref{fig:PottsvEffMagnonsAndCTheorem}
shows the c-function as a function $-\log T/M$ for the $D_{4}$ massless
model.
\selectlanguage{british}%

\section{Hydrodynamics of the tricritical to critical Ising flow\label{sec:A2Hydro}}

In this section we focus on the Euler scale hydrodynamics of the $A_{2}$
flow after a bipartite quench. The $A_{2}$ model, whose UV and IR
limiting theories are the tricritical and critical Ising CFTs \cite{AnZamo123,AnZamo2},
is the simplest massless flow with a RM and LM particle in the TBA
equations. By solving first the corresponding TBA equations \eqref{TBAAnDn}
with \eqref{DrivingTermTBAAnDn} for the left and right filling functions
and then the final form of the GHD equations \eqref{GHDBasicEq1XiSolution2}
and \eqref{vEffXi}, the ray-dependent density and current profiles
are easy to obtain from eqs. \eqref{qAv} and \eqref{jAv}. We start
our analysis by first discussing some peculiar features of these profiles
such as the existence of extended plateaux and the bounds on the currents
and densities which turn out to be quite generic for all massless
models studied in this work.

\subsection{Plateaux in the profiles and the dynamical central charges}

\begin{figure}[H]
\begin{centering}
\subfloat[\foreignlanguage{british}{$M^{-2}q_{e}(\xi)$}]{\begin{centering}
\includegraphics[width=0.4\columnwidth]{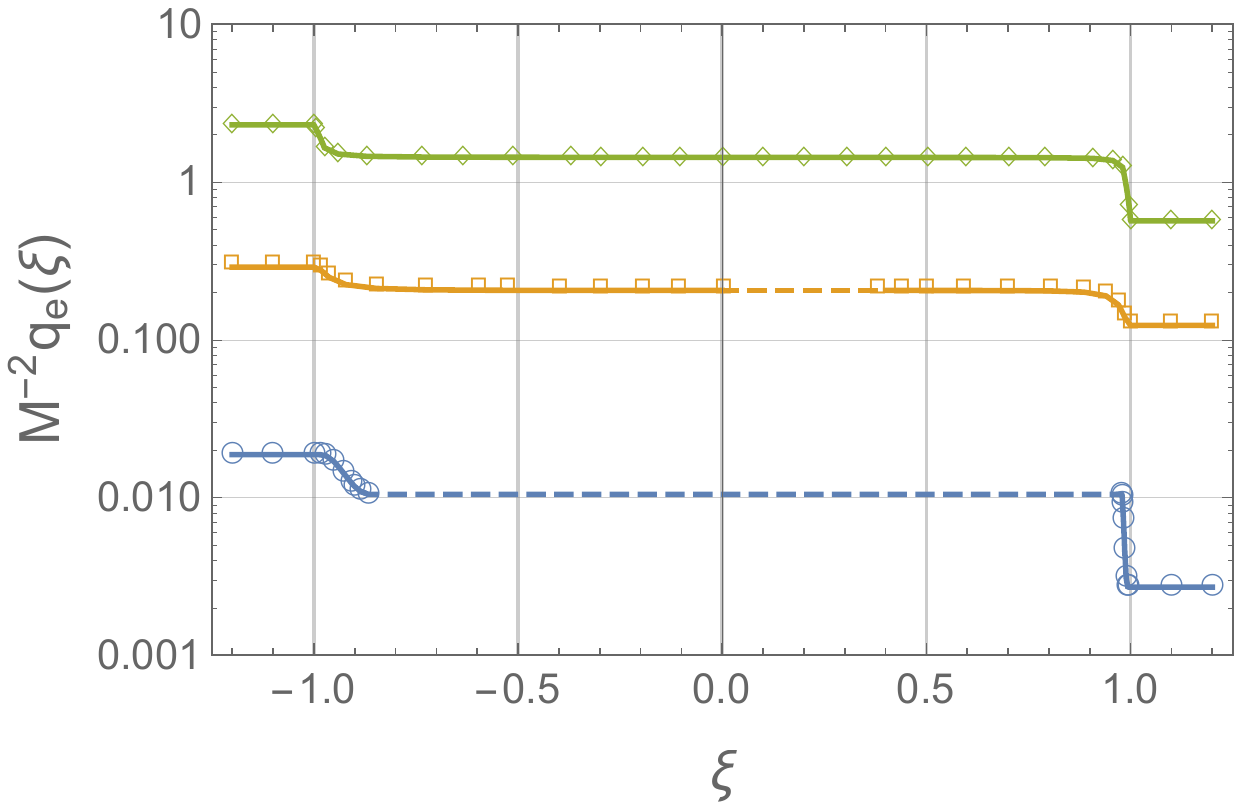} 
\par\end{centering}
}\subfloat[\foreignlanguage{british}{$M^{-2}j_{e}(\xi)$}]{\begin{centering}
\includegraphics[width=0.4\columnwidth]{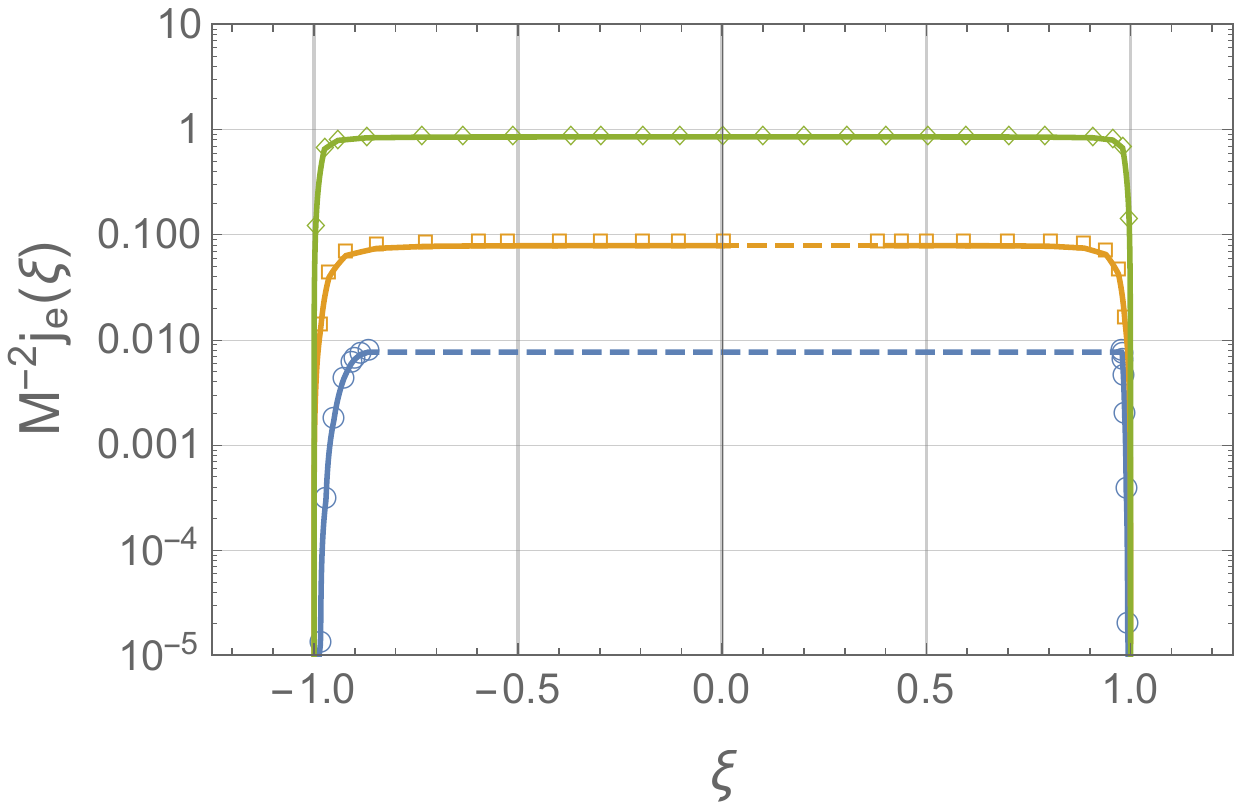} 
\par\end{centering}
}
\par\end{centering}
\begin{centering}
\subfloat[\foreignlanguage{british}{$M^{-2}q_{p}(\xi)$}]{\begin{centering}
\includegraphics[width=0.4\columnwidth]{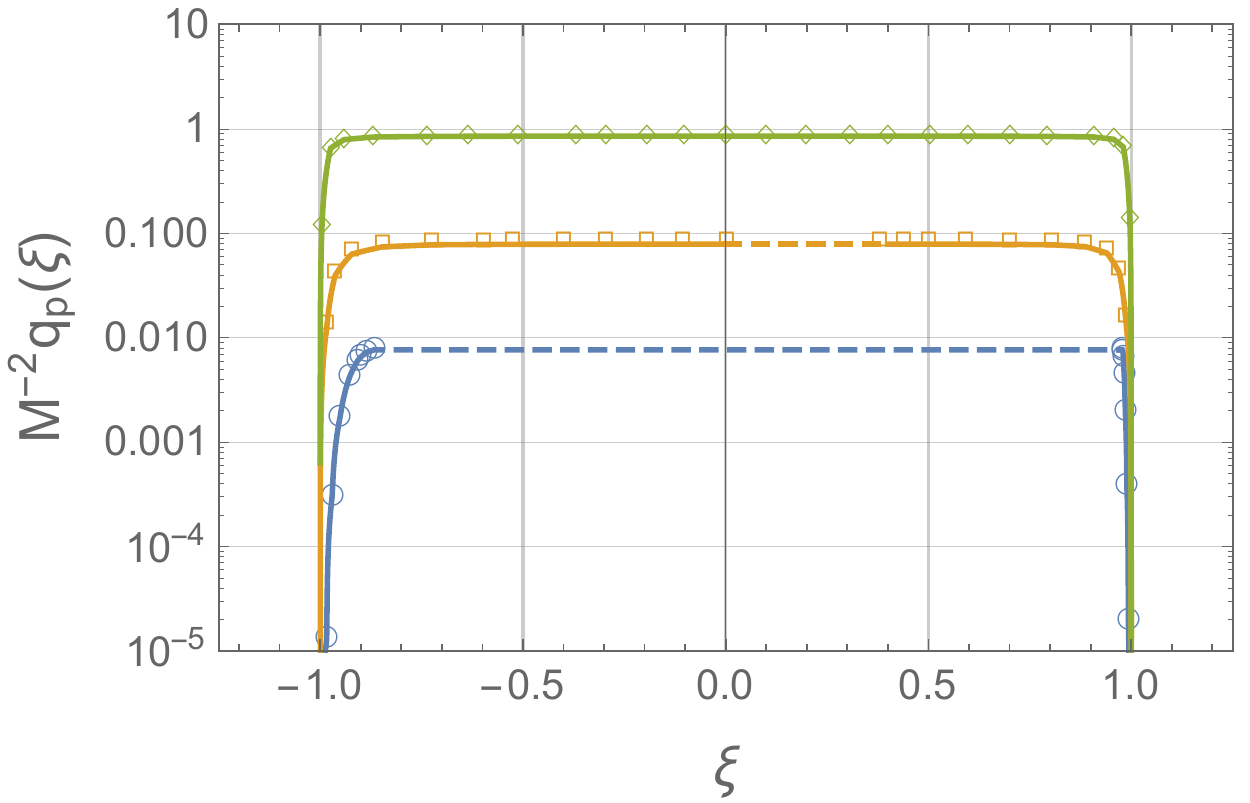} 
\par\end{centering}
}\subfloat[\foreignlanguage{british}{$M^{-2}j_{p}(\xi)$}]{\begin{centering}
\includegraphics[width=0.4\columnwidth]{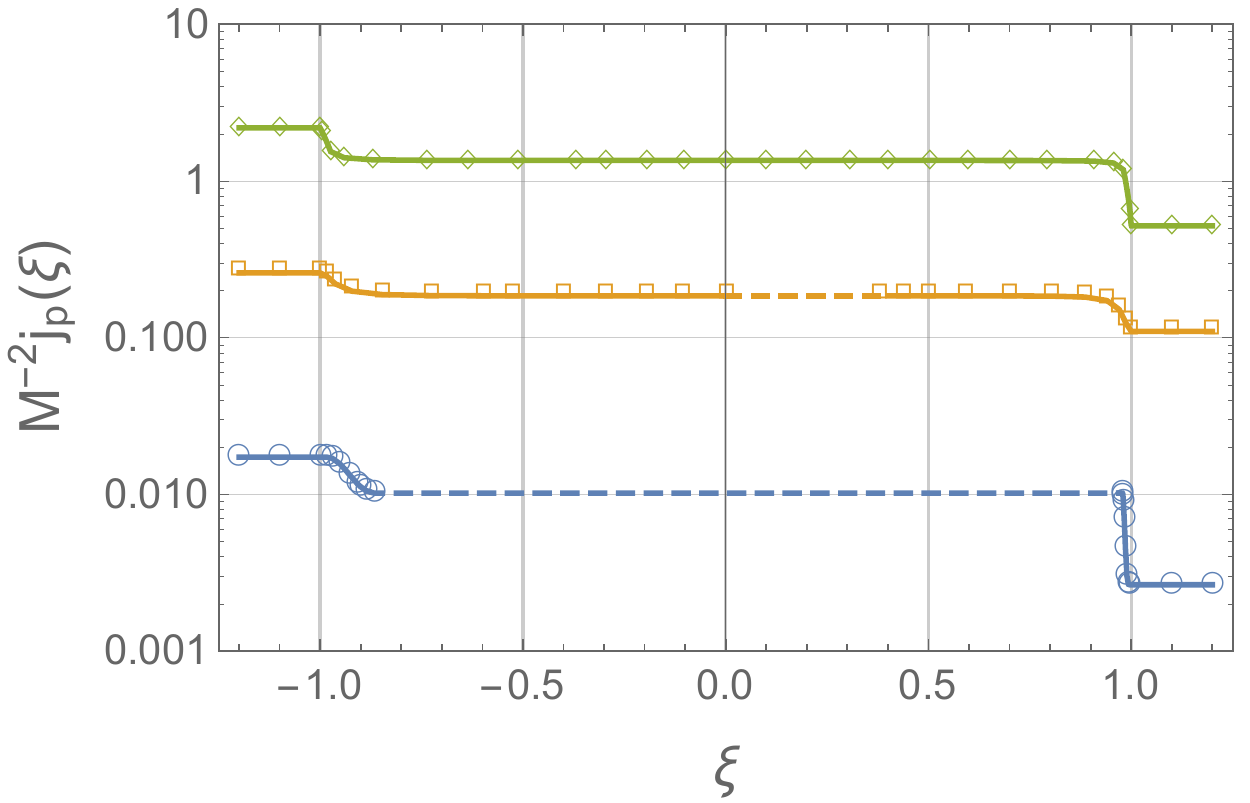} 
\par\end{centering}
}
\par\end{centering}
\caption{{\footnotesize{}\label{fig:A2EP}Ray-dependent (a) energy density
$q_{e}$, (b) energy current $j_{e}$, (c) momentum density $q_{p}$
and (d) momentum current $j_{p}$ in the $A_{2}$ tricritical to critical
Ising flow after bipartite quenches at the Euler scale. The green
curve with diamonds corresponds to left and right initial temperatures
$T_{l}=2.5M$ and $T_{r}=1.25M$, the orange curve with squares to
$T_{l}=0.9M$ and $T_{r}=0.6M$ and the blue curve with circles to
$T_{l}=0.25M$ and $T_{r}=0.1M$. The discrete points in the plots
indicated by the plotmarkers are obtained by the numerical solution
of GHD equations, the continuous curves are first order interpolations.
The dashed part of the curves indicates the region of constant densities
and currents. Due to relativistic invariance, $j_{e}=q_{p}$.}}
\end{figure}

\selectlanguage{american}%
Figure \ref{fig:A2EP} shows the ray-dependent energy and momentum
densities and currents and one peculiar feature of these quantities
are the extended plateaux in the currents and densities. As discussed
in the introduction, the Euler scale behaviour of these quantities
in CFTs are described by exactly flat plateaux and discontinuous jumps
at $\xi=\pm1$. Since both at low and high energies, i.e. low and
high temperatures the limiting theories of massless models are CFTs,
one should not be surprised about the emergence of similar plateaux
in the massless flows. It is slightly more interesting that the plateaux
still exist outside the conformal regimes where the left and right
temperatures are close to the inverse of the crossover scale $M^{-1}$,
corresponding to the massless flow being far away from the conformal
limits.

The existence of the plateaux in this regime can be understood from
the TBA equations using that in thermal states the filling functions
$n_{RM}$ and $n_{LM}$ are kinks. Whereas these kinks possess some
structure, they can be roughly regarded as Heaviside theta functions.
The transition from a non-zero value to zero takes place at a temperature
dependent rapidity, which we can denote as $\th_{RM/LM}^{(r)/(l)}$
depending on the species and its leading order temperature dependence
is logarithmic (c.f. eqs. \eqref{eq:FillingFunction}, \eqref{TBAAnDn}
and \eqref{DrivingTermTBAAnDn}). In the bipartite quench protocol
the left and right filling functions are joined together according
to \eqref{GHDBasicEq1XiSolution2} and the transition from the right
filling function to the left in $n_{RM}(\th,\xi)=n_{RM}(\th,\th_{RM})$
and $n_{LM}(\th,\xi)=n_{LM}(\th,\th_{LM})$ takes place at a rapidity
$\th_{RM}$ and $\th_{LM}$ respectively. When the rapidities $\th_{RM}$
and $\th_{LM}$ sweep through the real interval and the filling functions
are approximated by theta functions, $n_{RM}(\th,\th_{RM})$ and $n_{LM}(\th,\th_{LM})$
are either the left or right filling functions apart from the case
when $\th_{RM}\in[\th_{RM}^{(r)},\th_{RM}^{(l)}]$ and $\th_{LM}\in[\th_{LM}^{(l)},\th_{LM}^{(r)}]$.
For sufficiently high temperature with a not too large difference
between the left and right values these rapidity intervals are very
short. Although the ray $\xi$ is a generally function of both $\th_{RM}$
and $\th_{LM}$, the regions of $\xi$ where the right-left transition
occurs in $n_{RM}(\th,\xi)$ and $n_{LM}(\th,\xi)$ remain narrow.
These regions correspond to the sharp transitions in the density and
current profiles, whereas for the the left, middle and right plateaux
$n_{RM}(\th,\xi)=n_{RM}^{(l)}(\th)$ , $n_{LM}(\th,\xi)=n_{LM}^{(l)}(\th)$;
$n_{RM}(\th,\xi)=n_{RM}^{(l)}(\th)$ , $n_{LM}(\th,\xi)=n_{LM}^{(r)}(\th)$
and $n_{RM}(\th,\xi)=n_{RM}^{(r)}(\th)$ , $n_{LM}(\th,\xi)=n_{LM}^{(r)}(\th)$
respectively. Of course, the left and right filling functions are
not exact Heaviside theta functions and hence the middle plateaux
are not exactly flat. 

Our argument seems to be invalid when the difference between the left
and right temperatures is large. In such a case, however, the high
energy particles almost exclusively originate from the left side and
hence the UV limiting CFT dominates the dynamics as supported by table
\ref{tab:A2DynamicalC} discussed shortly. In summary, irrespectively
of the magnitude of the left and right temperatures, the plateaux
in the energy and momentum density and current profiles are a generic
feature the Euler scale hydrodynamics of the $A_{2}$ massless flow.
As demonstrated in the next section, the extended plateaux also occur
in other massless integrable theories.

\selectlanguage{british}%
\begin{table}[H]
\selectlanguage{american}%
\begin{tabular}{|c|c|c|c|c|c|c|c|}
\hline 
\selectlanguage{british}%
$T_{l}/M$\selectlanguage{american}%
 & \selectlanguage{british}%
$T_{r}/M$\selectlanguage{american}%
 & \selectlanguage{british}%
$\tilde{c}(T_{l})$\selectlanguage{american}%
 & \selectlanguage{british}%
$\tilde{c}(T_{r})$\selectlanguage{american}%
 & \selectlanguage{british}%
$\tilde{c}_{j_{e}}(T_{l},T_{r})$\selectlanguage{american}%
 & \selectlanguage{british}%
$\tilde{c}_{q_{e}}(T_{l},T_{r})$\selectlanguage{american}%
 & \selectlanguage{british}%
$\tilde{c}_{j_{p}}(T_{l},T_{r})$\selectlanguage{american}%
 & \selectlanguage{british}%
$\tilde{c}_{j_{e}}^{lb}(T_{l},T_{r})$\selectlanguage{american}%
\tabularnewline
\hline 
\hline 
\selectlanguage{british}%
0.25\selectlanguage{american}%
 & \selectlanguage{british}%
0.1\selectlanguage{american}%
 & \selectlanguage{british}%
0.527539\selectlanguage{american}%
 & \selectlanguage{british}%
0.505141\selectlanguage{american}%
 & \selectlanguage{british}%
0.555744\selectlanguage{american}%
 & \selectlanguage{british}%
0.551405\selectlanguage{american}%
 & \selectlanguage{british}%
0.535171\selectlanguage{american}%
 & \selectlanguage{british}%
0.531832\selectlanguage{american}%
\tabularnewline
\hline 
\selectlanguage{british}%
0.9\selectlanguage{american}%
 & \selectlanguage{british}%
0.1\selectlanguage{american}%
 & \selectlanguage{british}%
0.610694\selectlanguage{american}%
 & \selectlanguage{british}%
0.505141\selectlanguage{american}%
 & \selectlanguage{british}%
0.642663\selectlanguage{american}%
 & \selectlanguage{british}%
0.64557\selectlanguage{american}%
 & \selectlanguage{british}%
0.633447\selectlanguage{american}%
 & \selectlanguage{british}%
0.634259\selectlanguage{american}%
\tabularnewline
\hline 
\selectlanguage{british}%
0.9\selectlanguage{american}%
 & \selectlanguage{british}%
0.6\selectlanguage{american}%
 & \selectlanguage{british}%
0.610694\selectlanguage{american}%
 & \selectlanguage{british}%
0.581244\selectlanguage{american}%
 & \selectlanguage{british}%
0.666046\selectlanguage{american}%
 & \selectlanguage{british}%
0.670596\selectlanguage{american}%
 & \selectlanguage{british}%
0.603226\selectlanguage{american}%
 & \selectlanguage{british}%
0.612022\selectlanguage{american}%
\tabularnewline
\hline 
\selectlanguage{british}%
0.95\selectlanguage{american}%
 & \selectlanguage{british}%
0.9\selectlanguage{american}%
 & \selectlanguage{british}%
0.614468\selectlanguage{american}%
 & \selectlanguage{british}%
0.610694\selectlanguage{american}%
 & \selectlanguage{british}%
0.675198\selectlanguage{american}%
 & \selectlanguage{british}%
0.682408\selectlanguage{american}%
 & \selectlanguage{british}%
0.612715\selectlanguage{american}%
 & \selectlanguage{british}%
0.64752\selectlanguage{american}%
\tabularnewline
\hline 
\selectlanguage{british}%
1.4\selectlanguage{american}%
 & \selectlanguage{british}%
0.9\selectlanguage{american}%
 & \selectlanguage{british}%
0.639252\selectlanguage{american}%
 & \selectlanguage{british}%
0.610694\selectlanguage{american}%
 & \selectlanguage{british}%
0.682109\selectlanguage{american}%
 & \selectlanguage{british}%
0.691041\selectlanguage{american}%
 & \selectlanguage{british}%
0.631924\selectlanguage{american}%
 & \selectlanguage{british}%
0.659371\selectlanguage{american}%
\tabularnewline
\hline 
\selectlanguage{british}%
2.5\selectlanguage{american}%
 & \selectlanguage{british}%
1.25\selectlanguage{american}%
 & \selectlanguage{british}%
0.666775\selectlanguage{american}%
 & \selectlanguage{british}%
0.632483\selectlanguage{american}%
 & \selectlanguage{british}%
0.691766\selectlanguage{american}%
 & \selectlanguage{british}%
0.700926\selectlanguage{american}%
 & \selectlanguage{british}%
0.66099\selectlanguage{american}%
 & \selectlanguage{british}%
0.678208\selectlanguage{american}%
\tabularnewline
\hline 
\selectlanguage{british}%
6\selectlanguage{american}%
 & \selectlanguage{british}%
0.2\selectlanguage{american}%
 & \selectlanguage{british}%
0.68853\selectlanguage{american}%
 & \selectlanguage{british}%
0.519057\selectlanguage{american}%
 & \selectlanguage{british}%
0.695114\selectlanguage{american}%
 & \selectlanguage{british}%
0.698746\selectlanguage{american}%
 & \selectlanguage{british}%
0.69212\selectlanguage{american}%
 & \selectlanguage{british}%
0.68872\selectlanguage{american}%
\tabularnewline
\hline 
\selectlanguage{british}%
6\selectlanguage{american}%
 & \selectlanguage{british}%
5\selectlanguage{american}%
 & \selectlanguage{british}%
0.68853\selectlanguage{american}%
 & \selectlanguage{british}%
0.685515\selectlanguage{american}%
 & \selectlanguage{british}%
0.698716\selectlanguage{american}%
 & \selectlanguage{british}%
0.70355\selectlanguage{american}%
 & \selectlanguage{british}%
0.68731\selectlanguage{american}%
 & \selectlanguage{british}%
0.695382\selectlanguage{american}%
\tabularnewline
\hline 
\selectlanguage{british}%
6.5\selectlanguage{american}%
 & \selectlanguage{british}%
6\selectlanguage{american}%
 & \selectlanguage{british}%
0.689664\selectlanguage{american}%
 & \selectlanguage{british}%
0.68853\selectlanguage{american}%
 & \selectlanguage{british}%
0.69897\selectlanguage{american}%
 & \selectlanguage{british}%
0.703259\selectlanguage{american}%
 & \selectlanguage{british}%
0.689146\selectlanguage{american}%
 & \selectlanguage{british}%
0.696195\selectlanguage{american}%
\tabularnewline
\hline 
\selectlanguage{british}%
9\selectlanguage{american}%
 & \selectlanguage{british}%
6\selectlanguage{american}%
 & \selectlanguage{british}%
0.693292\selectlanguage{american}%
 & \selectlanguage{british}%
0.68853\selectlanguage{american}%
 & \selectlanguage{british}%
0.699239\selectlanguage{american}%
 & \selectlanguage{british}%
0.702619\selectlanguage{american}%
 & \selectlanguage{british}%
0.691869\selectlanguage{american}%
 & \selectlanguage{british}%
0.697102\selectlanguage{american}%
\tabularnewline
\hline 
\selectlanguage{british}%
9\selectlanguage{american}%
 & \selectlanguage{british}%
8.5\selectlanguage{american}%
 & \selectlanguage{british}%
0.693292\selectlanguage{american}%
 & \selectlanguage{british}%
0.692755\selectlanguage{american}%
 & \selectlanguage{british}%
0.69942\selectlanguage{american}%
 & \selectlanguage{british}%
0.702409\selectlanguage{american}%
 & \selectlanguage{british}%
0.693041\selectlanguage{american}%
 & \selectlanguage{british}%
0.697723\selectlanguage{american}%
\tabularnewline
\hline 
\end{tabular}

\selectlanguage{british}%
\caption{{\footnotesize{}\label{tab:A2DynamicalC}The effective central charges
for (left and right) thermal states and the dynamical central charges
for the partitioning protocol for various left and right temperatures
in the $A_{2}$ massless flow, where $c_{UV}=0.7$ and $c_{IR}=0.5$. }}
\end{table}

To discuss a second characteristic feature, we first define a dynamical
central charge inspired by \eqref{eq:EqCFT1} by writing

\begin{equation}
j_{e}(0)=q_{p}(0)=\frac{\tilde{c}_{j_{e}}(T_{l},T_{r})}{12\pi}\left(T_{l}^{2}-T_{r}^{2}\right)\,.\label{eq:DynamicalCentralCharge}
\end{equation}
The dynamical central charge $\tilde{c}_{j_{e}}$ (together with its
counterparts for the cumulants of the transferred energy) was already
defined and used in \cite{DoyonWrongHydro} for non-equilibrium situations.
Calculating the energy current (or momentum density) at the ray $\xi=0$
, $\tilde{c}_{j_{e}}$ can easily be obtained knowing the left and
right temperatures. Whereas in CFTs the energy current $j_{e}(\xi)$
is constant if $\xi\subset(-1,1)$, for the massless flows $j_{e}(\xi)$
usually exhibits a non-trivial $\xi$-dependence, therefore we must
specify at which ray $\xi$ $j_{e}$ is to take in order to define
$\tilde{c}_{j_{e}}$. It seems natural to choose $\xi$ at which $j_{e}$
is maximal, which occurs at $\xi=0$ as a consequence of the continuity
equation \eqref{GHDContEqXi}. Moreover, taking the limit $t\rightarrow\infty$
at any fixed position $x$ gives the NESS, which also corresponds
to $\xi=0$.

In table \ref{tab:A2DynamicalC} the values of $\tilde{c}_{j_{e}}$
are collected for various left and right temperatures together with
the effective central charges for the left and right thermal states.
Based on the data we propose the following conjecture for the dynamical
central charge in bipartite quenches with left and right initial thermal
states:

\begin{equation}
c_{IR}\leq\tilde{c}(\max\left(T_{r},T_{l}\right))\leq\tilde{c}_{j_{e}}(T_{l},T_{r})\leq c_{UV}\,.\label{eq:ConjectureDynamicalC}
\end{equation}
This is an interesting property of the dynamical central charge which
\eqref{eq:ConjectureDynamicalC} also implies upper and lower bounds
on the maximum of the ray-dependent energy current or equivalently
the current in the NESS, and an upper bound for the energy current
at any ray. In \cite{DoyonLowerBounds} a lower bound for the steady-state
energy current was given, which reads in our case

\begin{equation}
j_{e}(0)\geq\frac{\langle{\tt j}_{p}\rangle_{\beta_{l}}-\langle{\tt j}_{p}\rangle_{\beta_{r}}}{2}\,,\label{eq:DoyonLowerBound}
\end{equation}
where the averages of the momentum current are taken in the left and
right thermal states. When a lower bound on $j_{e}(0)$ is considered,
this inequality \eqref{eq:DoyonLowerBound} turns out to be more restrictive
than \eqref{eq:ConjectureDynamicalC} in almost all cases, which is
demonstrated by table \ref{tab:A2DynamicalC} too. For better transparency
we define

\begin{equation}
\tilde{c}_{j_{e}}^{lb}=\frac{6}{\pi}\frac{\langle{\tt j}_{p}\rangle_{\beta_{l}}-\langle{\tt j}_{p}\rangle_{\beta_{r}}}{\left(T_{l}^{2}-T_{r}^{2}\right)}
\end{equation}
to compare the dynamical central charge with. 

A trivial consequence of \eqref{eq:ConjectureDynamicalC} is the bound 

\begin{equation}
c_{IR}\leq\tilde{c}_{j_{e}}(T_{l},T_{r})\leq c_{UV}\,,\label{eq:ConjectureDynamicalC-1}
\end{equation}
where the bounds are now independent of the temperatures, and the
lower bound is less strict than the one obtained in \cite{DoyonLowerBounds}.
This means that a simple estimate for $j_{e}(0)$ is given by 

\begin{equation}
\frac{\pi c_{IR}}{12}\left(T_{l}^{2}-T_{r}^{2}\right)\leq j_{e}(0)=\max j_{e}\leq\frac{\pi c_{UV}}{12}\left(T_{l}^{2}-T_{r}^{2}\right)\,,
\end{equation}
and hence the the whole Euler scale energy current is always bounded
by $\frac{\pi c_{UV}}{12}\left(T_{l}^{2}-T_{r}^{2}\right)$. 

The dynamical central charge $\tilde{c}_{j_{e}}$ has another nice
property; similarly to the effective central charge, this quantity
is a monotonously increasing function of the energy scale. In more
precise terms 

\begin{equation}
\begin{split}\tilde{c}_{j_{e}}(T_{l}^{(1)},T_{r}^{(1)})\geq\tilde{c}_{j_{e}}(T_{l}^{(2)},T_{r}^{(2)})\: & \quad\text{ if }\quad\;\max\left(T_{l}^{(1)},T_{r}^{(1)}\right)\geq\max\left(T_{l}^{(2)},T_{r}^{(2)}\right)\\
 & \quad\text{ and }\;\min\left(T_{l}^{(1)},T_{r}^{(1)}\right)\geq\min\left(T_{l}^{(2)},T_{r}^{(2)}\right)
\end{split}
\label{eq:Monoton1}
\end{equation}
and

\begin{equation}
\begin{split}\tilde{c}_{j_{e}}(T_{l}^{(1)},T_{r}^{(1)})\geq\tilde{c}_{j_{e}}(T_{l}^{(2)},T_{r}^{(2)})\: & \quad\text{ if }\quad\;\min\left((T_{l}^{(1)},T_{r}^{(1)}\right)\geq\min\left(T_{l}^{(2)},T_{r}^{(2)}\right)\\
 & \quad\text{ and }\;\max\left(T_{l}^{(1)},T_{r}^{(1)}\right)=\max\left(T_{l}^{(2)},T_{r}^{(2)}\right)\,.
\end{split}
\label{eq:Monoton2}
\end{equation}

A monotonic behaviour of $\tilde{c}_{j_{e}}$ was first pointed out
in \cite{DoyonWrongHydro}. However, the validity of the approach
used to determine the NESS in that particular work was not justified
by later comparison with the GHD \cite{GHDFundation}, although the
difference between the numerical values was small. Our analysis relying
on the GHD approach now gives solid evidence for the monotonicity
of $\tilde{c}_{j_{e}}$ in terms of \eqref{eq:Monoton1} and \eqref{eq:Monoton2}.

Based on eq. \eqref{eq:EqCFT1}, nevertheless, an effective central
charge can be defined not only from $j_{e}$, but also from $q_{e}$
and $j_{p}$, which are not equal due to the lack of exact conformal
symmetry in massless models. Again, it seems reasonable to chose their
value at the ray $\xi=0$, which corresponds to the NESS, but it is
important to keep in mind that for these quantities their maximum
does not occur at $\xi=0$. Defining the corresponding dynamical central
charges denoted them by $\tilde{c}_{q_{e}}$ and $\tilde{c}_{j_{p}}$
their numerical values can be determined as seen in table \ref{tab:A2DynamicalC}.
As indicated by table \ref{tab:A2DynamicalC} similar observation
can be made to the case of $\tilde{c}_{j_{e}}$, which we summarise
as follows: for $\tilde{c}_{j_{p}}$

\begin{equation}
c_{IR}\leq\tilde{c}_{j_{p}}(T_{l},T_{r})\leq\tilde{c}_{j_{e}}(T_{l},T_{r})\label{eq:ConjectureDynamicalCJp}
\end{equation}
holds, from which it also follows that 

\begin{equation}
c_{IR}\leq\tilde{c}_{j_{p}}(T_{l},T_{r})\leq c_{UV}\,.
\end{equation}
On the contrary for $\tilde{c}_{q_{e}}$ we find that

\begin{equation}
c_{IR}\leq\tilde{c}(\max\left(T_{l},T_{r}\right))\leq\tilde{c}_{q_{e}}(T_{l},T_{r})\apprle c_{UV}\,.\label{eq:ConjectureDynamicalCQe}
\end{equation}
In this formula the symbol $\apprle$ before $c_{UV}$ seems very
unnatural first. Even though the problematic numerical values for
$\tilde{c}_{q_{e}}$ (i.e. those slightly large than $0.7$) seem
to be stable against varying the parameters in the numerical solutions
(such as the number of iterations or discretisation points etc.) we
cannot exclude the possibility of numerical errors and therefore leave
this problem open. Nevertheless it is important to remember that $q_{e}$
has no global extremum at $\xi=0$ and in the rest of the paper we
present an interesting argument stating that unlike for currents such
an anomalous behaviour is not prohibited by physical principles for
densities (apart from the $T^{(l)},T^{(r)}\rightarrow0$ and $T^{(l)},T^{(r)}\rightarrow\infty$
limits where the CFT description becomes exact).

In the next section it is also demonstrated that the conjectures \eqref{eq:ConjectureDynamicalC},
\eqref{eq:ConjectureDynamicalCJp} and \eqref{eq:ConjectureDynamicalCQe}
and the monotonicity property \eqref{eq:Monoton1}, \eqref{eq:Monoton2}
for $\tilde{c}_{j_{e}}$ remain valid for the other flows of the $A_{n}$
and $D_{n}$ family together with the extended plateaux in the current
and density profiles.

\subsection{Low energy behaviour and constant regions in the density/current
profiles in the $A_{2}$ flow}

As indicated by figure \ref{tab:A2DynamicalC}, for low and intermediate
left and right temperatures, there are regions of the ray $\xi$ where
the densities and currents have constant values. To understand the
emergence of such regions, it is useful to remember that the bare
velocities of the RM and LM particles are $\pm1$ and in macro-states
(either homogeneous or inhomogeneous) the interactions changes the
range of the effective velocities from $\pm1$ to $[v_{RM}^{min},1]$
and $[-1,v_{LM}^{max}]$. The appearance of the flat regions in the
density and current profiles is related the fact that at not too high
temperatures these intervals do not overlap. This means that, according
to our discussion at the end of section \ref{subsec:BriefGHD}, 

\begin{equation}
\begin{split}\begin{split}n_{RM}(\th,\xi)=n_{RM}^{(l)}(\th)\\
n_{LM}(\th,\xi)=n_{LM}^{(r)}(\th)
\end{split}
 & \qquad\text{ if }\xi\in[v_{LM}^{max},v_{RM}^{min}]\end{split}
\,,\label{eq:nFlatRegion}
\end{equation}
which are independent of $\xi$ and consequently the current and density
profiles have no $\xi$-dependence too. It is a notable observation
that if $[v_{RM}^{min},1]\cap[-1,v_{LM}^{max}]=\{\}$ then $0\in[v_{LM}^{max},v_{RM}^{min}]$
always holds.

At very low temperatures it is also possible to treat the problem
by analytical means at least in two different ways. An obvious approach
is the low temperature expansion of the thermal TBA equations first.
If one is interested in the flat regions of the profiles the values
of the densities and currents are straightforwardly obtained using
again a low-temperature expansion in \eqref{qAv}, \eqref{jAv} and
\eqref{eq:Dressing} since \eqref{eq:nFlatRegion} can be exploited. 

Another approach is describing the low energy behaviour of the $A_{2}$
massless flow as an irrelevant perturbation of the IR CFT by $\int\!d^{2}x\,T\bar{T}$
\cite{KastorMartinecTTBar,ZamolodchikovTTBar}. In particular for
the $A_{2},$ model the low-energy effective Lagrangian reads \cite{MasslessFlow,KastorMartinecTTBar}

\begin{equation}
\mathcal{L}_{eff}=\psi\bar{\partial}\psi+\bar{\psi}\partial\bar{\psi}-\frac{4}{M^{2}}\left(\psi\partial\psi\right)\left(\bar{\psi}\bar{\partial}\bar{\psi}\right)+\ldots\,,\label{eq:LEff}
\end{equation}
where $\psi\bar{\partial}\psi+\bar{\psi}\partial\bar{\psi}$ is the
Lagrangian of the critical Ising field theory with $c=\frac{1}{2}$
and

\begin{equation}
\begin{split}T= & -\frac{1}{2}\psi\partial\psi\\
\bar{T}= & -\frac{1}{2}\bar{\psi}\bar{\partial}\bar{\psi}\,.
\end{split}
\end{equation}
The bipartite quench with left and right thermal heat reservoirs was
analysed in \cite{TTBarCFT} for perturbed CFTs with the action

\begin{equation}
S=S_{CFT}+g\int d^{2}x\,T\,\bar{T}\,.
\end{equation}
 In particular, the average of the energy and momentum densities and
currents are modified by the perturbation \cite{TTBarCFT} from eq.
\eqref{eq:EqCFT1} to

\begin{equation}
\begin{split}q_{e} & =\frac{c\pi}{12}\left[T_{l}^{2}+T_{r}^{2}-\frac{gc\pi}{12}\left(T_{l}^{4}+T_{r}^{4}+T_{l}^{2}T_{r}^{2}\right)\right]+\mathcal{O}(g^{2})\\
j_{e}=q_{p} & =\frac{c\pi}{12}\left[T_{l}^{2}\left(1-\frac{gc\pi}{12}T_{l}^{2}\right)-T_{r}^{2}\left(1-\frac{gc\pi}{12}T_{r}^{2}\right)\right]+\mathcal{O}(g^{2})\\
j_{p} & =\frac{c\pi}{12}\left[T_{l}^{2}+T_{r}^{2}-\frac{gc\pi}{12}\left(T_{l}^{4}+T_{r}^{4}-T_{l}^{2}T_{r}^{2}\right)\right]+\mathcal{O}(g^{2})
\end{split}
\label{eq:EPDensCurrTTBarCFT}
\end{equation}
and the velocities for the shock waves $v_{LM}^{max}$ and $v_{RM}^{min}$
are

\begin{equation}
\begin{split}v_{LM}^{max}= & -1-\frac{gc\pi}{12}T_{l}^{2}+\mathcal{O}(g^{2})\\
v_{RM}^{min}= & 1+\frac{gc\pi}{12}T_{r}^{2}+\mathcal{O}(g^{2})\,.
\end{split}
\label{eq:VmaxVminTTBarCFT}
\end{equation}

Substituting $-\frac{16}{M^{2}}$ for $g$ according to eqs. \eqref{eq:EPDensCurrTTBarCFT}
and \eqref{eq:VmaxVminTTBarCFT} and trading the small parameter to
$T^{2}$, the results obtained describe the low temperature Euler
scale behaviour of the $A_{2}$ flow:

\begin{equation}
\begin{split}q_{e} & =\frac{\pi}{24}\left[T_{l}^{2}+T_{r}^{2}+\frac{2\pi}{3}\left(T_{l}^{4}+T_{r}^{4}+T_{l}^{2}T_{r}^{2}\right)\right]+\mathcal{O}(T^{6})\\
j_{e}=q_{p} & =\frac{\pi}{24}\left[T_{l}^{2}\left(1+\frac{2\pi}{3}T_{l}^{2}\right)-T_{r}^{2}\left(1+\frac{2\pi}{3}T_{r}^{2}\right)\right]+\mathcal{O}(T^{6})\\
j_{p} & =\frac{\pi}{24}\left[T_{l}^{2}+T_{r}^{2}+\frac{2\pi}{3}\left(T_{l}^{4}+T_{r}^{4}-T_{l}^{2}T_{r}^{2}\right)\right]+\mathcal{O}(T^{6})
\end{split}
\end{equation}
for $\xi\in[v_{LM}^{max},v_{RM}^{min}]$, where 

\begin{equation}
\begin{split}v_{LM}^{max}= & -1+\frac{2\pi}{3}T_{l}^{2}+\mathcal{O}(T^{4})\\
v_{RM}^{min}= & 1-\frac{2\pi}{3}T_{r}^{2}+\mathcal{O}(T^{4})\,.
\end{split}
\end{equation}
An alternative derivation of these results is given in appendix \ref{subsec:Appendix-A-MasslessFF}
by performing the low temperature expansion of the TBA equations \eqref{eq:Dressing},
\eqref{qAv}, \eqref{jAv} and \eqref{TBA}. 

Finally, it is important to note that for the particular case of the
$A_{2}$ flow no shocks develop due to integrability contrary to the
general case of the $T\bar{T}$ perturbation \cite{TTBarCFT}. 

\section{Hydrodynamics of the $A_{n}$ and $D_{n}$ flows\label{sec:AnDnHydro}}

Now we turn to studying higher members of the $A_{n}$ and also $D_{n}$
type massless models, which correspond to the Dynkin diagrams in figure
\ref{figDynkin}. 

\begin{figure}[H]
\begin{centering}
\subfloat[\foreignlanguage{british}{$M^{-2}q_{e}(\xi)$}]{\begin{centering}
\includegraphics[width=0.4\columnwidth]{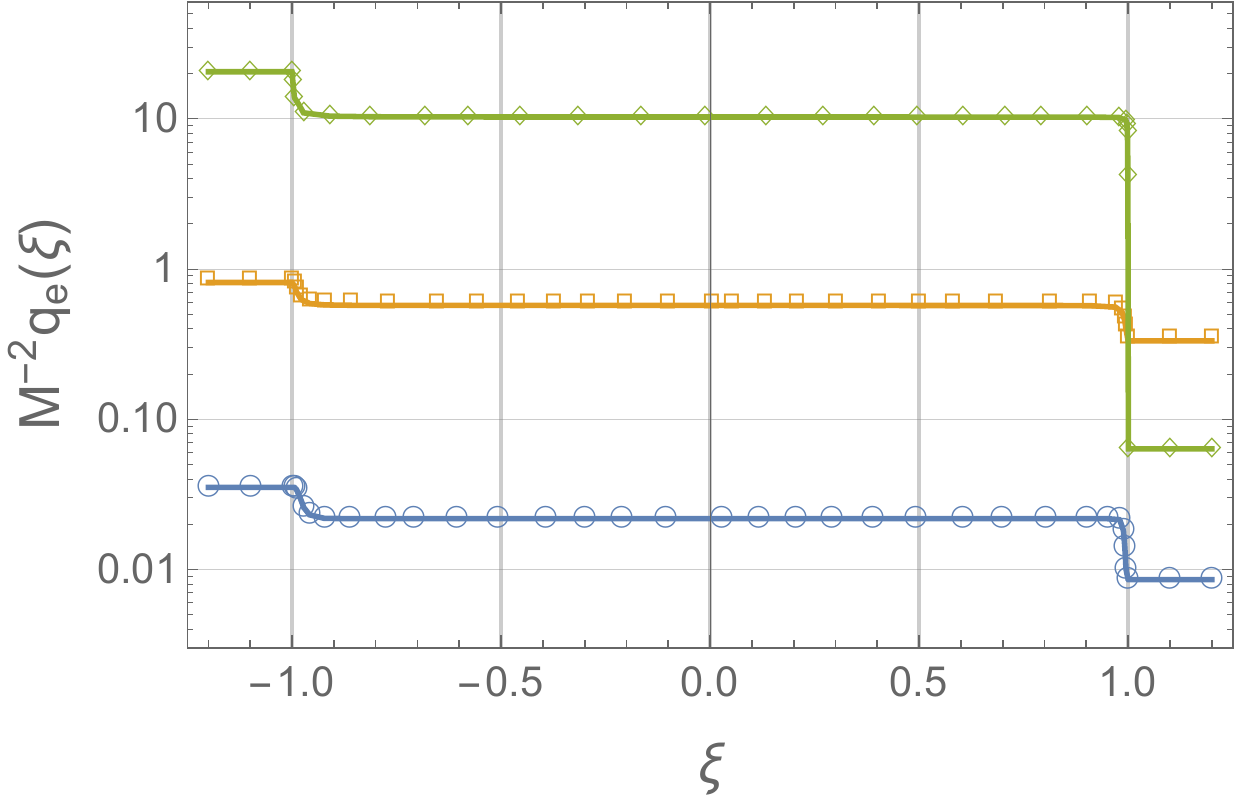} 
\par\end{centering}
}\subfloat[\foreignlanguage{british}{$M^{-2}j_{e}(\xi)$}]{\begin{centering}
\includegraphics[width=0.4083\columnwidth]{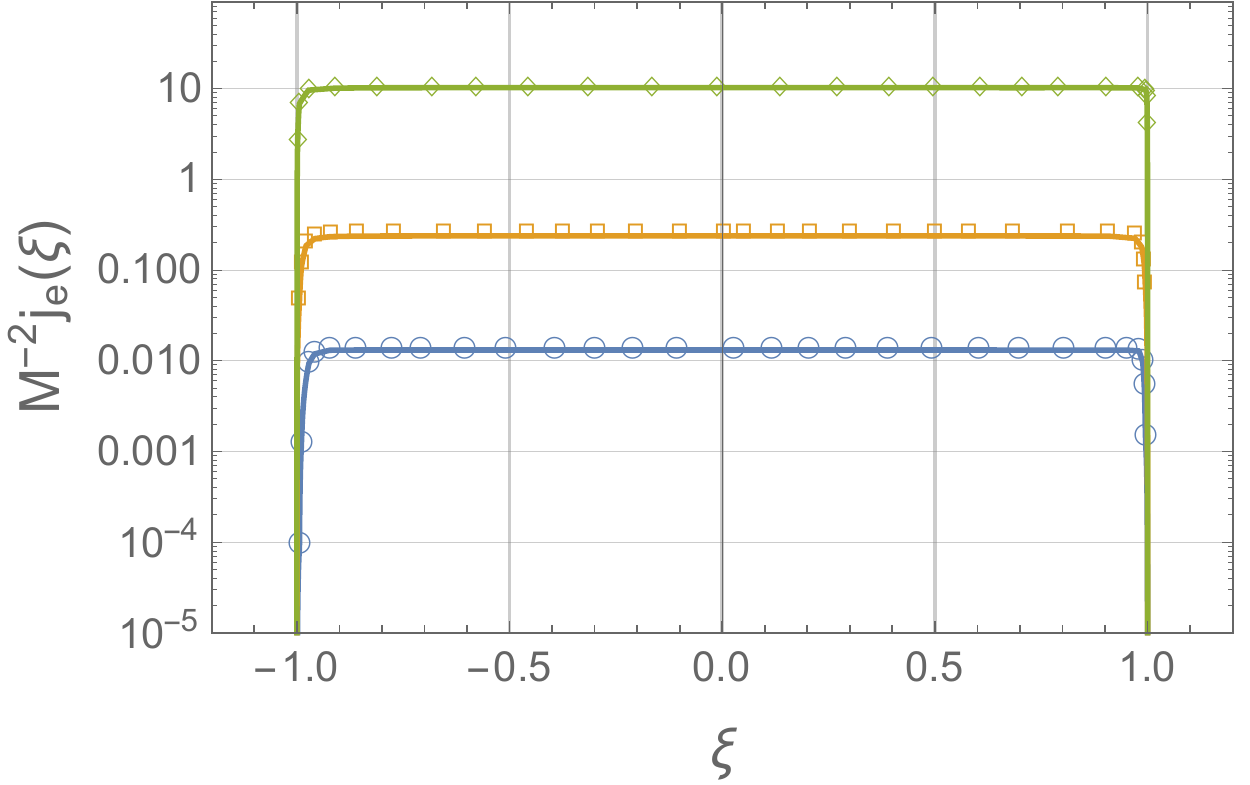} 
\par\end{centering}
}
\par\end{centering}
\begin{centering}
\subfloat[\foreignlanguage{british}{$M^{-2}q_{p}(\xi)$}]{\begin{centering}
\includegraphics[width=0.407\columnwidth]{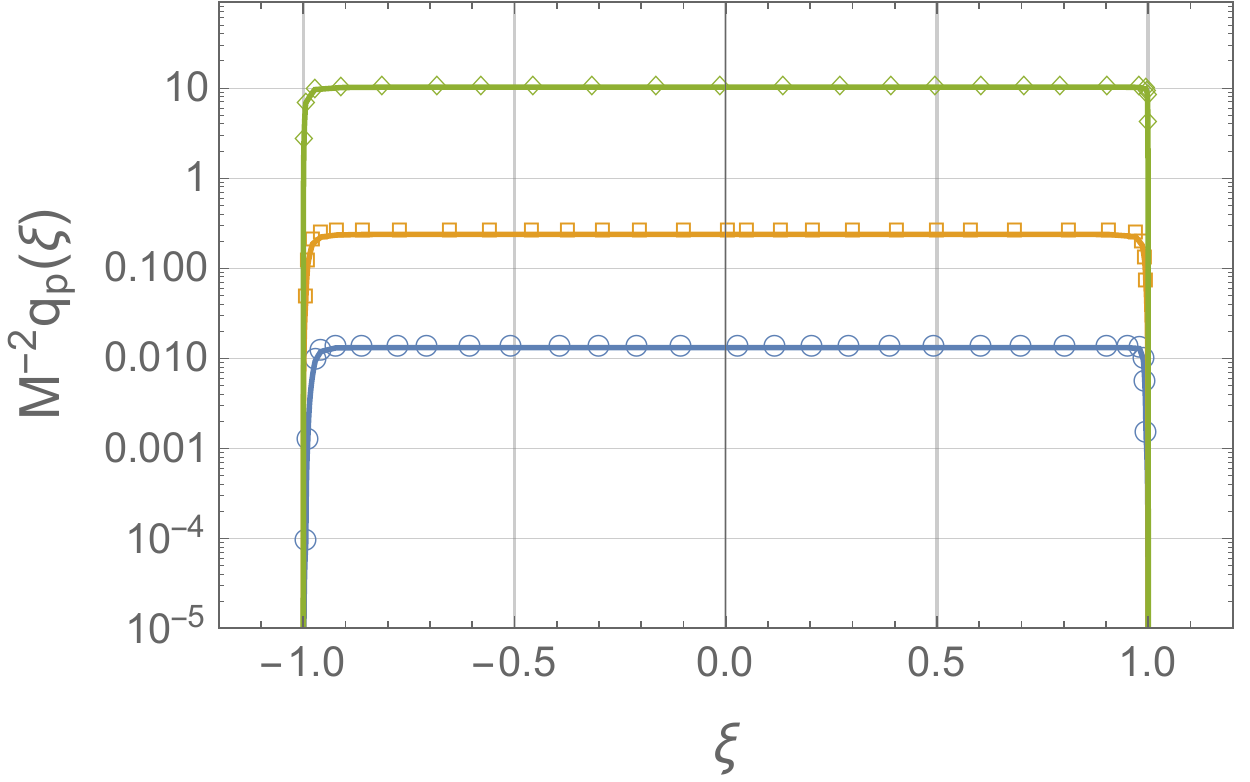} 
\par\end{centering}
}\subfloat[\foreignlanguage{british}{$M^{-2}j_{p}(\xi)$}]{\begin{centering}
\includegraphics[width=0.4\columnwidth]{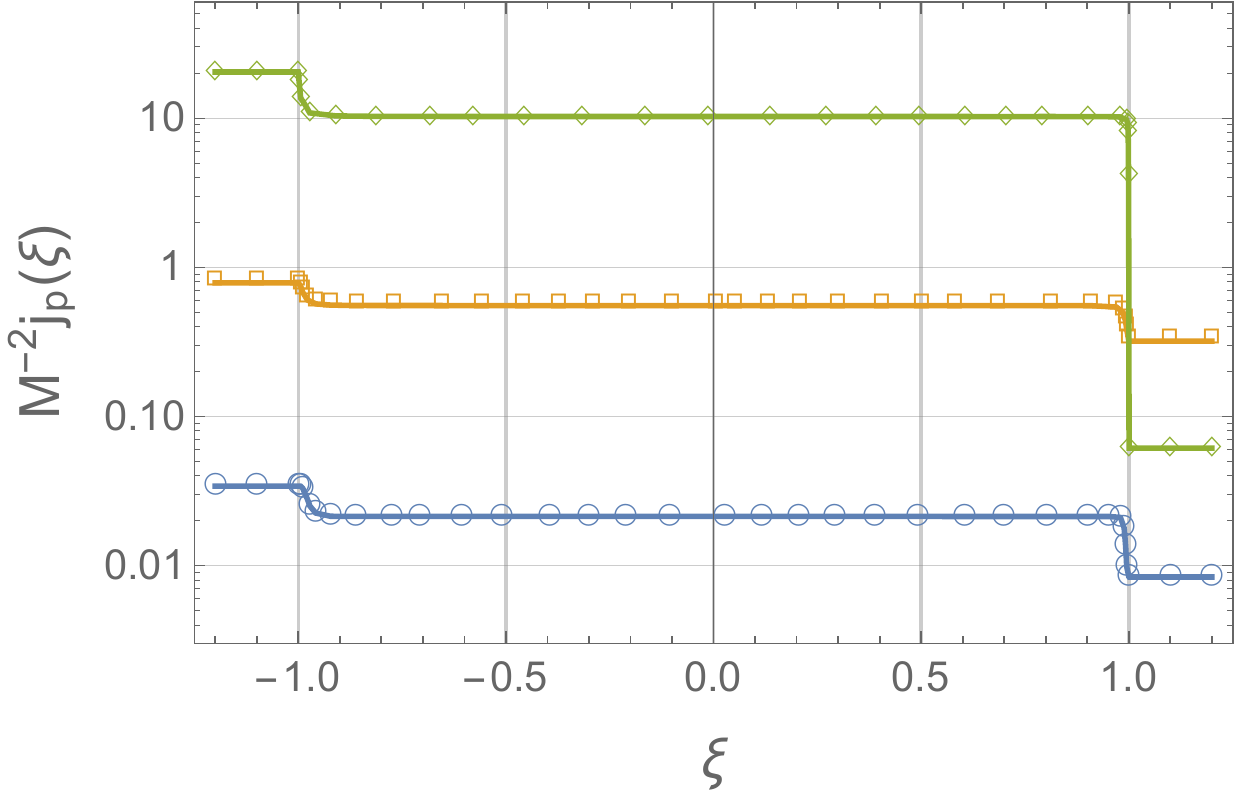} 
\par\end{centering}
}
\par\end{centering}
\caption{{\footnotesize{}\label{fig:A3EP}Ray-dependent (a) energy density
$q_{e}$, (b) energy current $j_{e}$, (c) momentum density $q_{p}$
and (d) momentum current $j_{p}$ in the }\foreignlanguage{british}{{\footnotesize{}$A_{3}$
(tetra-critical to tricritical Ising) and $D_{3}$}}{\footnotesize{}
flow after bipartite quenches at the Euler scale. The green curve
with diamonds corresponds to left and right initial temperatures $T_{l}=7M$
and $T_{r}=0.4M$, the orange curve with squares to $T_{l}=1.4M$
and $T_{r}=0.9M$ and the blue curve with circles to $T_{l}=0.3M$
and $T_{r}=0.15M$. The discrete points in the plots indicated by
the plotmarkers are obtained by the numerical solution of GHD equations,
the continuous curves are first order interpolations. Due to relativistic
invariance, $j_{e}=q_{p}$.}}
\end{figure}

In figure \ref{fig:A3EP} the density and current profiles for the
$A_{3}$ flow are displayed for various left and right initial temperatures.
The model can be regarded as the RG trajectory from the tetra-critical
to the tricritical Ising field theory with $c_{UV}=0.8$ and $c_{IR}=0.7$,
and for this model the TBA equations also include a magnonic particle.
Figure \ref{fig:A3EP} shows a behaviour similar to the case of the
$A_{2}$ flow: irrespectively of the left and right temperatures the
profiles include extended plateaux and in fact the flat regions are
even broader than in the previous case. For the $A_{2}$ model an
explanation for the plateaux was given based on the high- and low-temperature
conformal behaviour and also on the qualitative behaviour of GHD and
TBA equations. To give an explanation for the plateaux for this case
and eventually for all higher members of the $A_{n}$ flow it is sufficient
to note that the difference between the IR and UV central charges
decreases as $n$ increases \eqref{eq:CIRUVAn}. This means that the
UV and IR CFTs only slightly differ from each other as long as the
number of effective degrees of freedom or the transport properties
determined by the central charge are concerned. Thus the properties
of the interpolating flow are expected to be similar to the CFT case
including the appearance of the plateaux. In this respect, the $A_{2}$
case is the one expected to show the largest departure from the CFT
behaviour since it is for this case that the difference between the
UV and IR central charges is the largest. 

\selectlanguage{american}%

\selectlanguage{british}%
It is worth noting that unlike for the $A_{2}$ for the $A_{3}$ flow
(and also for higher members as one expects) the exactly flat constant
regions in the profiles do not appear. Although at low temperatures
the range of effective velocities for RM and LM particles do not overlap
and consequently their filling functions become independent of $\xi$,
the range of the effective velocity for the magnons spans this non-overlapping
region and $n_{M}(\th,\xi)$ has real ray-dependence. Even though
the contribution to the densities and currents originates from only
the RM and LM particles, the ray-dependence of $n_{M}(\th,\xi)$ is
transmitted to the densities and currents due to the dressing equation
\eqref{eq:Dressing}. This accounts for a weak ray-dependence in the
profiles in the non-overlapping region in the range of the RM and
LM effective velocities. At sufficiently small temperatures the effective
velocities for magnons also do not overlap with that of the RM or
LM particles. It is a general observation the even in this case the
range of the magnonic effective velocity (at least for one magnonic
species) always touches the lower and upper endpoints of the range
of the RM and LM effective velocities.

\begin{table}[H]
\selectlanguage{american}%
\begin{tabular}{|c|c|c|c|c|c|c|c|}
\hline 
\selectlanguage{british}%
$T_{l}/M$\selectlanguage{american}%
 & \selectlanguage{british}%
$T_{r}/M$\selectlanguage{american}%
 & \selectlanguage{british}%
$\tilde{c}(T_{l})$\selectlanguage{american}%
 & \selectlanguage{british}%
$\tilde{c}(T_{r})$\selectlanguage{american}%
 & \selectlanguage{british}%
$\tilde{c}_{j_{e}}(T_{l},T_{r})$\selectlanguage{american}%
 & \selectlanguage{british}%
$\tilde{c}_{q_{e}}(T_{l},T_{r})$\selectlanguage{american}%
 & \selectlanguage{british}%
$\tilde{c}_{j_{p}}(T_{l},T_{r})$\selectlanguage{american}%
 & \selectlanguage{british}%
$\tilde{c}_{j_{e}}^{lb}(T_{l},T_{r})$\selectlanguage{american}%
\tabularnewline
\hline 
\hline 
\selectlanguage{british}%
0.2\selectlanguage{american}%
 & \selectlanguage{british}%
0.1\selectlanguage{american}%
 & \selectlanguage{british}%
0.713424\selectlanguage{american}%
 & \selectlanguage{british}%
0.704842\selectlanguage{american}%
 & \selectlanguage{british}%
0.726069\selectlanguage{american}%
 & \selectlanguage{british}%
0.724252\selectlanguage{american}%
 & \selectlanguage{british}%
0.714633\selectlanguage{american}%
 & \selectlanguage{british}%
0.716494\selectlanguage{american}%
\tabularnewline
\hline 
\selectlanguage{british}%
0.25\selectlanguage{american}%
 & \selectlanguage{british}%
0.1\selectlanguage{american}%
 & \selectlanguage{british}%
0.717545\selectlanguage{american}%
 & \selectlanguage{british}%
0.704842\selectlanguage{american}%
 & \selectlanguage{british}%
0.730977\selectlanguage{american}%
 & \selectlanguage{british}%
0.729825\selectlanguage{american}%
 & \selectlanguage{british}%
0.720382\selectlanguage{american}%
 & \selectlanguage{british}%
0.720142\selectlanguage{american}%
\tabularnewline
\hline 
\selectlanguage{british}%
0.3\selectlanguage{american}%
 & \selectlanguage{british}%
0.15\selectlanguage{american}%
 & \selectlanguage{british}%
0.721433\selectlanguage{american}%
 & \selectlanguage{british}%
0.709139\selectlanguage{american}%
 & \selectlanguage{british}%
0.738176\selectlanguage{american}%
 & \selectlanguage{british}%
0.736503\selectlanguage{american}%
 & \selectlanguage{british}%
0.722257\selectlanguage{american}%
 & \selectlanguage{british}%
0.725665\selectlanguage{american}%
\tabularnewline
\hline 
\selectlanguage{british}%
0.9\selectlanguage{american}%
 & \selectlanguage{british}%
0.1\selectlanguage{american}%
 & \selectlanguage{british}%
0.751467\selectlanguage{american}%
 & \selectlanguage{british}%
0.704842\selectlanguage{american}%
 & \selectlanguage{british}%
0.766027\selectlanguage{american}%
 & \selectlanguage{british}%
0.767382\selectlanguage{american}%
 & \selectlanguage{british}%
0.761988\selectlanguage{american}%
 & \selectlanguage{british}%
0.752054\selectlanguage{american}%
\tabularnewline
\hline 
\selectlanguage{british}%
0.9\selectlanguage{american}%
 & \selectlanguage{british}%
0.85\selectlanguage{american}%
 & \selectlanguage{british}%
0.751467\selectlanguage{american}%
 & \selectlanguage{british}%
0.749811\selectlanguage{american}%
 & \selectlanguage{british}%
0.778166\selectlanguage{american}%
 & \selectlanguage{british}%
0.779638\selectlanguage{american}%
 & \selectlanguage{british}%
0.750716\selectlanguage{american}%
 & \selectlanguage{british}%
0.765082\selectlanguage{american}%
\tabularnewline
\hline 
\selectlanguage{british}%
0.95\selectlanguage{american}%
 & \selectlanguage{british}%
0.9\selectlanguage{american}%
 & \selectlanguage{british}%
0.753023\selectlanguage{american}%
 & \selectlanguage{british}%
0.751467\selectlanguage{american}%
 & \selectlanguage{british}%
0.779363\selectlanguage{american}%
 & \selectlanguage{british}%
0.781042\selectlanguage{american}%
 & \selectlanguage{british}%
0.752307\selectlanguage{american}%
 & \selectlanguage{british}%
0.76658\selectlanguage{american}%
\tabularnewline
\hline 
\selectlanguage{british}%
1.4\selectlanguage{american}%
 & \selectlanguage{british}%
0.9\selectlanguage{american}%
 & \selectlanguage{british}%
0.763667\selectlanguage{american}%
 & \selectlanguage{british}%
0.751467\selectlanguage{american}%
 & \selectlanguage{british}%
0.783582\selectlanguage{american}%
 & \selectlanguage{british}%
0.786193\selectlanguage{american}%
 & \selectlanguage{british}%
0.76077\selectlanguage{american}%
 & \selectlanguage{british}%
0.772155\selectlanguage{american}%
\tabularnewline
\hline 
\selectlanguage{british}%
2.9\selectlanguage{american}%
 & \selectlanguage{british}%
1.1\selectlanguage{american}%
 & \selectlanguage{british}%
0.779635\selectlanguage{american}%
 & \selectlanguage{british}%
0.75717\selectlanguage{american}%
 & \selectlanguage{british}%
0.790933\selectlanguage{american}%
 & \selectlanguage{british}%
0.794089\selectlanguage{american}%
 & \selectlanguage{british}%
0.778236\selectlanguage{american}%
 & \selectlanguage{british}%
0.783189\selectlanguage{american}%
\tabularnewline
\hline 
\selectlanguage{british}%
6.5\selectlanguage{american}%
 & \selectlanguage{british}%
6\selectlanguage{american}%
 & \selectlanguage{british}%
0.790221\selectlanguage{american}%
 & \selectlanguage{british}%
0.789469\selectlanguage{american}%
 & \selectlanguage{british}%
0.796616\selectlanguage{american}%
 & \selectlanguage{british}%
0.799242\selectlanguage{american}%
 & \selectlanguage{british}%
0.789463\selectlanguage{american}%
 & \selectlanguage{british}%
0.793952\selectlanguage{american}%
\tabularnewline
\hline 
\selectlanguage{british}%
7\selectlanguage{american}%
 & \selectlanguage{british}%
0.4\selectlanguage{american}%
 & \selectlanguage{british}%
0.790867\selectlanguage{american}%
 & \selectlanguage{british}%
0.728439\selectlanguage{american}%
 & \selectlanguage{british}%
0.794791\selectlanguage{american}%
 & \selectlanguage{british}%
0.796047\selectlanguage{american}%
 & \selectlanguage{british}%
0.792994\selectlanguage{american}%
 & \selectlanguage{british}%
0.790611\selectlanguage{american}%
\tabularnewline
\hline 
\selectlanguage{british}%
7\selectlanguage{american}%
 & \selectlanguage{british}%
6\selectlanguage{american}%
 & \selectlanguage{british}%
0.790867\selectlanguage{american}%
 & \selectlanguage{british}%
0.789469\selectlanguage{american}%
 & \selectlanguage{british}%
0.796669\selectlanguage{american}%
 & \selectlanguage{british}%
0.799237\selectlanguage{american}%
 & \selectlanguage{british}%
0.789854\selectlanguage{american}%
 & \selectlanguage{british}%
0.794116\selectlanguage{american}%
\tabularnewline
\hline 
\selectlanguage{british}%
9\selectlanguage{american}%
 & \selectlanguage{british}%
8.5\selectlanguage{american}%
 & \selectlanguage{british}%
0.792724\selectlanguage{american}%
 & \selectlanguage{british}%
0.792344\selectlanguage{american}%
 & \selectlanguage{british}%
0.796894\selectlanguage{american}%
 & \selectlanguage{british}%
0.799172\selectlanguage{american}%
 & \selectlanguage{british}%
0.791988\selectlanguage{american}%
 & \selectlanguage{british}%
0.795059\selectlanguage{american}%
\tabularnewline
\hline 
\end{tabular}

\selectlanguage{british}%
\caption{{\footnotesize{}\label{tab:A3DynamicalC}The effective central charges
for (left and right) thermal states and the dynamical central charges
for the partitioning protocol for various left and right temperatures
in the $A_{3}$ and $D_{3}$ massless flows, where $c_{UV}=0.8$ and
$c_{IR}=0.7$.}}
\end{table}

In table \ref{tab:A3DynamicalC}, the effective and dynamical central
charges defined in eq. \eqref{eq:DynamicalCentralCharge} are collected
for different left and right temperatures in the $A_{3}$ interpolating
flow, and the data confirm that the conjecture \eqref{eq:ConjectureDynamicalC}
proposed for the $A_{2}$ massless flow remains valid in the present
case. Therefore it is plausible to assume that together with the appearance
of the plateaux in the profile, the conjectures \eqref{eq:ConjectureDynamicalC},
\eqref{eq:ConjectureDynamicalCJp} and \eqref{eq:ConjectureDynamicalCQe}
for the dynamical central charges and the bounds for the currents
are a general property of the $A_{n}$ models in the thermal partitioning
protocol. 

In fact the broad plateaux in the profiles and the conjectures \eqref{eq:ConjectureDynamicalC}-\eqref{eq:ConjectureDynamicalCQe}
seem to be a generic property of not only the $A_{n}$ massless models,
but also of the $D_{n}$ series and eventually of all unitary massless
integrable flows. This is confirmed by studying the first two members
of of the $D_{n}$ series. What concerns the $D_{3}$ flow it is to
mention that its TBA system is equivalent with that of the $A_{3}$
model. Though the operator content and consequently the models themselves
are different, the energy and momentum densities and currents are
given by the same equations and figure \ref{fig:A3EP} and table \ref{tab:A3DynamicalC}
describes both the $A_{3}$ and $D_{3}$ flows. Another notable remark
is that the $\mathbb{Z}_{3}$ parafermion model, i.e. the UV limiting
theory of the $D_{3}$ model is the critical three-state Potts model.

\selectlanguage{american}%
\begin{figure}[H]
\begin{centering}
\subfloat[\foreignlanguage{british}{$M^{-2}q_{e}(\xi)$}]{\begin{centering}
\includegraphics[width=0.4\columnwidth]{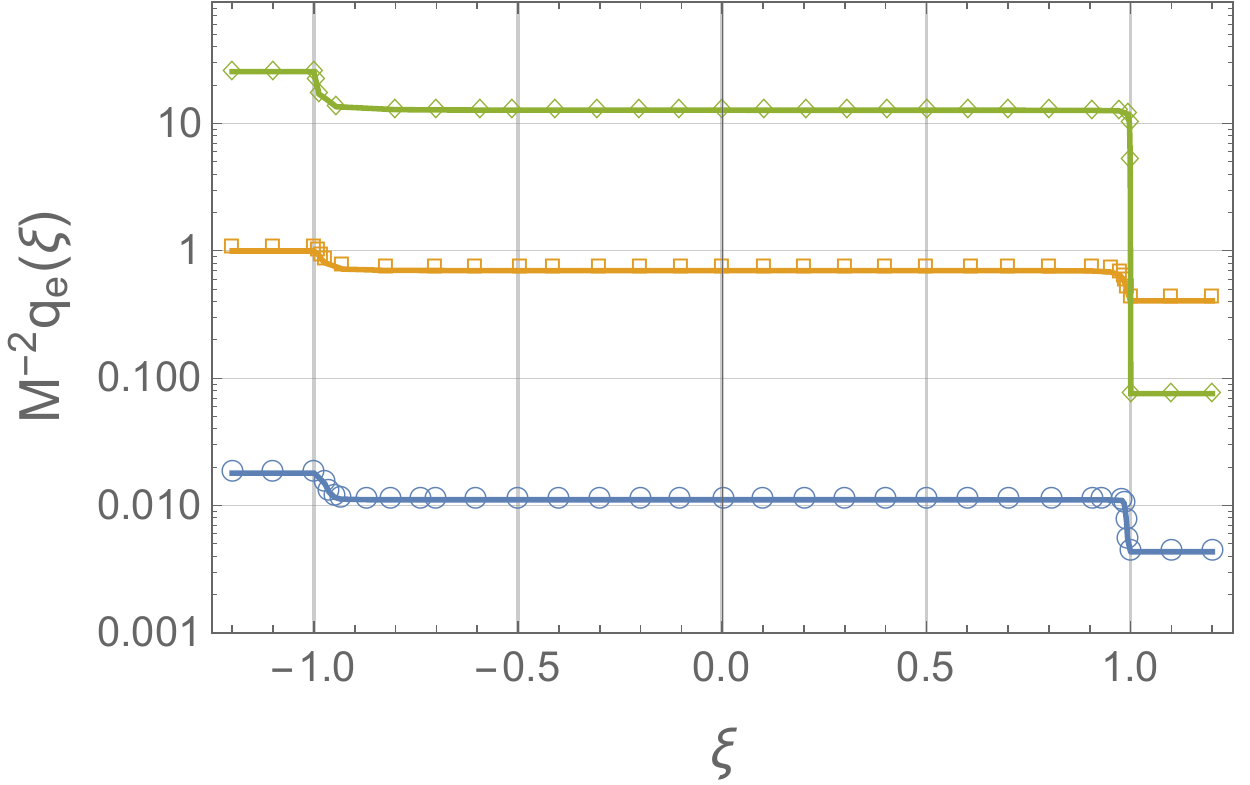} 
\par\end{centering}
}\subfloat[\foreignlanguage{british}{$M^{-2}j_{e}(\xi)$}]{\begin{centering}
\includegraphics[width=0.4\columnwidth]{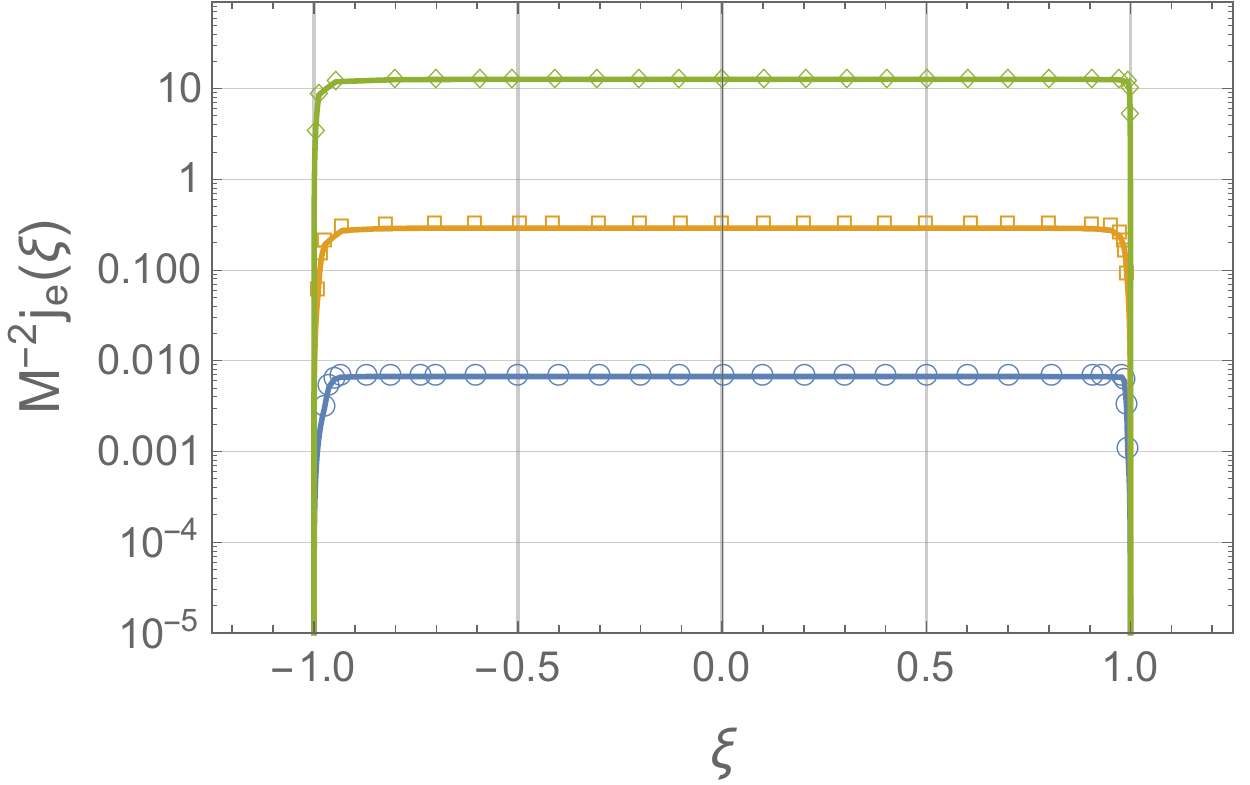} 
\par\end{centering}
}
\par\end{centering}
\begin{centering}
\subfloat[\foreignlanguage{british}{$M^{-2}q_{p}(\xi)$}]{\begin{centering}
\includegraphics[width=0.4\columnwidth]{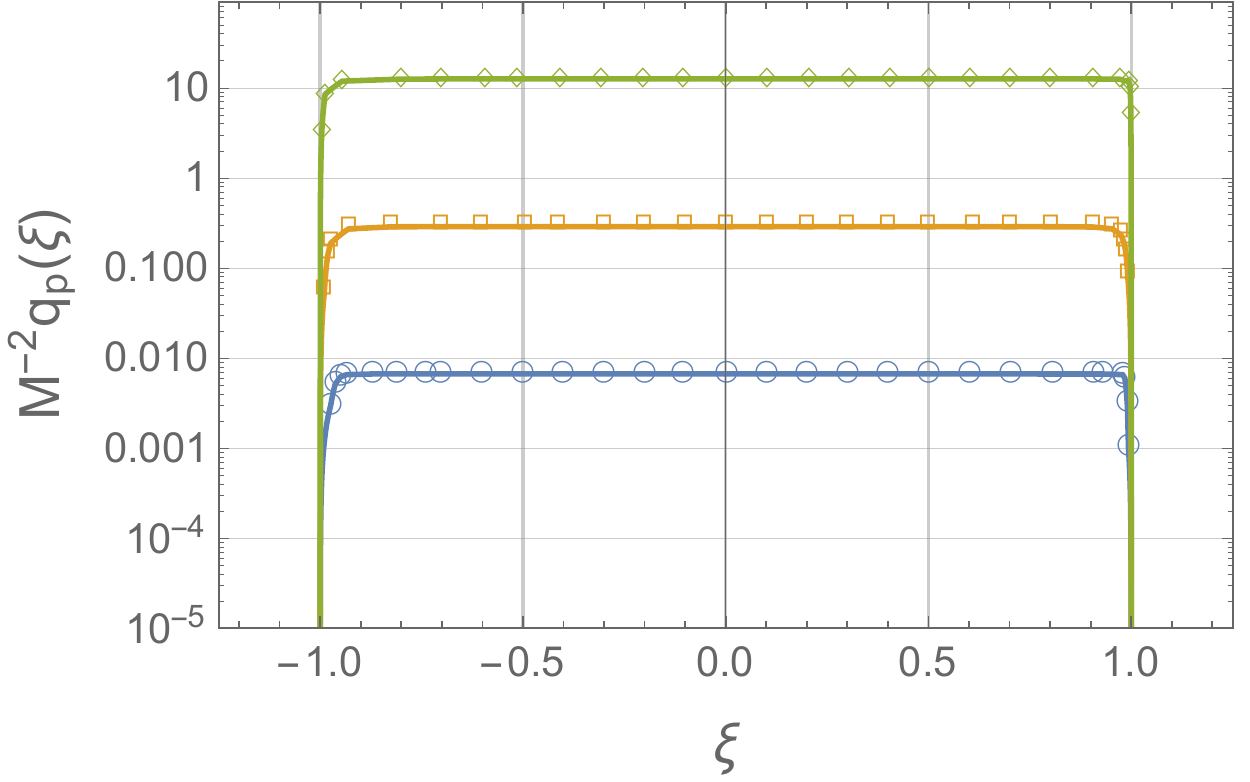} 
\par\end{centering}
}\subfloat[\foreignlanguage{british}{$M^{-2}j_{p}(\xi)$}]{\begin{centering}
\includegraphics[width=0.4\columnwidth]{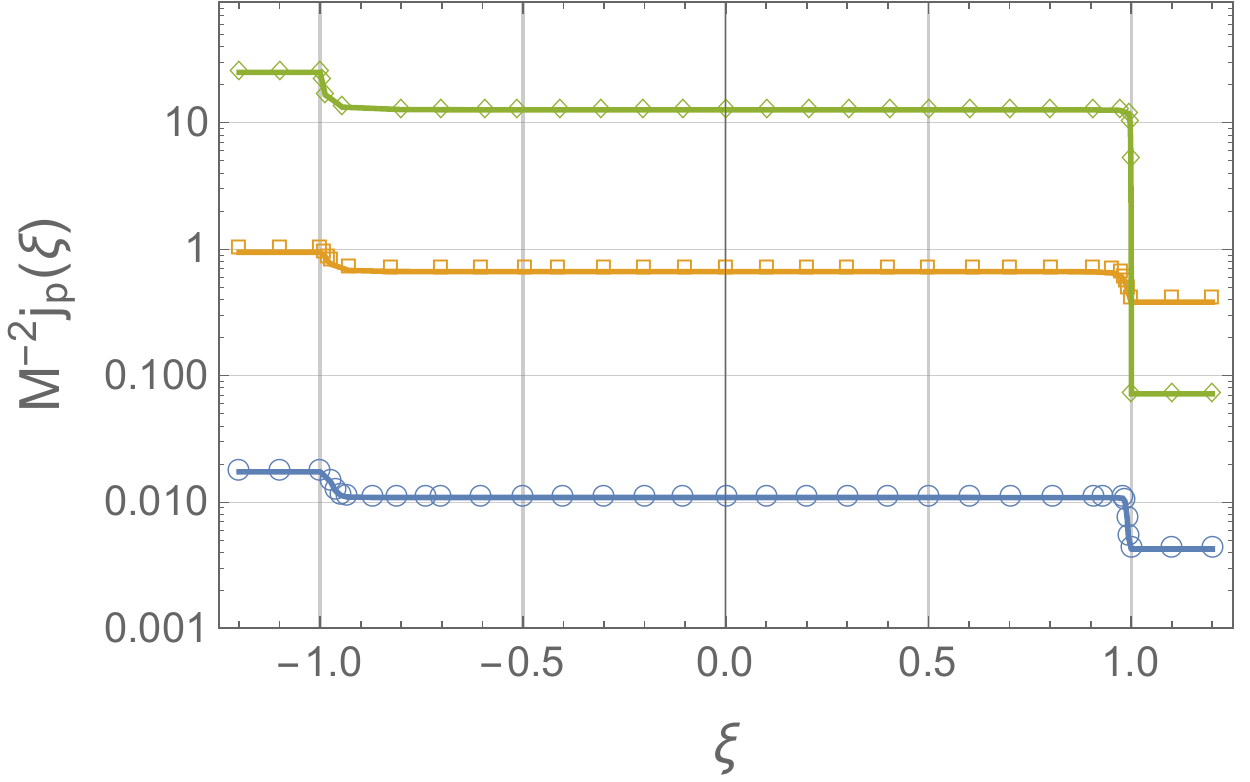} 
\par\end{centering}
}
\par\end{centering}
\caption{{\footnotesize{}\label{fig:D4EP}Ray-dependent (a) energy density
$q_{e}$, (b) energy current $j_{e}$, (c) momentum density $q_{p}$
and (d) momentum current $j_{p}$ in the $D_{4}$ flow after bipartite
quenches at the Euler scale. The green curve with diamonds corresponds
to left and right initial temperatures $T_{l}=7M$ and $T_{r}=0.4M$,
the orange curve with squares to $T_{l}=1.4M$ and $T_{r}=0.9M$ and
the blue curve with circles to $T_{l}=0.2M$ and $T_{r}=0.1M$. The
discrete points in the plots indicated by the plotmarkers are obtained
by the numerical solution of GHD equations, the continuous curves
are first order interpolations. Due to relativistic invariance, $j_{e}=q_{p}$.}}
\end{figure}

\selectlanguage{british}%
Figure \ref{fig:D4EP} and table \ref{tab:D4DynamicalC} show the
ray-dependent energy and momentum density and current profiles and
the values of effective and dynamical central charges. For the $D_{n}$
series the $n\rightarrow\infty$ limit the difference of the UV and
IR central charges is now 1 \eqref{eq:CIRUVDn} and arguing that the
UV and IR CFTs are 'similar' is not possible. Nevertheless, the presence
of the extended plateaux and validity of the conjectures \eqref{eq:ConjectureDynamicalC}-\eqref{eq:ConjectureDynamicalCQe}
in the $D_{4}$ model besides the $D_{3}$ case suggest that these
features emerge in any member of the $D_{n}$ flows.

Whereas tables \ref{tab:A3DynamicalC} and \ref{tab:D4DynamicalC}
suggest that the approximate inequality in \eqref{eq:ConjectureDynamicalCQe}
is rather a strict one, studying the hydrodynamics of massless perturbation
of the $W_{5}^{3}$ CFT (carried out in appendix \ref{sec:W3p}) slightly
larger than $c_{UV}$ values for $\tilde{c}_{q_{e}}$ can be seen
again. Our analysis is therefore not conclusive in the question whether
$c_{UV}\geq\tilde{c}_{q_{e}}$ or $c_{UV}\apprge\tilde{c}_{q_{e}}$
holds for finite and non-zero left and right temperatures but we revisit
this issue in the next section.

\begin{table}[H]
\selectlanguage{american}%
\begin{tabular}{|c|c|c|c|c|c|c|c|}
\hline 
\selectlanguage{british}%
$T_{l}/M$\selectlanguage{american}%
 & \selectlanguage{british}%
$T_{r}/M$\selectlanguage{american}%
 & \selectlanguage{british}%
$\tilde{c}(T_{l})$\selectlanguage{american}%
 & \selectlanguage{british}%
$\tilde{c}(T_{r})$\selectlanguage{american}%
 & \selectlanguage{british}%
$\tilde{c}_{j_{e}}(T_{l},T_{r})$\selectlanguage{american}%
 & \selectlanguage{british}%
$\tilde{c}_{q_{e}}(T_{l},T_{r})$\selectlanguage{american}%
 & \selectlanguage{british}%
$\tilde{c}_{j_{p}}(T_{l},T_{r})$\selectlanguage{american}%
 & \selectlanguage{british}%
$\tilde{c}_{j_{e}}^{lb}(T_{l},T_{r})$\selectlanguage{american}%
\tabularnewline
\hline 
\hline 
\selectlanguage{british}%
0.2\selectlanguage{american}%
 & \selectlanguage{british}%
0.1\selectlanguage{american}%
 & \selectlanguage{british}%
0.824632\selectlanguage{american}%
 & \selectlanguage{british}%
0.808665\selectlanguage{american}%
 & \selectlanguage{british}%
0.847828\selectlanguage{american}%
 & \selectlanguage{british}%
0.844655\selectlanguage{american}%
 & \selectlanguage{british}%
0.826961\selectlanguage{american}%
 & \selectlanguage{british}%
0.830546\selectlanguage{american}%
\tabularnewline
\hline 
\selectlanguage{british}%
0.6\selectlanguage{american}%
 & \selectlanguage{british}%
0.1\selectlanguage{american}%
 & \selectlanguage{british}%
0.87268\selectlanguage{american}%
 & \selectlanguage{british}%
0.808665\selectlanguage{american}%
 & \selectlanguage{british}%
0.900474\selectlanguage{american}%
 & \selectlanguage{british}%
0.902396\selectlanguage{american}%
 & \selectlanguage{british}%
0.889841\selectlanguage{american}%
 & \selectlanguage{british}%
0.874614\selectlanguage{american}%
\tabularnewline
\hline 
\selectlanguage{british}%
1.1\selectlanguage{american}%
 & \selectlanguage{british}%
0.1\selectlanguage{american}%
 & \selectlanguage{british}%
0.905374\selectlanguage{american}%
 & \selectlanguage{british}%
0.808665\selectlanguage{american}%
 & \selectlanguage{british}%
0.93072\selectlanguage{american}%
 & \selectlanguage{british}%
0.933312\selectlanguage{american}%
 & \selectlanguage{british}%
0.924594\selectlanguage{american}%
 & \selectlanguage{british}%
0.905972\selectlanguage{american}%
\tabularnewline
\hline 
\selectlanguage{british}%
1.1\selectlanguage{american}%
 & \selectlanguage{british}%
0.9\selectlanguage{american}%
 & \selectlanguage{british}%
0.905374\selectlanguage{american}%
 & \selectlanguage{british}%
0.894632\selectlanguage{american}%
 & \selectlanguage{british}%
0.950457\selectlanguage{american}%
 & \selectlanguage{british}%
0.954026\selectlanguage{american}%
 & \selectlanguage{british}%
0.901267\selectlanguage{american}%
 & \selectlanguage{british}%
0.926701\selectlanguage{american}%
\tabularnewline
\hline 
\selectlanguage{british}%
1.4\selectlanguage{american}%
 & \selectlanguage{british}%
0.9\selectlanguage{american}%
 & \selectlanguage{british}%
0.917764\selectlanguage{american}%
 & \selectlanguage{british}%
0.894632\selectlanguage{american}%
 & \selectlanguage{british}%
0.955841\selectlanguage{american}%
 & \selectlanguage{british}%
0.960591\selectlanguage{american}%
 & \selectlanguage{british}%
0.912188\selectlanguage{american}%
 & \selectlanguage{british}%
0.933531\selectlanguage{american}%
\tabularnewline
\hline 
\selectlanguage{british}%
2.9\selectlanguage{american}%
 & \selectlanguage{british}%
1.1\selectlanguage{american}%
 & \selectlanguage{british}%
0.949132\selectlanguage{american}%
 & \selectlanguage{british}%
0.905346\selectlanguage{american}%
 & \selectlanguage{british}%
0.971613\selectlanguage{american}%
 & \selectlanguage{british}%
0.977795\selectlanguage{american}%
 & \selectlanguage{british}%
0.946437\selectlanguage{american}%
 & \selectlanguage{british}%
0.955505\selectlanguage{american}%
\tabularnewline
\hline 
\selectlanguage{british}%
6.5\selectlanguage{american}%
 & \selectlanguage{british}%
6.0\selectlanguage{american}%
 & \selectlanguage{british}%
0.971295\selectlanguage{american}%
 & \selectlanguage{british}%
0.969662\selectlanguage{american}%
 & \selectlanguage{british}%
0.985085\selectlanguage{american}%
 & \selectlanguage{british}%
0.991036\selectlanguage{american}%
 & \selectlanguage{british}%
0.968856\selectlanguage{american}%
 & \selectlanguage{british}%
0.978339\selectlanguage{american}%
\tabularnewline
\hline 
\selectlanguage{british}%
7\selectlanguage{american}%
 & \selectlanguage{british}%
0.4\selectlanguage{american}%
 & \selectlanguage{british}%
0.972897\selectlanguage{american}%
 & \selectlanguage{british}%
0.852085\selectlanguage{american}%
 & \selectlanguage{british}%
0.980875\selectlanguage{american}%
 & \selectlanguage{british}%
0.983581\selectlanguage{american}%
 & \selectlanguage{british}%
0.977281\selectlanguage{american}%
 & \selectlanguage{british}%
0.971403\selectlanguage{american}%
\tabularnewline
\hline 
\selectlanguage{british}%
7\selectlanguage{american}%
 & \selectlanguage{british}%
6\selectlanguage{american}%
 & \selectlanguage{british}%
0.972707\selectlanguage{american}%
 & \selectlanguage{british}%
0.969662\selectlanguage{american}%
 & \selectlanguage{british}%
0.985222\selectlanguage{american}%
 & \selectlanguage{british}%
0.991092\selectlanguage{american}%
 & \selectlanguage{british}%
0.969695\selectlanguage{american}%
 & \selectlanguage{british}%
0.978704\selectlanguage{american}%
\tabularnewline
\hline 
\selectlanguage{british}%
9\selectlanguage{american}%
 & \selectlanguage{british}%
8.5\selectlanguage{american}%
 & \selectlanguage{british}%
0.976816\selectlanguage{american}%
 & \selectlanguage{british}%
0.975969\selectlanguage{american}%
 & \selectlanguage{british}%
0.98579\selectlanguage{american}%
 & \selectlanguage{british}%
0.991373\selectlanguage{american}%
 & \selectlanguage{british}%
0.974226\selectlanguage{american}%
 & \selectlanguage{british}%
0.980793\selectlanguage{american}%
\tabularnewline
\hline 
\end{tabular}

\selectlanguage{british}%
\caption{{\footnotesize{}\label{tab:D4DynamicalC}The effective central charges
for (left and right) thermal states and the dynamical central charges
for the partitioning protocol for various left and right temperatures
in the $D_{4}$ massless flow, where $c_{UV}=1$ and $c_{IR}=0.8$.}}
\end{table}

\section{The Landauer-Büttiker picture and its implications on transport\label{sec:LBF}}

In this section, we would like to discuss a simple physical picture
that gives a natural interpretation for the observed bounds for the
dynamical central charges and provides a possible explanation for
the anomalous behaviour of $\tilde{c}_{q_{e}}$. This picture is based
on the Landauer-Büttiker (LB) formalism \cite{Landauer,Buttiker1,Buttiker2,Landauer2}
of electronic systems, which claims that in the simplest case the
total electric conductance for 1D systems with ballistic electron
propagation is the number of open channels for the electrons times
the conductance quantum $2e^{2}/h$. The prototypical case where this
formalism is used is a clean quantum wire supporting coherent propagation
of electronic excitations attached to a source and a drain with an
applied voltage emitting and absorbing electrons according to Fermi-Dirac
statistics. It is easy to use the LB formalism to describe thermal
transport when the temperatures of the drain and the source are different.
For the energy current of relativistic and spinless fermions with
mass $m$ elementary calculations give

\begin{equation}
\mathcal{J}(T_{l})-\mathcal{J}(T_{r})\label{eq:LBCurrent}
\end{equation}
with

\begin{equation}
\mathcal{J}(T)=\frac{T^{2}\left[\pi^{2}-3\frac{m^{2}}{T^{2}}+\frac{6m}{T}\log\left(1+e^{m/T}\right)+6\text{Li}_{2}\left(-e^{m/T}\right)\right]}{12\pi}\label{eq:LBCurrentLong}
\end{equation}
whose $m\rightarrow0$ limit, which is the conformal limit, yields

\begin{equation}
\frac{1}{2}\frac{\pi}{12}\left(T_{l}^{2}-T_{r}^{2}\right)\,,
\end{equation}
i.e. the CFT result. Whereas this result is obtained for the case
of non-interacting quasi-particles and it is not obvious how the LB
formalism could be applied for general CFTs, the connection between
the LB formalism and the CFT treatment in the free case inspires the
following interpretation for the CFT current formula: the central
charge simply counts the number of open channels for the energy transport.
This statement is in agreement with usual interpretation of the central
charge counting the number of effective degrees of freedom in the
theory.

If one accepts this picture linking the central charge and the number
of open channels for the energy transport, the existence of both an
upper and a lower bound for the dynamical central charge $\tilde{c}_{j_{e}}$
in massless flows comes naturally. The number of effective degrees
of freedom in massless models cannot be larger than $c_{UV}$ and
smaller than $c_{IR}$ and consequently the number of open channels
available for energy transport must be bounded by $c_{UV}$ and $c_{IR}$
too. Clearly, the same argument holds also for the current of the
momentum.

Even if the transport is ballistic and an upper bound for the conductance
is expected, the accumulation of charge is not prohibited, at least
by our simple picture. In our analysis the problematic quantity with
respect to an upper bound was the average energy density at $\xi=0$,
which is a density of a conserved charge and hence is compatible with
the previous statements. Due to relativistic invariance $j_{e}=q_{p}$,
which explains why anomalous behaviour could be detected only for
$q_{e}$.

The formulae \eqref{eq:LBCurrent} and \eqref{eq:LBCurrentLong} were
derived for massive relativistic fermions, which means that $c_{IR}$
equals zero. Analysing the behaviour of the dynamical central charge
$\tilde{c}_{j_{e}}$ for the energy current \eqref{eq:LBCurrent},
it is easy to check that it is indeed bounded by $0$ and $1/2$ in
accordance with the above interpretation and that $c_{IR}=0$. Due
to the exact expression for $\tilde{c}_{j_{e}}$, nevertheless, its
monotonicity properties can also be explicitly investigated and \eqref{eq:Monoton1}
and \eqref{eq:Monoton2} are satisfied. For this property, which is
therefore valid for massless flows as well as relativistic fermions
in the LB picture, the interpretation is also at hand. Increasing
the temperature in any of the initial subsystems results in higher
energy densities and a larger number of open channels available for
the energy transport since in unitary theories the number of effective
degrees of freedom is increased with increased energy density. Whereas
it remains an assumption that the number of open channels for transport
is a monotonous function of the number of effective degrees of freedom
as it happens for free fermions, thinking in terms of the stable quasi-particles
of integrable systems with ballistic propagation makes this link very
plausible.

Having seen how the LB picture can explain some of the observed bounds
for the dynamical central charges together with the monotonicity of
$\tilde{c}_{j_{e}}$ in massless integrable models and massive free
fermions, it is natural to speculate about the case of massive integrable
and eventually general 1+1D QFTs. For this reason we also analyse
the dynamical central charges in the sinh-Gordon model using the GHD
and the partitioning protocol again. Without explaining too many details
about the model (in connection with GHD this model was investigated
in \cite{GHDFundation} reviewing its definition) we merely quote
the corresponding TBA equation containing one single particle species:

\begin{equation}
\eps(\th)=\frac{m\text{cosh}(\th)}{T}-\left(\fii_{shG}\star\ln\left(1+e^{-\eps}\right)\right)(\th)\,,\label{TBAshG}
\end{equation}
with

\begin{equation}
\fii_{shG}(\th)=-\frac{4\cosh(\th)\sin(\frac{B\pi}{2})}{\cos(B\pi)-\cosh(2\th)}\,,
\end{equation}
where $m$ is particle mass and the parameter $B$ takes values from
$[0,2]$ and is related to the strength of the interaction. 

\begin{table}[H]
\selectlanguage{american}%
\begin{tabular}{|c|c|c|c|c|c|c|c|}
\hline 
\selectlanguage{british}%
$T_{l}/M$\selectlanguage{american}%
 & \selectlanguage{british}%
$T_{r}/M$\selectlanguage{american}%
 & \selectlanguage{british}%
$\tilde{c}(T_{l})$\selectlanguage{american}%
 & \selectlanguage{british}%
$\tilde{c}(T_{r})$\selectlanguage{american}%
 & \selectlanguage{british}%
$\tilde{c}_{j_{e}}(T_{l},T_{r})$\selectlanguage{american}%
 & \selectlanguage{british}%
$\tilde{c}_{q_{e}}(T_{l},T_{r})$\selectlanguage{american}%
 & \selectlanguage{british}%
$\tilde{c}_{j_{p}}(T_{l},T_{r})$\selectlanguage{american}%
 & \selectlanguage{british}%
$\tilde{c}_{j_{e}}^{lb}(T_{l},T_{r})$\selectlanguage{american}%
\tabularnewline
\hline 
\hline 
\selectlanguage{british}%
0.2\selectlanguage{american}%
 & \selectlanguage{british}%
0.04\selectlanguage{american}%
 & \selectlanguage{british}%
0.0122958\selectlanguage{american}%
 & \selectlanguage{british}%
0\selectlanguage{american}%
 & \selectlanguage{british}%
0.0255886\selectlanguage{american}%
 & \selectlanguage{british}%
0.0681836\selectlanguage{american}%
 & \selectlanguage{british}%
0.0118277\selectlanguage{american}%
 & \selectlanguage{british}%
0.0128082\selectlanguage{american}%
\tabularnewline
\hline 
\selectlanguage{british}%
0.2\selectlanguage{american}%
 & \selectlanguage{british}%
0.1\selectlanguage{american}%
 & \selectlanguage{british}%
0.0122958\selectlanguage{american}%
 & \selectlanguage{british}%
0.00011\selectlanguage{american}%
 & \selectlanguage{british}%
0.0326524\selectlanguage{american}%
 & \selectlanguage{british}%
0.0569543\selectlanguage{american}%
 & \selectlanguage{british}%
0.00986332\selectlanguage{american}%
 & \selectlanguage{british}%
0.0163566\selectlanguage{american}%
\tabularnewline
\hline 
\selectlanguage{british}%
0.5\selectlanguage{american}%
 & \selectlanguage{british}%
0.1\selectlanguage{american}%
 & \selectlanguage{british}%
0.172818\selectlanguage{american}%
 & \selectlanguage{british}%
0.00011\selectlanguage{american}%
 & \selectlanguage{british}%
0.263287\selectlanguage{american}%
 & \selectlanguage{british}%
0.444249\selectlanguage{american}%
 & \selectlanguage{british}%
0.167334\selectlanguage{american}%
 & \selectlanguage{british}%
0.180014\selectlanguage{american}%
\tabularnewline
\hline 
\selectlanguage{british}%
0.5\selectlanguage{american}%
 & \selectlanguage{british}%
0.3\selectlanguage{american}%
 & \selectlanguage{british}%
0.172818\selectlanguage{american}%
 & \selectlanguage{british}%
0.054912\selectlanguage{american}%
 & \selectlanguage{british}%
0.342032\selectlanguage{american}%
 & \selectlanguage{british}%
0.397678\selectlanguage{american}%
 & \selectlanguage{british}%
0.142199\selectlanguage{american}%
 & \selectlanguage{british}%
0.23914\selectlanguage{american}%
\tabularnewline
\hline 
\selectlanguage{british}%
0.6\selectlanguage{american}%
 & \selectlanguage{british}%
0.1\selectlanguage{american}%
 & \selectlanguage{british}%
0.227963\selectlanguage{american}%
 & \selectlanguage{british}%
0.000113\selectlanguage{american}%
 & \selectlanguage{british}%
0.326837\selectlanguage{american}%
 & \selectlanguage{british}%
0.521414\selectlanguage{american}%
 & \selectlanguage{british}%
0.223859\selectlanguage{american}%
 & \selectlanguage{british}%
0.234473\selectlanguage{american}%
\tabularnewline
\hline 
\selectlanguage{british}%
0.6\selectlanguage{american}%
 & \selectlanguage{british}%
0.3\selectlanguage{american}%
 & \selectlanguage{british}%
0.227963\selectlanguage{american}%
 & \selectlanguage{british}%
0.054912\selectlanguage{american}%
 & \selectlanguage{british}%
0.392442\selectlanguage{american}%
 & \selectlanguage{british}%
0.473548\selectlanguage{american}%
 & \selectlanguage{british}%
0.194677\selectlanguage{american}%
 & \selectlanguage{british}%
0.285647\selectlanguage{american}%
\tabularnewline
\hline 
\selectlanguage{british}%
1.2\selectlanguage{american}%
 & \selectlanguage{british}%
0.1\selectlanguage{american}%
 & \selectlanguage{british}%
0.447465\selectlanguage{american}%
 & \selectlanguage{british}%
0.000113\selectlanguage{american}%
 & \selectlanguage{british}%
0.546621\selectlanguage{american}%
 & \selectlanguage{british}%
0.721098\selectlanguage{american}%
 & \selectlanguage{british}%
0.452063\selectlanguage{american}%
 & \selectlanguage{british}%
0.450593\selectlanguage{american}%
\tabularnewline
\hline 
\selectlanguage{british}%
1.2\selectlanguage{american}%
 & \selectlanguage{british}%
0.6\selectlanguage{american}%
 & \selectlanguage{british}%
0.447465\selectlanguage{american}%
 & \selectlanguage{british}%
0.227963\selectlanguage{american}%
 & \selectlanguage{british}%
0.619103\selectlanguage{american}%
 & \selectlanguage{british}%
0.698749\selectlanguage{american}%
 & \selectlanguage{british}%
0.406992\selectlanguage{american}%
 & \selectlanguage{british}%
0.520632\selectlanguage{american}%
\tabularnewline
\hline 
\selectlanguage{british}%
7\selectlanguage{american}%
 & \selectlanguage{british}%
3.5\selectlanguage{american}%
 & \selectlanguage{british}%
0.789094\selectlanguage{american}%
 & \selectlanguage{british}%
0.695874\selectlanguage{american}%
 & \selectlanguage{british}%
0.855191\selectlanguage{american}%
 & \selectlanguage{british}%
0.884953\selectlanguage{american}%
 & \selectlanguage{british}%
0.773724\selectlanguage{american}%
 & \selectlanguage{british}%
0.820188\selectlanguage{american}%
\tabularnewline
\hline 
\selectlanguage{british}%
7\selectlanguage{american}%
 & \selectlanguage{british}%
6\selectlanguage{american}%
 & \selectlanguage{british}%
0.789094\selectlanguage{american}%
 & \selectlanguage{british}%
0.771895\selectlanguage{american}%
 & \selectlanguage{british}%
0.867738\selectlanguage{american}%
 & \selectlanguage{british}%
0.892567\selectlanguage{american}%
 & \selectlanguage{british}%
0.782\selectlanguage{american}%
 & \selectlanguage{british}%
0.836757\selectlanguage{american}%
\tabularnewline
\hline 
\selectlanguage{british}%
8\selectlanguage{american}%
 & \selectlanguage{british}%
0.6\selectlanguage{american}%
 & \selectlanguage{british}%
0.802629\selectlanguage{american}%
 & \selectlanguage{british}%
0.227963\selectlanguage{american}%
 & \selectlanguage{british}%
0.842828\selectlanguage{american}%
 & \selectlanguage{british}%
0.878939\selectlanguage{american}%
 & \selectlanguage{british}%
0.813516\selectlanguage{american}%
 & \selectlanguage{british}%
0.805906\selectlanguage{american}%
\tabularnewline
\hline 
\selectlanguage{british}%
8\selectlanguage{american}%
 & \selectlanguage{british}%
6\selectlanguage{american}%
 & \selectlanguage{british}%
0.802629\selectlanguage{american}%
 & \selectlanguage{british}%
0.771895\selectlanguage{american}%
 & \selectlanguage{british}%
0.871823\selectlanguage{american}%
 & \selectlanguage{british}%
0.895377\selectlanguage{american}%
 & \selectlanguage{british}%
0.792172\selectlanguage{american}%
 & \selectlanguage{british}%
0.842194\selectlanguage{american}%
\tabularnewline
\hline 
\end{tabular}

\selectlanguage{british}%
\caption{{\footnotesize{}\label{tab:shGDynamicalC}The effective central charges
for (left and right) thermal states and the dynamical central charges
for the partitioning protocol for various left and right temperatures
in the}\foreignlanguage{american}{{\footnotesize{} massive sinh-Gordon
theory with $B=0.5$}}{\footnotesize{}, where $c_{UV}=1$ and $c_{IR}=0$.}}
\end{table}

Table \ref{tab:shGDynamicalC} shows the data for the effective and
dynamical central charges for the sinh-Gordon model and the data are
in perfect agreement with the predictions of the LB picture. All the
dynamical central charges are bounded by $c_{IR}$ from below and
for the currents (and in this case also for $q_{e}$) the data are
consistent with an upper bound given by $c_{UV}$, and the monotonicity
property of $\tilde{c}_{j_{e}}$ is also satisfied. Moreover, the
conjectures \eqref{eq:ConjectureDynamicalC}, \eqref{eq:ConjectureDynamicalCJp}
and \eqref{eq:ConjectureDynamicalCQe} proposed for the massless models
also hold in this case together with \eqref{eq:DoyonLowerBound},
whose validity is independent of the massless or massive nature of
the theory.

Although a systematic treatment of the hydrodynamics in massive integrable
QFTs is out of the scope of this paper, based on the above results
it is plausible to expect that the bounds for the dynamical central
charges of currents given by $c_{IR}$ and $c_{UV}$, the monotonicity
of $\tilde{c}_{j_{e}}$ and the conjectures \eqref{eq:ConjectureDynamicalC},
\eqref{eq:ConjectureDynamicalCJp} and \eqref{eq:ConjectureDynamicalCQe}
are valid in any relativistic and unitary integrable QFT and the interpretation
inspired by the LB picture is correct. The monotonic behaviour of
$\tilde{c}_{j_{e}}$ is an especially interesting finding, since it
corresponds to an out-of-equilibrium version of the well-known c-theorem.

In fact, considering a generic unitary and relativistic 1+1 D QFT
with homogeneous bulk action it is reasonable to assume that the dynamical
central charges of the energy and momentum current are still bounded
by $c_{UV}$ from above and $c_{IR}$ from below. In non-integrable
cases there are usually no additional conserved quantities to the
energy and the momentum and the ray-dependent profiles for the currents
and densities of these conserved quantities can exhibit discontinuities
at the Euler scale. Nevertheless the NESS corresponding to $\xi=0$
at the Euler scale is expected to be well defined and therefore the
dynamical central charges can be introduced similarly to the integrable
case. Whether the strict monotonicity property of $\tilde{c}_{j_{e}}$
alias a non-equilibrium c-theorem holds in non-integrable cases is
a more subtle issue. Whereas the c-function \cite{cTheorem} in eq.
\eqref{eq:EffectiveC} was defined via the TBA and hence exists only
for Bethe Ansatz integrable models, a c-function with the same monotonicity
property was first defined from the two-point function of the stress-energy
tensor \cite{OriginalCTheorem}. This function can be ascribed to
the same interpretation as the TBA c-function i.e. counting the number
of effective degrees of freedom but it can be defined for any 1+1
D perturbed CFT and accordingly the equilibrium c-theorem is valid
in any field theory of this type. Although its monotonicity suggests
a monotonic behaviour for $\tilde{c}_{j_{e}}$, due to the inelastic
scattering between particles, the possible particle production processes
and the absence of ballistic propagation it is difficult to give evidence
and to reduce the uncertainty left. We therefore leave this question
open, which we hope to be the subject of further studies.

\section{Conclusions\label{sec:Conclusions}}

In this paper we studied the Euler scale hydrodynamics of massless
integrable quantum field theories (QFTs) in the partitioning protocol
using the recently developed framework of generalised hydrodynamics
(GHD) \cite{GHDFundation,GHDFundationB,GHDFoundationC}. The peculiarity
of these models is that they interpolate between two non-trivial renormalisation
group fixed points corresponding to conformal field theories (CFTs)
with a crossover from one CFT to another characterised by an energy
scale called crossover scale. In particular we calculated the Euler
scale current and density profiles for the energy and momentum after
joining the semi-infinite left and right halves of the massless systems
initially prepared to different thermal states. Focusing on the first
few members of the $A_{n}$ and $D_{n}$ massless flows, we carried
out a systematic treatment of the $A_{n}$ and $D_{n}$ series. 

Our analysis identified some general characteristic features regarding
the transport properties of these massless flows. Irrespectively of
the magnitude of the left and right temperatures with respect to the
crossover scale of the massless theory, the density and current profiles
exhibit extended plateaux in the non-trivial region $\xi\in[-1,1]$,
where $\xi$ stands for $x/t$ sending $x$ and $t$ to infinity.
This behaviour is similar to the case of CFTs, whose profiles contain
constant regions separated by discontinuous jumps at $\xi=\pm1$ in
the Euler limit \cite{NonEqCFT1,NonEqCFT2}, but integrability prevents
the formation of such discontinuities and shock-waves in massless
flows. The presence of discontinuities in the CFT profiles and their
absence in the integrable massless case might seem surprising given
that rational CFTs are eventually integrable models having infinitely
many local conserved quantities. The main difference between these
theories accounting for the different behaviour is the existence of
particles with non-trivial dispersion in integrable cases, whereas
in CFTs the propagation of modes is dispersionless, which can thus
accumulate to form shock-waves.

We analysed the magnitude of the currents and densities and constructed
a series of bounds for them at ray $\xi=0$, which corresponds to
the non-equilibrium stationary state (NESS) that is when $t\rightarrow\infty$
but $x$ remains finite. Simple upper and lower bounds for the energy
current can be given using the CFT result \eqref{eq:EqCFT1} with
central charge of the UV and IR limiting CFTs respectively and the
initial left and right temperatures. We also showed the the dynamical
central charge $\tilde{c}_{j_{e}}$ defined in eq. \eqref{eq:DynamicalCentralCharge}
inspired by writing the maximum of the energy current in form of the
CFT energy current is bounded from below by not only the central charge
of the IR limiting CFT but by the effective central charges \eqref{eq:EffectiveC}
of the massless model corresponding to the left and right temperatures.
In \cite{DoyonLowerBounds} a different lower bound was given for
the energy current in the NESS or for the corresponding dynamical
central charge. Our data are consistent with this lower bound, which
in almost all cases turned out to be even more restrictive than the
one resulting from the effective central charges. It is worth mentioning
that the value of the energy current in the NESS is equal to its absolute
maximum, therefore the upper bound for $j_{e}(0)$ is an upper bound
for $j_{e}$ at any ray. We also verified, that the dynamical central
charge $\tilde{c}_{j_{e}}$ satisfies a particular monotonicity property
first discussed in \cite{DoyonWrongHydro}, which resembles the usual
effective central charge.

Similarly to the energy current we defined dynamical central charges
$\tilde{c}_{q_{e}}$ and $\tilde{c}_{j_{p}}$ for the energy density
$q_{e}$ and for the momentum current $j_{p}$ using their values
in the NESS and corresponding CFT formula \eqref{eq:EqCFT1}. For
$\tilde{c}_{j_{p}}$ a lower bound is given by $c_{IR}$ and an upper
bound by $\tilde{c}_{j_{e}}$ whereas for $\tilde{c}_{q_{e}}$ only
a lower bound could be given by $\tilde{c}_{j_{e}}$, as its value
can occasionally be larger than $c_{UV}$.

All these observations are based on the explicit investigation of
the $A_{2}$, $A_{3}$ $D_{3}$ and $D_{4}$ models and on the massless
flow from the $W_{5}^{3}$ to the $W_{4}^{3}$ CFTs and their validity
is very plausible for any member of the $A_{n}$ and $D_{n}$ flow
and eventually in any unitary massless models. The emergence of the
extended plateaux is easy to understand in the low- and high-temperature
limits as in these cases the CFT description of the massless model
becomes valid and for intermediate temperatures a simple argument
was given based on the TBA equations and neglecting magnonic particles
in section \ref{sec:A2Hydro}. 

Regarding the peculiar properties of the currents and densities we
discussed the Landauer-Büttiker formalism of ballistic transport and
its applicability to the free fermion CFT in the partitioning protocol.
This connection suggest that the dynamical central charges for the
currents count the number of open channels for transport and also
accounts for the extremal upper and lower bounds given by $c_{IR}$
and $c_{UV}$. This picture, however, does not prohibit the accumulation
of charge and can be compatible with the observed behaviour of $\tilde{c}_{q_{e}}$.
Using this picture it is also possible to argue for the monotonicity
property of $\tilde{c}_{j_{e}}$. These predictions do not depend
on the massless and massive nature of the integrable theory and were
indeed found valid for the massive sinh-Gordon model together with
the other bounds discussed in the previous paragraphs. The bounds,
and the predictions coming from the Landauer-Büttiker picture such
as the extremal bounds and the monotonicity of $\tilde{c}_{j_{e}}$\textcolor{red}{{}
}corresponding to a non-equilibrium c-theorem are therefore likely
to be valid in any 1+1 D relativistic integrable QFT. A further notable
implication of the argument based upon the Landauer-Büttiker formalism
is that for any low dimensional perturbed conformal field theory universal
upper and lower bounds exist for the energy and momentum currents
in the NESS. These bounds are given by $c_{UV}$ and $c_{IR}$ i.e.
the central charge of the UV limiting CFT and the IR limiting CFT
when $c_{IR}\neq0$. The $c_{IR}=0$ case corresponds to massive theories,
whose low-energy properties are not described by a CFT. This statement
should also hold in near critical systems as long as the temperatures
are not too large to spoil the effective field theory description.

Finally, for the particular case of the $A_{2}$ flow, which corresponds
to the massless flow from the tricritical to the critical Ising model,
interesting low-temperature transport properties were observed: in
this case exactly constant regions appear in the profiles although
at the same time they become smooth functions contrary to the CFT
limit. The origin of this phenomenon is related to the existence of
additional (temperature-dependent) bounds for the effective velocities
of massless particles besides the speed of light, and the exactly
flat regions emerge in any massless integrable models without magnons
as demonstrated by the $W_{5}^{3}\rightarrow W_{4}^{3}$ flow in appendix
B. For the particular case of the $A_{2}$ whose low-energy limit
can be described by a $T\bar{T}$ perturbation of the IR CFT, we checked
our results against the prediction of \cite{TTBarCFT}.

Our work leaves many interesting open directions for further exploration.
The current formulation of GHD uses the language of TBA but for many
integrable models the non-linear integral equation (NLIE) \cite{ZamoNLIE,DestriVegaNLIE,DestriVegaNLIE2,MasslessNLIE,TakiNLIE1,TakiNLIE2,TakiNLIE3,TakiNLIE4}
provides an alternative way to extract thermal (or equivalently finite
volume) properties. It is therefore tempting to try to generalise
the NLIE to be able to describe inhomogeneous situations. The applicability
of a generalised NLIE would be especially valuable for models with
a nested Bethe Ansatz. An important example is the sine-Gordon theory
for general coupling, whose TBA system has infinitely many magnons,
whereas the its NLIE description is much simpler. As for many massless
models, such as the $A_{n}$ series, the NLIEs are known they provide
an ideal playground to develop an 'inhomogeneous NLIE' because of
their simple transport properties to crosscheck the NLIE results with.

The NLIE would be also helpful when analysing the flows associated
with the exceptional Lie algebras as their TBA system contains several
magnons and the numerical solution of the GHD equations can become
very tedious. Another interesting potential application is to non-unitary
flows such as the $T_{n}$ series of models. For many non-unitary
flows that the effective central charge has a non-monotonic behaviour
\cite{MasslessNLIE}, which may have interesting implications for
the transport properties of the system.

Finally, it would be interesting to study if the Landauer-Büttiker
picture can explain the other observed bounds for the dynamical central
charges and to explore if some of its implications together with the
out-of-equilibrium version of the c-theorem remain valid in non-integrable
relativistic and low dimensional quantum field theories.

\subsection*{Acknowledgements}

I am grateful to Benjamin Doyon, Bruno Bertini and Márton Kormos for
useful discussions and to Gábor Takács for his valuable comments and
suggestions such as to investigate the connection with the Landauer-Büttiker
picture. I am also indebted to Márton Kormos for useful comments and
to Gábor Takács for carefully reading the manuscript. 

This research was supported by the National Research Development and
Innovation Office (NKFIH) under K-2016 grant no. 119204 and by the
BME-Nanotechnology FIKP grant of EMMI (BME FIKP-NAT).

\appendix
%dummy comment inserted by tex2lyx to ensure that this paragraph is not empty

\section{Low temperature expansion of the GHD equations for the $A_{2}$ flow\label{subsec:Appendix-A-MasslessFF}}

In the following we briefly review the main steps of deriving the
low temperature behaviour of the $A_{2}$ model via the low temperature
expansion of the GHD equations \eqref{eq:Dressing}, \eqref{qAv},
\eqref{jAv} and \eqref{TBA}. Considering first the RM and LM pseudo-energies
in thermal states they are determined by 

\begin{equation}
\begin{split}\eps_{RM}= & \frac{M}{2T}e^{\th}-\frac{1}{2\pi}\int\ud\th'\frac{1}{\cosh\left(\th-\th'\right)}\ln\left(1+e^{\eps_{LM}(\th')}\right)\\
\eps_{LM}= & \frac{M}{2T}e^{-\th}-\frac{1}{2\pi}\int\ud\th'\frac{1}{\cosh\left(\th-\th'\right)}\ln\left(1+e^{\eps_{LM}(\th')}\right)\,,
\end{split}
\label{eq:A2TBA}
\end{equation}
which are the rewriting of eqs. \eqref{TBAAnDn} and \eqref{DrivingTermTBAAnDn}.
In the lowest order in $T$, the pseudo-energies $\eps_{RM}^{(0)}$
and $\eps_{LM}^{(0)}$ equal the driving terms in \eqref{eq:A2TBA}
and the convolutions are neglected. The next order contributions are
obtained by iterating \eqref{eq:A2TBA} with $\eps_{RM}^{(0)}$ and
$\eps_{LM}^{(0)}$ using expansion of the logarithmic expression in
the rewriting of eq. \eqref{eq:A2TBA}

\begin{equation}
\begin{split}\eps_{RM}^{(1)}= & \frac{M}{2T}e^{\th}-\frac{1}{2\pi}\int\ud\tilde{\th}\frac{1}{\cosh\tilde{\th}}\ln\left(1+\exp\left[\left(\frac{M}{2T}e^{-\th}\right)e^{-\tilde{\th}}\right]\right)\\
\eps_{LM}^{\text{(1)}}= & \frac{M}{2T}e^{-\th}-\frac{1}{2\pi}\int\ud\tilde{\th}\frac{1}{\cosh\tilde{\th}}\ln\left(1+\exp\left[\left(\frac{M}{2T}e^{\th}\right)e^{\tilde{\th}}\right]\right)\,,
\end{split}
\label{eq:A2TBA-1}
\end{equation}
which gives

\begin{equation}
\begin{split}\eps_{RM}^{(1)}= & \frac{M}{2T}e^{\th}-\frac{1}{2\pi}\left(\frac{\pi^{2}}{6}\frac{2T}{M}e^{\th}-\frac{49\pi^{4}}{1260}\left(\frac{2T}{M}e^{\th}\right)^{3}+\ldots\right)\\
\eps_{LM}^{\text{(1)}}= & \frac{M}{2T}e^{-\th}-\frac{1}{2\pi}\left(\frac{\pi^{2}}{6}\frac{2T}{M}e^{-\th}-\frac{49\pi^{4}}{1260}\left(\frac{2T}{M}e^{-\th}\right)^{3}+\ldots\right)\,.
\end{split}
\label{eq:A2TBA-1-1}
\end{equation}

\begin{equation}
\end{equation}
The leading order corrections for the pseudo-energies are hence

\begin{equation}
\begin{split}\eps_{RM}^{(1)} & =\frac{M}{2T}e^{\th}\left(1-\frac{\pi}{3}\left(\frac{T}{M}\right)^{2}+\mathcal{O}(T^{4})\right)\\
 & =\frac{M}{2\tilde{T}}e^{\th}+\mathcal{O}(T^{3})\\
\eps_{LM}^{(1)} & =\frac{M}{2T}e^{-\th}\left(1-\frac{\pi}{3}\left(\frac{T}{M}\right)^{2}+\mathcal{O}(T^{4})\right)\\
 & =\frac{M}{2\tilde{T}}e^{-\th}+\mathcal{O}(T^{3})\,,
\end{split}
\end{equation}
where to simplify the formulas $\tilde{T}$ was introduced as

\begin{equation}
\tilde{T}=T\left(1+\frac{\pi}{3}\left(\frac{T}{M}\right)^{2}\right)\,.
\end{equation}
We need to determine the minimal and maximal values of the RM and
LM effective velocities in pure thermal states. The effective velocities
are defined in eq. \eqref{vEff} and for their extremal values it
is first necessary to calculate $\left(e_{j}'(\th)\right)^{\text{dr}}$
and $\left(p_{j}'(\th)\right)^{\text{dr}}$ in the limits $\th\rightarrow\mp\infty$
for RM and LM particles respectively. For this goal we assume that
in these limits the following Ansatz is valid

\begin{equation}
\begin{split}\left(e'_{RM}(\th)\right)^{\text{dr}} & \approx c_{RM}^{e'}e^{\th}\\
\left(p'_{RM}(\th)\right)^{\text{dr}} & \approx c_{RM}^{\text{p'}}e^{\th}\\
\left(e'_{LM}(\th)\right)^{\text{dr}} & \approx-c_{LM}^{e'}e^{-\th}\\
\left(p'_{LM}(\th)\right)^{\text{dr}} & \approx c_{LM}^{p'}e^{-\th}\,.
\end{split}
\label{eq:EPDressedApprox}
\end{equation}
Then the unknown coefficients are determined by the dressing equation
\eqref{eq:Dressing}, which read for $\left(e'(\th)\right)^{\text{dr}}$
as

\begin{equation}
\begin{split}c_{RM}^{e'}e^{\th}= & \frac{M}{2}e^{\th}-\frac{1}{2\pi}\int\ud\th'\frac{1}{\cosh\left(\th-\th'\right)}n_{LM}^{(1)}(\th')c_{LM}^{e'}e^{-\th'}\\
-c_{LM}^{e'}e^{-\th}= & -\frac{M}{2}e^{-\th}+\frac{1}{2\pi}\int\ud\th'\frac{1}{\cosh\left(\th-\th'\right)}n_{RM}^{(1)}(\th')c_{RM}^{e'}e^{\th'}\,.
\end{split}
\label{eq:EPDressedCs}
\end{equation}
eq. \eqref{eq:EPDressedApprox} is indeed a consistent approximation
of $\left(e_{j}'(\th)\right)^{\text{dr}}$ and $\left(p_{j}'(\th)\right)^{\text{dr}}$
in the limits $\th\rightarrow\mp\infty$ for RM and LM particles.
In the first line of eq. \eqref{eq:EPDressedCs} the region where
\eqref{eq:EPDressedApprox} is not valid for $\left(e'_{LM}(\th)\right)^{\text{dr}}$
is suppressed by $n_{LM}^{(1)}(\th')$ in a super-exponential way.
Expanding $\frac{1}{\cosh\left(\th-\th'\right)}$ in eq. \eqref{eq:EPDressedCs}
and performing the integration, the following equations are obtained
for \eqref{eq:EPDressedApprox}:

\begin{equation}
\begin{split}c_{RM}^{e'}e^{\th}= & \frac{M}{2}e^{\th}-\frac{\pi}{12}c_{LM}^{e'}e^{\th}\left(\frac{2\tilde{T}}{M}\right)^{2}+\mathcal{O}(T^{4}e^{3\th})\\
c_{LM}^{e'}e^{-\th}= & \frac{M}{2}e^{-\th}-\frac{\pi}{12}c_{RM}^{e'}e^{-\th}\left(\frac{2\tilde{T}}{M}\right)^{2}+\mathcal{O}(T^{4}e^{3\th})
\end{split}
\label{eq:EDressedCsLin}
\end{equation}
and after similar manipulations for $\left(p_{j}'(\th)\right)^{\text{dr}}$

\begin{equation}
\begin{split}c_{RM}^{p'}e^{\th}= & \frac{M}{2}e^{\th}+\frac{\pi}{12}c_{LM}^{p'}e^{\th}\left(\frac{2\tilde{T}}{M}\right)^{2}+\mathcal{O}(T^{4}e^{3\th})\\
c_{LM}^{p'}e^{-\th}= & \frac{M}{2}e^{-\th}+\frac{\pi}{12}c_{RM}^{p'}e^{-\th}\left(\frac{2\tilde{T}}{M}\right)^{2}+\mathcal{O}(T^{4}e^{3\th})\,,
\end{split}
\label{eq:PDressedCsLin-1}
\end{equation}
from which the coefficients in eq. \eqref{eq:EPDressedApprox} can
be uniquely determined:

\begin{equation}
\begin{split}c_{RM}^{e'} & =c_{LM}^{e'}=\frac{M}{2}\frac{1-\frac{\pi}{3}\left(\frac{\tilde{T}}{M}\right)^{2}}{1-\left(\frac{\pi}{3}\left(\frac{\tilde{T}}{M}\right)^{2}\right)^{2}}+\mathcal{O}(T^{4})\\
c_{RM}^{p'} & =c_{LM}^{p'}=\frac{M}{2}\frac{1+\frac{\pi}{3}\left(\frac{\tilde{T}}{M}\right)^{2}}{1-\left(\frac{\pi}{3}\left(\frac{\tilde{T}}{M}\right)^{2}\right)^{2}}+\mathcal{O}(T^{4})
\end{split}
\label{eq:EPDressedApproxSolution}
\end{equation}
which are correct up to the second order in $T$, that is $\tilde{T}\rightarrow T$
and from the denominators only $T^{2}$ terms are to keep. The minimal
and maximal values for the RM and LM effective velocities are hence

\begin{equation}
\begin{split}v_{RM}^{min} & =\frac{1-\frac{\pi}{3}\left(\frac{T}{M}\right)^{2}}{1-\frac{\pi}{3}\left(\frac{T}{M}\right)^{2}}+\mathcal{O}(T^{4})\\
 & =1-\frac{2\pi}{3}\left(\frac{T}{M}\right)^{2}+\mathcal{O}(T^{4})\\
v_{LM}^{max} & =-\frac{1-\frac{\pi}{3}\left(\frac{T}{M}\right)^{2}}{1-\frac{\pi}{3}\left(\frac{T}{M}\right)^{2}}+\mathcal{O}(T^{4})\\
 & =-\left(1-\frac{2\pi}{3}\left(\frac{T}{M}\right)^{2}\right)+\mathcal{O}(T^{4})\,.
\end{split}
\label{eq:vMinvMaxPureThermal}
\end{equation}
It is easy to apply the above calculation for the partitioning protocol
we are primarily interested in. At low temperatures $v_{RM}^{min}$
and $v_{LM}^{max}$ are the endpoints of the interval of $\xi$ where
the density profiles are constant where $n_{RM}(\th,\xi)=n_{RM}^{(l)}(\th)$
and $n_{LM}(\th,\xi)=n_{LM}^{(r)}(\th)$. As a consequence, the temperature
$T$ in \eqref{eq:vMinvMaxPureThermal} has to be modified; for the
RM particles $T^{(r)}$ and for the LM particles $T^{(l)}$ has to
be used leading to 

\begin{equation}
\begin{split}c_{RM}^{e'} & =\frac{M}{2}\frac{1-\frac{\pi}{3}\left(\frac{\tilde{T}_{r}}{M}\right)^{2}}{1-\left(\frac{\pi}{3}\left(\frac{\tilde{T}_{r}}{M}\right)^{2}\right)\left(\frac{\pi}{3}\left(\frac{\tilde{T}_{l}}{M}\right)^{2}\right)}+\mathcal{O}(T^{4})\\
c_{RM}^{p'} & =\frac{M}{2}\frac{1+\frac{\pi}{3}\left(\frac{\tilde{T}_{r}}{M}\right)^{2}}{1-\left(\frac{\pi}{3}\left(\frac{\tilde{T}_{r}}{M}\right)^{2}\right)\left(\frac{\pi}{3}\left(\frac{\tilde{T}_{l}}{M}\right)^{2}\right)}+\mathcal{O}(T^{4})\\
c_{LM}^{e'} & =\frac{M}{2}\frac{1-\frac{\pi}{3}\left(\frac{\tilde{T}_{l}}{M}\right)^{2}}{1-\left(\frac{\pi}{3}\left(\frac{\tilde{T}_{r}}{M}\right)^{2}\right)\left(\frac{\pi}{3}\left(\frac{\tilde{T}_{l}}{M}\right)^{2}\right)}+\mathcal{O}(T^{4})\\
c_{LM}^{p'} & =\frac{M}{2}\frac{1+\frac{\pi}{3}\left(\frac{\tilde{T}_{l}}{M}\right)^{2}}{1-\left(\frac{\pi}{3}\left(\frac{\tilde{T}_{r}}{M}\right)^{2}\right)\left(\frac{\pi}{3}\left(\frac{\tilde{T}_{l}}{M}\right)^{2}\right)}+\mathcal{O}(T^{4})
\end{split}
\label{eq:EPDressedApproxSolutionGHD}
\end{equation}
which are correct only up to 2nd order in the temperature and hence
the velocity bounds read 

\begin{equation}
\begin{split}v_{RM}^{min} & =1-\frac{2\pi}{3}\left(\frac{T_{r}}{M}\right)^{2}+\mathcal{O}(T^{4})\\
v_{LM}^{max} & =-\left(1-\frac{2\pi}{3}\left(\frac{T_{l}}{M}\right)^{2}\right)+\mathcal{O}(T^{4})\,,
\end{split}
\label{eq:vMinvMaxGHD}
\end{equation}
which equal \cite{TTBarCFT}.

Thanks to the known expressions \eqref{eq:EPDressedApproxSolutionGHD}
for $\left(e_{j}'(\th)\right)^{\text{dr}}$ and $\left(p_{j}'(\th)\right)^{\text{dr}}$we
can easily calculate the energy and momentum densities and currents
when $\xi\in[v_{LM}^{max},v_{RM}^{min}]$ using the second lines in
eqs. \eqref{qAv} and \eqref{jAv}. The corresponding formulas, which
are correct up to 4th order in the temperature, are as follows:

\begin{equation}
\begin{split}q_{e} & =\frac{\pi}{24}\left[\frac{M}{2}c_{RM}^{p'}\left(\frac{2\tilde{T}_{l}}{M}\right)^{2}+\frac{M}{2}c_{LM}^{p'}\left(\frac{2\tilde{T}_{r}}{M}\right)^{2}\right]+\mathcal{O}(T^{6})\\
j_{e} & =\frac{\pi}{24}\left[\frac{M}{2}c_{RM}^{e'}\left(\frac{2\tilde{T}_{l}}{M}\right)^{2}-\frac{M}{2}c_{LM}^{e'}\left(\frac{2\tilde{T}_{r}}{M}\right)^{2}\right]+\mathcal{O}(T^{6})\\
q_{p} & =\frac{\pi}{24}\left[\frac{M}{2}c_{RM}^{p'}\left(\frac{2\tilde{T}_{l}}{M}\right)^{2}-\frac{M}{2}c_{LM}^{p'}\left(\frac{2\tilde{T}_{r}}{M}\right)^{2}\right]+\mathcal{O}(T^{6})\\
j_{p} & =\frac{\pi}{24}\left[\frac{M}{2}c_{RM}^{e'}\left(\frac{2\tilde{T}_{l}}{M}\right)^{2}+\frac{M}{2}c_{LM}^{e'}\left(\frac{2\tilde{T}_{r}}{M}\right)^{2}\right]+\mathcal{O}(T^{6})\,,
\end{split}
\end{equation}
which we can rewrite as

\begin{equation}
\begin{split}q_{e} & =\frac{\pi}{24}\left[T_{l}^{2}+T_{r}^{2}+\frac{2\pi M^{2}}{3}\left\{ \left(\frac{T_{l}}{M}\right)^{4}+\left(\frac{T_{l}}{M}\right)^{2}\left(\frac{T_{r}}{M}\right)^{2}+\left(\frac{T_{r}}{M}\right)^{4}\right\} \right]+\mathcal{O}(T^{6})\\
j_{e} & =\frac{\pi}{24}\left[T_{l}^{2}\left(1+\frac{2\pi}{3}\left(\frac{T_{l}}{M}\right)^{2}\right)-T_{r}^{2}\left(1+\frac{2\pi}{3}\left(\frac{T_{r}}{M}\right)^{2}\right)\right]+\mathcal{O}(T^{6})\\
q_{p} & =\frac{\pi}{24}\left[T_{l}^{2}\left(1+\frac{2\pi}{3}\left(\frac{T_{l}}{M}\right)^{2}\right)-T_{r}^{2}\left(1+\frac{2\pi}{3}\left(\frac{T_{r}}{M}\right)^{2}\right)\right]+\mathcal{O}(T^{6})\\
j_{p} & =\frac{\pi}{24}\left[T_{l}^{2}+T_{r}^{2}+\frac{2\pi M^{2}}{3}\left\{ \left(\frac{T_{l}}{M}\right)^{4}-\left(\frac{T_{l}}{M}\right)^{2}\left(\frac{T_{r}}{M}\right)^{2}+\left(\frac{T_{r}}{M}\right)^{4}\right\} \right]+\mathcal{O}(T^{6})\,.
\end{split}
\end{equation}

\section{Hydrodynamics of the $W_{5}^{3}\rightarrow W_{4}^{3}$ massless model\label{sec:W3p}}

The $W_{p}^{3}$ models \cite{W3p} are CFTs with an extended symmetry
algebra, whose generators are spin-3 currents. These models possess
a $\mathbb{Z}_{3}$ symmetry and can be regarded as certain generalisations
of the critical 3-states Potts model, which is recovered for $p=4$.
The central charge is given by

\begin{equation}
c=2\left(1-\frac{12}{p\left(p+1\right)}\right)\,,
\end{equation}
with $p\geq4$. A particular perturbation of these models by the field
$\Phi_{3}$ with $\Delta=1-\frac{3}{p+1}$ results in integrable massless
flows from $W_{p}^{3}$ to $W_{p-1}^{3}$.

\begin{figure}[H]
\begin{centering}
\subfloat[\foreignlanguage{british}{$M^{-2}q_{e}(\xi)$}]{\begin{centering}
\includegraphics[width=0.4\columnwidth]{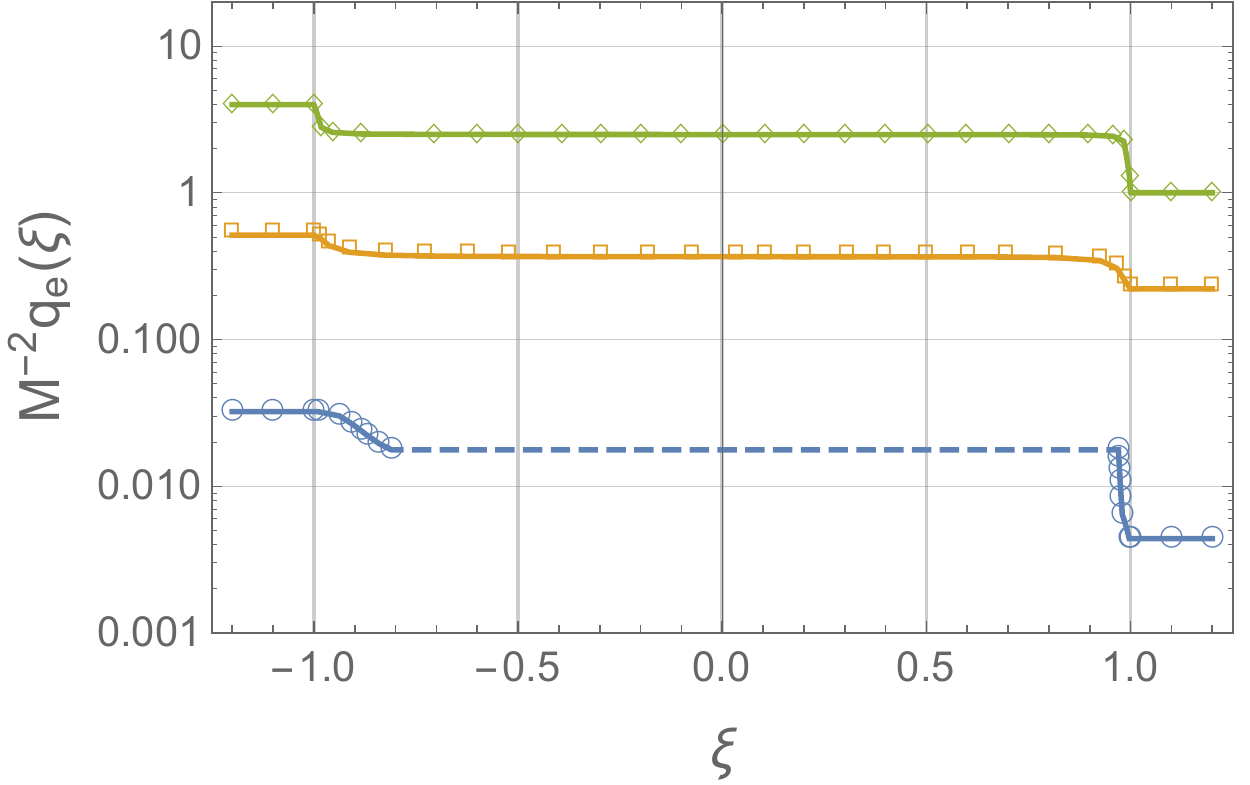} 
\par\end{centering}
}\subfloat[\foreignlanguage{british}{$M^{-2}j_{e}(\xi)$}]{\begin{centering}
\includegraphics[width=0.4\columnwidth]{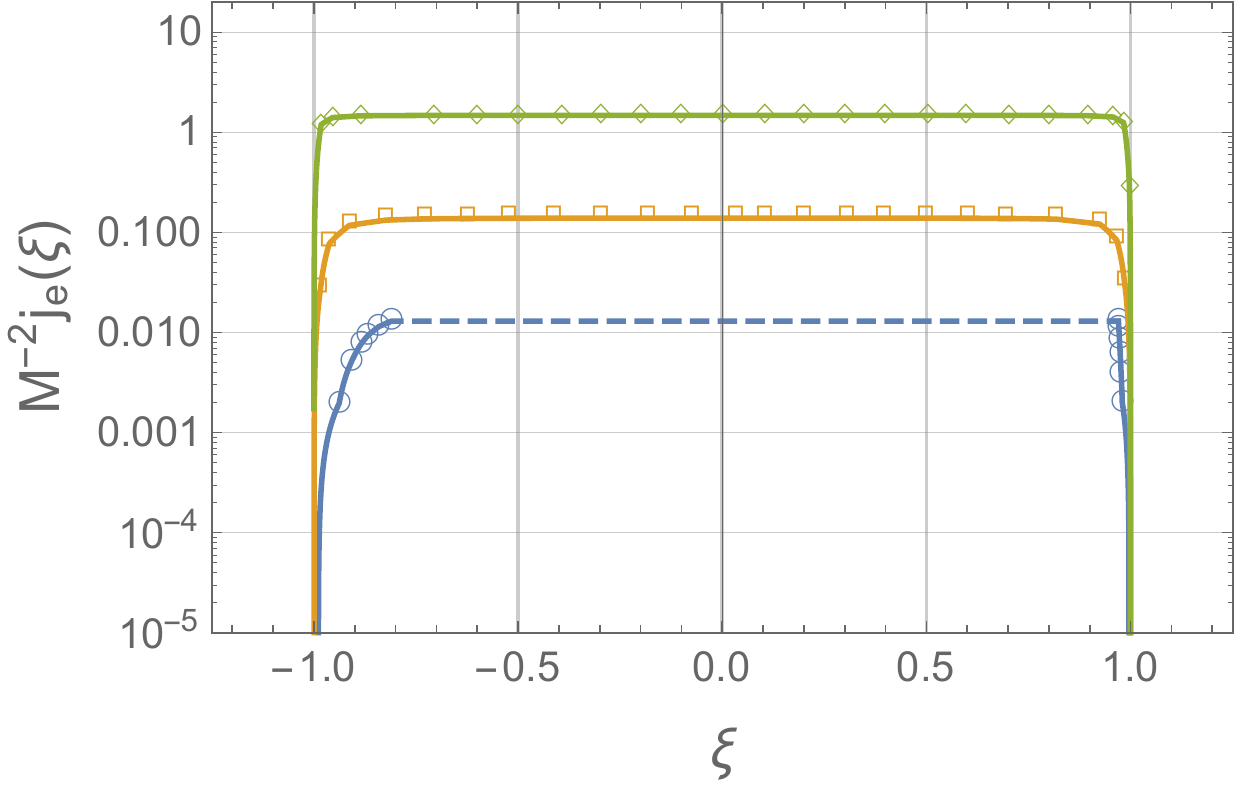} 
\par\end{centering}
}
\par\end{centering}
\begin{centering}
\subfloat[\foreignlanguage{british}{$M^{-2}q_{p}(\xi)$}]{\begin{centering}
\includegraphics[width=0.4\columnwidth]{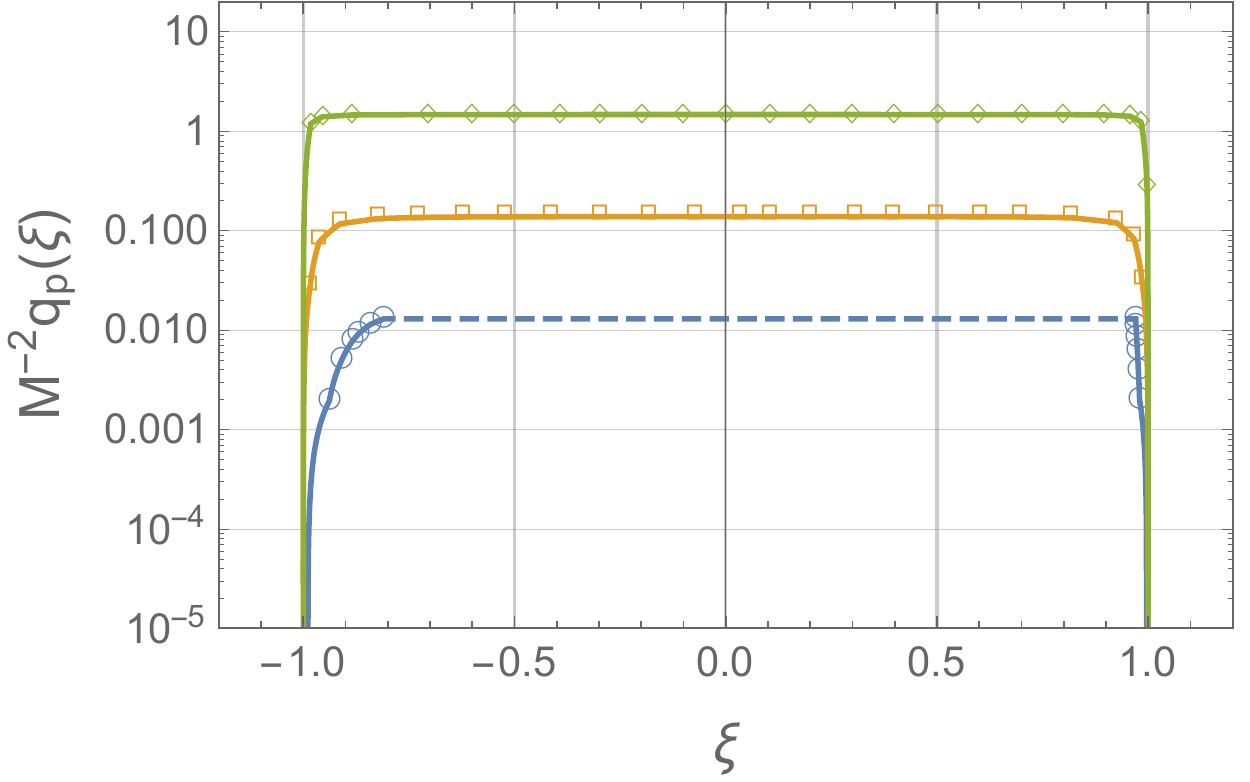} 
\par\end{centering}
}\subfloat[\foreignlanguage{british}{$M^{-2}j_{p}(\xi)$}]{\begin{centering}
\includegraphics[width=0.4\columnwidth]{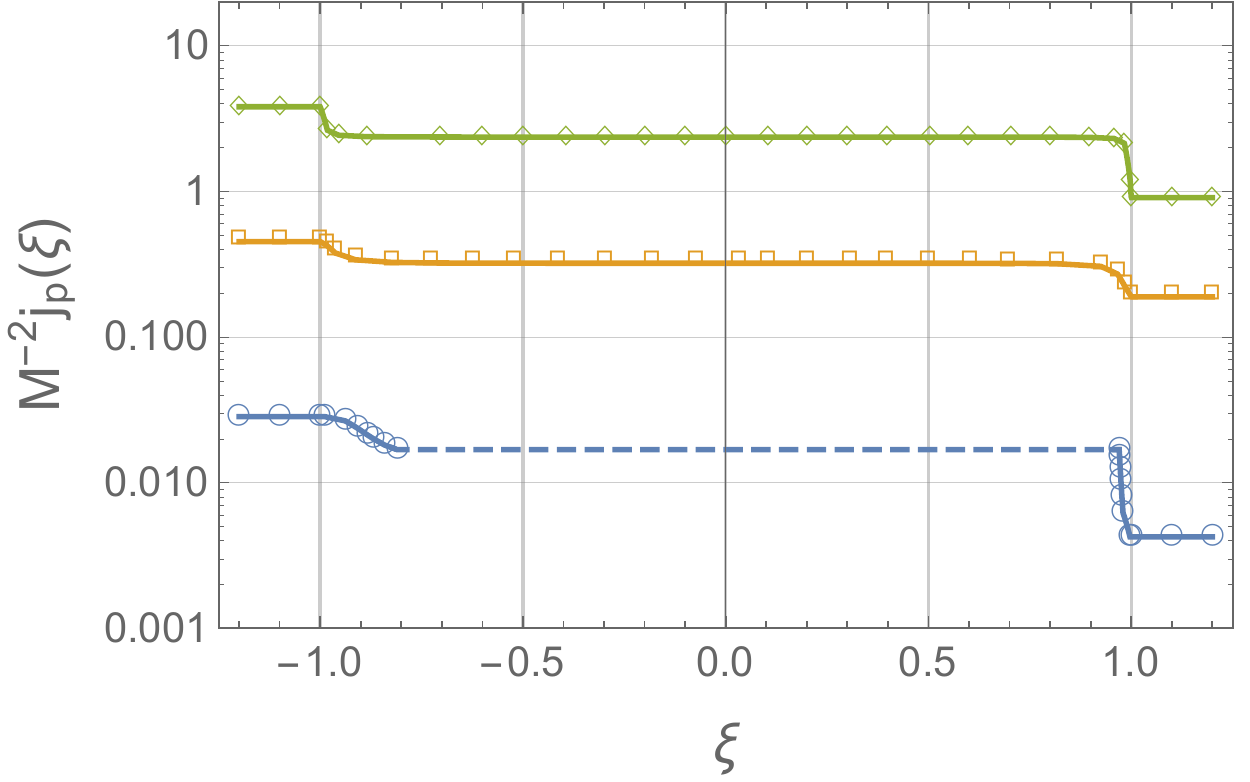} 
\par\end{centering}
}
\par\end{centering}
\caption{{\footnotesize{}\label{fig:W3p}Ray-dependent (a) energy density $q_{e}$,
(b) energy current $j_{e}$, (c) momentum density $q_{p}$ and (d)
momentum current $j_{p}$ in the $W_{5}^{3}\rightarrow W_{4}^{3}$
massless flow after bipartite quenches at the Euler scale. The green
curve with diamonds corresponds to left and right initial temperatures
$T_{l}=2.5M$ and $T_{r}=1.25M$, the orange curve with squares to
$T_{l}=0.9M$ and $T_{r}=0.6M$ and the blue curve with circles to
$T_{l}=0.25M$ and $T_{r}=0.1M$. The discrete points in the plots
indicated by the plotmarkers are obtained by the numerical solution
of GHD equations, the continuous curves are first order interpolations.
The dashed part of the curves indicates the region of constant densities
and currents. Due to relativistic invariance, $j_{e}=q_{p}$.}}
\end{figure}

A TBA description for the $W_{p}^{3}$ massless flows (and also for
massive perturbations) was proposed in \cite{W3pBetheA} which are
related to the $A_{2}$ Dynkin diagram in a non-trivial way. For the
particular case of the $W_{5}^{3}\rightarrow W_{4}^{3}$ flow the
TBA equations contain only RM and LM species but each species is doubled.
The corresponding TBA system reads

\begin{equation}
\begin{split}\eps_{RM}^{a}(\th)= & \frac{M}{2T}e^{\th}-\left(K_{ab}\star\ln\left(1+e^{-\eps_{RM}^{b}}\right)\right)(\th)\\
 & \quad\,\:+\left(K_{ab}\star\ln\left(1+e^{-\eps_{LM}^{b}}\right)\right)(\th)\\
\eps_{LM}^{a}(\th)= & \frac{M}{2T}e^{-\th}-\left(K_{ab}\star\ln\left(1+e^{-\eps_{LM}^{b}}\right)\right)(\th)\\
 & \quad\,\:+\left(K_{ab}\star\ln\left(1+e^{-\eps_{RM}^{b}}\right)\right)(\th)\,,
\end{split}
\label{TBAW3p}
\end{equation}
where $a,b=1,2$ and with the kernels

\begin{equation}
\begin{split}K_{11}(\th)=K_{22}(\th) & =-\frac{\sqrt{3}}{2\cosh\th-1}\\
K_{12}(\th)=K_{21}(\th) & =-\frac{\sqrt{3}}{2\cosh\th+1}\,.
\end{split}
\end{equation}
The one-particle energies and momentum are 

\begin{equation}
\begin{split}e_{RM/LM}^{1,2} & =\frac{M}{2}e^{\pm\th}\\
p_{RM/LM}^{1,2} & =\mp\frac{M}{2}e^{\pm\th}\,.
\end{split}
\end{equation}
The Euler scale hydrodynamics of this model is easy to obtain. Figure
\ref{fig:W3p} shows the profiles for various left and right temperatures
and table \ref{tab:W3pDynamicalC} displays the values of the effective
and dynamical central charges. These are in accordance with our previous
findings, such as the broad plateaux in the profiles, the bound on
the dynamical central charges and the monotonicity of $\tilde{c}_{j_{e}}$.
Moreover, regions of constant densities are currents appear in the
profiles for low temperatures as a consequence of the lack of magnonic
particles in the TBA system.

\selectlanguage{american}%

\selectlanguage{british}%
\begin{table}[H]
\selectlanguage{american}%
\begin{tabular}{|c|c|c|c|c|c|c|c|}
\hline 
\selectlanguage{british}%
$T_{l}/M$\selectlanguage{american}%
 & \selectlanguage{british}%
$T_{r}/M$\selectlanguage{american}%
 & \selectlanguage{british}%
$\tilde{c}(T_{l})$\selectlanguage{american}%
 & \selectlanguage{british}%
$\tilde{c}(T_{r})$\selectlanguage{american}%
 & \selectlanguage{british}%
$\tilde{c}_{j_{e}}(T_{l},T_{r})$\selectlanguage{american}%
 & \selectlanguage{british}%
$\tilde{c}_{q_{e}}(T_{l},T_{r})$\selectlanguage{american}%
 & \selectlanguage{british}%
$\tilde{c}_{j_{p}}(T_{l},T_{r})$\selectlanguage{american}%
 & \selectlanguage{british}%
$\tilde{c}_{j_{e}}^{lb}(T_{l},T_{r})$\selectlanguage{american}%
\tabularnewline
\hline 
\hline 
\selectlanguage{british}%
0.25\selectlanguage{american}%
 & \selectlanguage{british}%
0.1\selectlanguage{american}%
 & \selectlanguage{british}%
0.868088\selectlanguage{american}%
 & \selectlanguage{british}%
0.811868\selectlanguage{american}%
 & \selectlanguage{british}%
0.93966\selectlanguage{american}%
 & \selectlanguage{british}%
0.928449\selectlanguage{american}%
 & \selectlanguage{british}%
0.88748\selectlanguage{american}%
 & \selectlanguage{british}%
0.878801\selectlanguage{american}%
\tabularnewline
\hline 
\selectlanguage{british}%
0.6\selectlanguage{american}%
 & \selectlanguage{british}%
0.3\selectlanguage{american}%
 & \selectlanguage{british}%
1.0011\selectlanguage{american}%
 & \selectlanguage{british}%
0.890578\selectlanguage{american}%
 & \selectlanguage{british}%
1.1193\selectlanguage{american}%
 & \selectlanguage{british}%
1.12183\selectlanguage{american}%
 & \selectlanguage{british}%
0.991055\selectlanguage{american}%
 & \selectlanguage{british}%
1.03793\selectlanguage{american}%
\tabularnewline
\hline 
\selectlanguage{british}%
0.9\selectlanguage{american}%
 & \selectlanguage{british}%
0.6\selectlanguage{american}%
 & \selectlanguage{british}%
1.06489\selectlanguage{american}%
 & \selectlanguage{british}%
1.0011\selectlanguage{american}%
 & \selectlanguage{british}%
1.17196\selectlanguage{american}%
 & \selectlanguage{british}%
1.19334\selectlanguage{american}%
 & \selectlanguage{british}%
1.047\selectlanguage{american}%
 & \selectlanguage{british}%
1.11592\selectlanguage{american}%
\tabularnewline
\hline 
\selectlanguage{british}%
0.9\selectlanguage{american}%
 & \selectlanguage{british}%
0.85\selectlanguage{american}%
 & \selectlanguage{british}%
1.06489\selectlanguage{american}%
 & \selectlanguage{british}%
1.05658\selectlanguage{american}%
 & \selectlanguage{british}%
1.1807\selectlanguage{american}%
 & \selectlanguage{british}%
1.20618\selectlanguage{american}%
 & \selectlanguage{british}%
1.06099\selectlanguage{american}%
 & \selectlanguage{british}%
1.1335\selectlanguage{american}%
\tabularnewline
\hline 
\selectlanguage{british}%
0.95\selectlanguage{american}%
 & \selectlanguage{british}%
0.9\selectlanguage{american}%
 & \selectlanguage{british}%
1.07249\selectlanguage{american}%
 & \selectlanguage{british}%
1.06489\selectlanguage{american}%
 & \selectlanguage{british}%
1.18307\selectlanguage{american}%
 & \selectlanguage{british}%
1.20934\selectlanguage{american}%
 & \selectlanguage{british}%
1.06891\selectlanguage{american}%
 & \selectlanguage{british}%
1.13906\selectlanguage{american}%
\tabularnewline
\hline 
\selectlanguage{british}%
1.1\selectlanguage{american}%
 & \selectlanguage{british}%
0.1\selectlanguage{american}%
 & \selectlanguage{british}%
1.09176\selectlanguage{american}%
 & \selectlanguage{british}%
0.811868\selectlanguage{american}%
 & \selectlanguage{british}%
1.14362\selectlanguage{american}%
 & \selectlanguage{british}%
1.16375\selectlanguage{american}%
 & \selectlanguage{british}%
1.12339\selectlanguage{american}%
 & \selectlanguage{british}%
1.09408\selectlanguage{american}%
\tabularnewline
\hline 
\selectlanguage{british}%
1.4\selectlanguage{american}%
 & \selectlanguage{british}%
0.9\selectlanguage{american}%
 & \selectlanguage{british}%
1.11894\selectlanguage{american}%
 & \selectlanguage{british}%
1.06489\selectlanguage{american}%
 & \selectlanguage{british}%
1.18975\selectlanguage{american}%
 & \selectlanguage{british}%
1.21554\selectlanguage{american}%
 & \selectlanguage{british}%
1.10385\selectlanguage{american}%
 & \selectlanguage{british}%
1.157\selectlanguage{american}%
\tabularnewline
\hline 
\selectlanguage{british}%
2.5\selectlanguage{american}%
 & \selectlanguage{british}%
1.25\selectlanguage{american}%
 & \selectlanguage{british}%
1.16275\selectlanguage{american}%
 & \selectlanguage{british}%
1.10687\selectlanguage{american}%
 & \selectlanguage{british}%
1.19694\selectlanguage{american}%
 & \selectlanguage{british}%
1.21587\selectlanguage{american}%
 & \selectlanguage{british}%
1.15202\selectlanguage{american}%
 & \selectlanguage{british}%
1.18136\selectlanguage{american}%
\tabularnewline
\hline 
\selectlanguage{british}%
5.5\selectlanguage{american}%
 & \selectlanguage{british}%
5.0\selectlanguage{american}%
 & \selectlanguage{british}%
1.18864\selectlanguage{american}%
 & \selectlanguage{british}%
1.18681\selectlanguage{american}%
 & \selectlanguage{british}%
1.19973\selectlanguage{american}%
 & \selectlanguage{british}%
1.20689\selectlanguage{american}%
 & \selectlanguage{british}%
1.18777\selectlanguage{american}%
 & \selectlanguage{british}%
1.19729\selectlanguage{american}%
\tabularnewline
\hline 
\selectlanguage{british}%
6\selectlanguage{american}%
 & \selectlanguage{british}%
0.6\selectlanguage{american}%
 & \selectlanguage{british}%
1.19009\selectlanguage{american}%
 & \selectlanguage{british}%
1.0011\selectlanguage{american}%
 & \selectlanguage{british}%
1.19854\selectlanguage{american}%
 & \selectlanguage{british}%
1.20418\selectlanguage{american}%
 & \selectlanguage{british}%
1.18918\selectlanguage{american}%
 & \selectlanguage{british}%
1.19196\selectlanguage{american}%
\tabularnewline
\hline 
\selectlanguage{british}%
6\selectlanguage{american}%
 & \selectlanguage{british}%
5\selectlanguage{american}%
 & \selectlanguage{british}%
1.19009\selectlanguage{american}%
 & \selectlanguage{british}%
1.18681\selectlanguage{american}%
 & \selectlanguage{british}%
1.19974\selectlanguage{american}%
 & \selectlanguage{british}%
1.20642\selectlanguage{american}%
 & \selectlanguage{british}%
1.18871\selectlanguage{american}%
 & \selectlanguage{british}%
1.19751\selectlanguage{american}%
\tabularnewline
\hline 
\selectlanguage{british}%
9\selectlanguage{american}%
 & \selectlanguage{british}%
8.5\selectlanguage{american}%
 & \selectlanguage{british}%
1.19482\selectlanguage{american}%
 & \selectlanguage{british}%
1.19432\selectlanguage{american}%
 & \selectlanguage{british}%
1.19976\selectlanguage{american}%
 & \selectlanguage{british}%
1.2033\selectlanguage{american}%
 & \selectlanguage{british}%
1.19453\selectlanguage{american}%
 & \selectlanguage{british}%
1.19888\selectlanguage{american}%
\tabularnewline
\hline 
\end{tabular}

\selectlanguage{british}%
\caption{{\footnotesize{}\label{tab:W3pDynamicalC}The effective central charges
for (left and right) thermal states and the dynamical central charges
for the partitioning protocol for various left and right temperatures
in the}\foreignlanguage{american}{{\footnotesize{} $W_{5}^{3}\rightarrow W_{4}^{3}$}}{\footnotesize{}
massless flow, where $c_{UV}=1.2$ and $c_{IR}=0.8$.}}
\end{table}


\begin{thebibliography}{100}
\bibitem{NewtonCradle}T. Kinoshita, T. Wenger, and D. S. Weiss,
\emph{Nature} \textbf{440} (2006) 900\textendash 903.

\bibitem{ExperimentalNoThermalization1}S. Trotzky, Y.-A. Chen, A.
Flesch, I. P. McCulloch, U. Schollwöck, J. Eisert and I. Bloch, \emph{Nature
Phys.} \textbf{8} (2012) 325-330, arXiv:1101.2659 {[}cond-mat.quant-gas{]}. 

\bibitem{ExperimentalNoThermalization3}M. Gring, M. Kuhnert, T. Langen,
T. Kitagawa, B. Rauer, M. Schreitl, I. Mazets, D. A. Smith, E. Demler
and J. Schmiedmayer, \emph{Science} \textbf{337} (2012) 1318-1322,
arXiv:1112.0013 {[}cond-mat.quant-gas{]}.

\bibitem{GGEExperimental}T. Langen, S. Erne, R. Geiger, B. Rauer,
T. Schweigler, M. Kuhnert, W. Rohringer, I. E. Mazets, T. Gasenzer
and J. Schmiedmayer,  \emph{Science} \textbf{348} (2015) 207\textendash 211,
arXiv:1411.7185 {[}cond-mat.quant-gas{]}.

\bibitem{ColdAtomSchm1}S. Hofferberth, I. Lesanovsky, B. Fischer,
T. Schumm and J. Schmiedmayer,  \textit{Nature} \textbf{449}, 324
(2007), arXiv:0706.2259 {[}cond-mat.other{]}.

\bibitem{ColdAtomSchm2}T. Langen, R. Geiger, M. Kuhnert, B. Rauer
and J. Schmiedmayer, \textit{Nature Physics} \textbf{9}, 640 (2013),
arXiv:1305.3708 {[}cond-mat.quant-gas{]}.

\bibitem{Nagerl}F. Meinert, M. J. Mark, E. Kirilov, K. Lauber, P.
Weinmann, A. J. Daley and H.-C. Nägerl, \emph{Phys. Rev. Lett.} \textbf{111}
(2013) 053003, arXiv:1304.2628 {[}cond-mat.quant-gas{]}.

\bibitem{Fukuhara}T. Fukuhara, P. Schauß, M. Endres, S. Hild, M.
Cheneau, I. Bloch and C. Gross,  \textit{Nature} \textbf{502} (2013)
76, arXiv:1305.6598 {[}cond-mat.quant-gas{]}.

\bibitem{Kaufman}A. M. Kaufman, M. E. Tai, A. Lukin, M. Rispoli,
R. Schittko, P. M. Preiss and M. Greiner, \texttt{}\textit{Science}
\textbf{353} (2016) 794, arXiv:1603.04409 {[}quant-ph{]}.

\bibitem{ExperimentalNoThermalization2}M. Cheneau, P. Barmettler,
D. Poletti, M. Endres, P. Schauss, T. Fukuhara, C. Gross, I. Bloch,
C. Kollath and S. Kuhr, \emph{Nature} \textbf{481} (2012) 484-487,
arXiv:1111.0776 {[}cond-mat.quant-gas{]}. 

\bibitem{CardyCalabrese}P. Calabrese and J. Cardy, \emph{Phys. Rev.
Lett.} \textbf{96} (2006) 136801, arXiv:cond-mat/0601225.

\bibitem{CardyCalabrese2}P. Calabrese and J. Cardy, \emph{J. Stat.
Mech.} \textbf{0706} (2007) P06008, arXiv:0704.1880 {[}cond-mat.stat-mech{]}. 

\bibitem{GGEProposal}M. Rigol, V. Dunjko, V. Yurovsky and M. Olshanii,
\emph{Phys. Rev. Lett.} \textbf{98} (2007) 050405, arXiv:cond-mat/0604476
{[}cond-mat.other{]}.

\bibitem{GETH}A. C. Cassidy, C. W. Clark and M. Rigol, \textit{Phys.
Rev. Lett.} \textbf{106} (2011) 140405, arXiv:1008.4794 {[}cond-mat.stat-mech{]}.

\selectlanguage{english}%
\bibitem{ExcitedBetheStates}\foreignlanguage{british}{B. Pozsgay,
\emph{J. Stat. Mech.} \textbf{1101} (2011) P01011, arXiv:1009.4662
{[}hep-th{]}.}

\selectlanguage{british}%
\bibitem{CauxNoGGE}B. Wouters, J. De Nardis, M. Brockmann, D. Fioretto,
M. Rigol and J.-S. Caux\emph{, Phys. Rev. Lett.} \textbf{113} (2014)
117202, arXiv:1405.0172, arXiv:1405.0172 {[}cond-mat.str-el{]}. 

\bibitem{PozsgayNoGGE}B. Pozsgay, M. Mestyán, M. A. Werner, M. Kormos,
G. Zaránd and G. Takács\emph{, Phys. Rev. Lett.} \textbf{113} (2014)
117203, arXiv:1405.2843 {[}cond-mat.stat-mech{]}.

\bibitem{Pozsi2}B. Pozsgay,  \emph{J. Stat. Mech.} \textbf{9} (2014)
09026, arXiv:1406.4613 {[}cond-mat.stat-mech{]}.

\bibitem{Goldstein}G. Goldstein and N. Andrei,  \emph{Phys. Rev.
}\textbf{\emph{A}}\textbf{ 90} (2014) 043625, arXiv:1405.4224 {[}cond-mat.quant-gas{]}.

\bibitem{EsslerMussardo}F. H. L. Essler, G. Mussardo and M. Panfil,
\emph{Phys. Rev. }\textbf{\emph{A}}\textbf{ 91} (2015) 051602, arXiv:1411.5352
{[}cond-mat.quant-gas{]}.

\bibitem{ProsenCGGE}E. Ilievski, J. De Nardis, B. Wouters, J.-S.
Caux, F. H. L. Essler and T. Prosen\emph{, Phys. Rev. Lett.} \textbf{115}
(2015) 157201, arXiv:1507.02993 {[}quant-ph{]}.

\selectlanguage{english}%
\bibitem{MarciNoGGE}\foreignlanguage{british}{M. Mestyan, B. Pozsgay,
G. Takacs and M. A. Werner, \textit{J. Stat. Mech.} \textbf{1504}
(2015) 04001 arXiv:1412.4787 {[}cond-mat.stat-mech{]}.}

\selectlanguage{british}%
\bibitem{IlievskiStringChargeDuality}E. Ilievski, E. Quinn, J. De
Nardis and M. Brockmann, \textit{J. Stat. Mech.} \textbf{1606} (2016)
063101, arXiv:1512.04454 {[}cond-mat.stat-mech{]}.

\selectlanguage{english}%
\bibitem{PozsgaytGGE}\foreignlanguage{british}{B. Pozsgay, E. Vernier
and M. A. Werner, \textit{J. Stat. Mech. }\textbf{2017 }(2017) 093103,
arXiv:1703.09516 {[}cond-mat.stat-mech{]}.}

\bibitem{IlievskiCauxQPPicture}\foreignlanguage{british}{E. Ilievski,
E. Quinn and J.-S. Caux,  \textit{Phys. Rev.} \textbf{B 95} (2017)
115128, arXiv:1610.06911 {[}cond-mat.stat-mech{]}.}

\bibitem{IntegrableQuench}\foreignlanguage{british}{L. Piroli, B.
Pozsgay and E. Vernier, \textit{Nucl. Phys.} \textbf{B 925}(2017)
362, arXiv:1709.04796{[}cond-mat.stat-mech{]}.}

\selectlanguage{british}%
\bibitem{IlievskiQuasiLocal}E. Ilievski, M. Medenjak, T. Prosen and
L. Zadnik, \emph{J. Stat. Mech.} \textbf{1606} (2016) 064008, arXiv:1603.00440
{[}cond-mat.stat-mech{]}.

\bibitem{2005PlatiniIsingChain}T. Platini and D. Karevski, \textit{Eur.
Phys. J}. \textbf{B 48} (2005) 225, arXiv:cond-mat/0509594 {[}cond-mat.stat-mech{]}.

\bibitem{DeLucaVitiBernardDoyonIsing}A. De Luca, J. Viti, D. Bernard
and B. Doyon, \textit{Phys. Rev.} \textbf{B 88} (2013) 134301, arXiv:1305.4984
{[}cond-mat.str-el{]}.

\bibitem{ColluraKarevski}M. Collura and D. Karevski, \textit{Phys.
Rev.} \textbf{B 89} (2014) 214308, arXiv:1402.1944 {[}cond-mat.stat-mech{]}.

\bibitem{JeromeFreeFermion}J. Viti, J.-M. Stéphan, J. Dubail and
M. Haque, \textit{EPL} \textbf{115} (2016) 40011, arXiv:1507.08132
{[}cond-mat.stat-mech{]}.

\bibitem{DeLuca1}A. Biella, A. De Luca, J. Viti, D. Rossini, L. Mazza
and Rosario Fazio, \textit{Phys. Rev.} \textbf{B 93} (2016) 205121,
arXiv:1602.05357 {[}cond-mat.stat-mech{]}.

\bibitem{FagottiSpinChains}M. Fagotti,  \textit{Phys. Rev.} \textbf{B
96} (2017) 220302, arXiv:1708.05383 {[}cond-mat.stat-mech{]}.

\bibitem{MarciInHomQQ}M. Kormos, \textit{SciPost Phys.} \textbf{3}
(2017) 020, arXiv:1704.03744 {[}cond-mat.stat-mech{]}.

\bibitem{PerfettoGambassi}G. Perfetto and A. Gambassi, \textit{Phys.
Rev.} \textbf{E 96} (2017) 012138, arXiv:1704.03437 {[}cond-mat.stat-mech{]}.

\bibitem{MarciSemiClInHom}M. Kormos, C. P. Moca and G. Zaránd, \textit{Phys.
Rev.} \textbf{E 98} (2018) 032105, arXiv:1712.09466 {[}cond-mat.stat-mech{]}. 

\bibitem{TakatoGHDCheckedWithAbacus}B. Doyon, J. Dubail, R. Konik
and T. Yoshimura, \textit{ Phys. Rev. Lett.} \textbf{119} (2017)
195301, arXiv:1704.04151 {[}cond-mat.stat-mech{]}.

\bibitem{VasseurKarraschMooreHydroCheckDMRG}V. B. Bulchandani, R.
Vasseur, C. Karrasch and J. E. Moore, \textit{Phys. Rev. Lett.} \textbf{119}
(2017) 220604, arXiv:1704.03466 {[}cond-mat.stat-mech{]}.

\bibitem{GHDNewtonCraddleFirst}J.-S. Caux, B. Doyon, J. Dubail, R.
Konik and T. Yoshimura, \texttt{Hydrodynamics of the interacting Bose
gas in the Quantum Newton Cradle setup} Preprint, arXiv:1711.00873
{[}cond-mat.stat-mech{]}.

\bibitem{GHDNewtonCraddle}M. Schemmer, I. Bouchoule, B. Doyon and
J. Dubail, \textit{Phys. Rev. Lett. }\textbf{122} (2019) 090601,
arXiv:1810.07170 {[}cond-mat.quant-gas{]}.

\bibitem{DubailInHomCFT}P. Ruggiero, Y. Brun and J. Dubail, \textit{SciPost
Phys.} \textbf{6} (2019) 051, arXiv:1901.08132 {[}cond-mat.stat-mech{]}. 

\bibitem{GHDFundation}\foreignlanguage{english}{O. A. Castro-Alvaredo,
B. Doyon and T. Yoshimura, \textit{Phys. Rev.} \textbf{X 6} (2016)
041065, arXiv:1605.07331 {[}cond-mat.stat-mech{]}.}

\selectlanguage{english}%
\bibitem{GHDFundationB}B. Bertini, M. Collura, J. De Nardis and M.
Fagotti,  \textit{Phys. Rev. Lett.} \textbf{117} (2016) 207201, arXiv:1605.09790
{[}cond-mat.stat-mech{]}.

\selectlanguage{british}%
\bibitem{GHDFoundationC}B. Doyon and T. Yoshimura, \textit{SciPost
Phys.} \textbf{2} (2017) 014, arXiv:1611.08225 {[}cond-mat.stat-mech{]}.

\bibitem{GHDPiroli}\foreignlanguage{english}{L. Piroli, J. De Nardis,
M. Collura, B. Bertini and M. Fagotti \textit{Phys. Rev. }\textbf{B
96} (11) (2017) 115124,  arXiv:1706.00413 {[}cond-mat.stat-mech{]}.}

\bibitem{GHDDeLuca}A. De Luca, M. Collura and J. De Nardis, \textit{Phys.
Rev.} \textbf{B 96}(2) (2017) 020403, arXiv:1612.07265 {[}cond-mat.str-el{]}.

\bibitem{GHDHubbard}E. Ilievski and J. De Nardis, \textit{Phys.
Rev.} \textbf{B 96} (2017) 081118, arXiv:1706.05931 {[}cond-mat.stat-mech{]}.

\bibitem{GHDDomainWall}M. Collura, A. De Luca and J. Viti, \textit{Phys.
Rev. }\textbf{B 97} (2018) 081111, arXiv:1707.06218 {[}cond-mat.stat-mech{]}.

\bibitem{GHDAlba}V. Alba, \textit{Phys. Rev.} \textbf{B 97} (2018)
245135, arXiv:1706.00020 {[}cond-mat.stat-mech{]}.

\bibitem{GHDLowTXXZ}B. Bertini and L. Piroli, \textit{J. Stat. Mech.}
\textbf{1803} (2018) 033104, arXiv:1711.00519 {[}cond-mat.stat-mech{]}.

\bibitem{GHDKarrasch}V. B. Bulchandani, R. Vasseur, C. Karrasch and
J. Moore, \textit{Phys. Rev.} \textbf{B 97}(4) (2018) 045407, arXiv:1702.06146
{[}cond-mat.stat-mech{]}.

\bibitem{GHDBastianello}A. Bastianello and A. De Luca: \texttt{Integrability-protected
adiabatic reversibility in quantum spin chains} (2018) Preprint, arXiv:1811.07922
{[}cond-mat.stat-mech{]}.

\bibitem{GHDDoyonSpohn}B. Doyon and H. Spohn, \textit{J. Stat. Mech.}
\textbf{1707} (2017) 073210, arXiv:1703.05971 {[}cond-mat.stat-mech{]}.

\bibitem{GHDCao}X. Cao, V. B. Bulchandani and J. Moore, \textit{Phys.
Rev. Lett.} \textbf{120}(16) (2018)164101, arXiv:1710.09330 {[}cond-mat.stat-mech{]}.

\bibitem{GHDClassFT}A. Bastianello, B. Doyon, G. Watts and T. Yoshimura,
 \textit{SciPost Phys.} \textbf{4} (2018) 045, arXiv:1712.05687 {[}cond-mat.stat-mech{]}.

\bibitem{GHDClassToda}B. Doyon: \texttt{Generalised hydrodynamics
of the classical Toda system} (2019) Preprint, arXiv:1902.07624 {[}cond-mat.stat-mech{]}.

\bibitem{GHDMarci}M. Márton, B. Bertini, L. Piroli and P. Calabrese,
\textit{Phys. Rev.} \textbf{B 99} (2019) 014305, arXiv:1810.01089
{[}cond-mat.quant-gas{]}.

\bibitem{GHDSineG}B. Bertini, L. Piroli and M. Kormos: T\texttt{ransport
in the sine-Gordon field theory: from generalized hydrodynamics to
semiclassics} (2019) Preprint, arXiv:1904.02696 {[}cond-mat.stat-mech{]}.

\bibitem{GHDGeometric}B. Doyon, H. Spohn and T. Yoshimura, \textit{Nucl.
Phys.} \textbf{B 926} (2017) 570-582, arXiv:1704.04409 {[}cond-mat.stat-mech{]}.

\bibitem{DoyonSolitonGas}B. Doyon, T. Yoshimura and J.-S. Caux, \textit{Phys.
Rev. Lett.} \textbf{120} (2018) 045301, arXiv:1704.05482 {[}cond-mat.stat-mech{]}

\bibitem{MarciFleeGas}M. Mestyán and V. Alba: \texttt{Molecular dynamics
simulation of entanglement spreading in generalized hydrodynamics
}(2019) Preprint, arXiv:1905.03206 {[}cond-mat.stat-mech{]}.

\bibitem{GHDFluctuations1}J. Myers, M. J. Bhaseen, R. J. Harris and
B. Doyon: \texttt{Transport fluctuations in integrable models out
of equilibrium }(2018) Preprint, arXiv:1812.02082 {[}cond-mat.stat-mech{]}.

\bibitem{GHDFluctuations2}B. Doyon and J. Myers: \texttt{Fluctuations
in ballistic transport from Euler hydrodynamics }(2019) Preprint,
arXiv:1902.00320 {[}cond-mat.stat-mech{]}.

\bibitem{GHDDiff1}J. De Nardis, D. Bernard and B. Doyon, \textit{Phys.
Rev. Lett.} \textbf{121} (2018) 160603, arXiv:1807.02414 {[}cond-mat.stat-mech{]}.

\bibitem{GHDDiff2}S. Gopalakrishnan and R. Vasseur, \textit{Phys.
Rev. Lett.} \textbf{122} (2019) 127202, arXiv:1812.02701 {[}cond-mat.stat-mech{]}.

\bibitem{GHDDiff3}J. De Nardis, D. Bernard and B. Doyon, \textit{SciPost
Phys. }\textbf{6 }(2019) 049, arXiv:1812.00767 {[}cond-mat.stat-mech{]}.

\bibitem{BPZ}\foreignlanguage{american}{A. A. Belavin, A.M. Polyakov
and A.B. Zamolodchikov, \textit{Nucl. Phys.} \textbf{B 241} (1984)
333-380.}

\bibitem{NonEqCFTSpyros}S. Sotiriadis and J. Cardy, \textit{J. Stat.
Mech.} \textbf{0811} (2008) 11003, arXiv:0808.0116 {[}cond-mat.stat-mech{]}.

\bibitem{NonEqCFT1}D. Bernard and B. Doyon,  \textit{J. Phys.} \textbf{A
45} (2012)362001, arXiv:1202.0239 {[}cond-mat.str-el{]}.

\bibitem{NonEqCFT2}D. Bernard and B. Doyon,  \textit{Ann. Henri
Poincare} \textbf{16} (2015) 113-161, arXiv:1302.3125 {[}math-ph{]}.

\bibitem{NonEqCFT3}B. Doyon, M. Hoogeveen and D. Bernard,  \textit{J.
Stat. Mech.} \textbf{1403} (2014) 03002, arXiv:1306.3192 {[}cond-mat.stat-mech{]}.

\bibitem{TTBarCFT}D. Bernard and B. Doyon \textit{J. Stat. Mech.}
\textbf{1602} (2016) 033104, arXiv:1507.07474 {[}cond-mat.stat-mech{]}.

\bibitem{NonEqCFT5}A. Lucas, K. Schalm, B. Doyon and M. J. Bhaseen,
\textit{Phys. Rev.} \textbf{D 94} (2016) 025004, arXiv:1512.09037
{[}hep-th{]}.

\bibitem{NonEqCFTReview}D. Bernard and B. Doyon, \textit{J. Stat.
Mech.} \textbf{1606} (2016) 064005, arXiv:1603.07765 {[}cond-mat.stat-mech{]}.

\bibitem{AnZamo123}Al. B. Zamolodchikov, \textit{Nucl. Phys.} \textbf{B
358} (1991) 497.

\bibitem{AnZamo2}Al. B. Zamolodchikov, \textit{Nucl. Phys.} \textbf{B
358} (1991) 524.

\bibitem{AnZamo3} Al. B. Zamolodchikov, \textit{Nucl. Phys.} \textbf{B
366} (1991) 122.

\bibitem{DnParafermions}V. A.Fateev and Al. B.Zamolodchikov, \textit{Phys.
Lett.} \textbf{B 271} (1991) 91.

\bibitem{DynkinTBAs}F. Ravanini, R. Tateo and A. Valeriani, \textit{Int.
J. Mod .Phys.} \textbf{A 8} (1993) 1707-1728, arXiv:hep-th/9207040.

\bibitem{MasslessFlow} G. Delfino, G. Mussardo and P. Simonetti,
\texttt{} \emph{Phys. Rev.} \textbf{D 51} (1995) 6620-6624, arXiv:hep-th/9410117.

\bibitem{RefRoaming} P. Dorey, G. Siviour and G. Takács, \texttt{}
\emph{JHEP} \textbf{1503} (2015) 054, arXiv:1412.8442 {[}hep-th{]}.

\bibitem{RoamingFF}D. X. Horváth, P.E. Dorey and G. Takacs, \textit{JHEP}
\textbf{1607} (2016) 051, arXiv:1604.05635 {[}hep-th{]}.

\bibitem{RefZamo}Al. B. Zamolodchikov, \texttt{}\emph{J. Phys.}
\textbf{A 39} (2006) 12847-12862.

\bibitem{RGFlow2} M. J. Martins, \texttt{} \emph{Phys. Rev. Lett.}
\textbf{69} (1992) 2461-2464, arXiv:hep-th/9205024.

\bibitem{RGFlow3} P. E. Dorey and F. Ravanini, \emph{Int. J. Mod.
Phys.} \textbf{A 8} (1993) 873-894, arXiv:hep-th/9206052.

\bibitem{RGFlow4} M.J. Martins, \texttt{}\emph{Phys. Lett.} \textbf{B
304} (1993) 111-114.

\bibitem{RGFlow5} P. Dorey and F. Ravanini, \emph{Nucl. Phys.} \textbf{B
406} (1993) 708, arXiv:hep-th/9211115.

\bibitem{OriginalCTheorem}A. B. Zamolodchikov,  J\textit{ETP Lett.}
\textbf{43} (1986) 730\textendash 732.

\bibitem{cTheorem}Al. Zamolodchikov,  \textit{Nucl. Phys.} \textbf{B
342} (1990) 695\textendash 720.

\bibitem{TakatoProofVEff}D.-L. Vu and T. Yoshimura, \textit{SciPost
Phys.} \textbf{6} (2019) 023, arXiv:1809.03197 {[}cond-mat.stat-mech{]}.

\bibitem{PiroliLL}B. Bertini, L. Piroli and Pasquale Calabrese, \textit{Phys.
Rev. Lett.} \textbf{120} (2018)176801, arXiv:1709.10096 {[}cond-mat.stat-mech{]}.

\bibitem{DoyonWrongHydro}O. Castro-Alvaredo, Y. Chen, B. Doyon and
M. Hoogeveen, \textit{J. Stat. Mech.} \textbf{1403} (2014) 03011,
arXiv:1310.4779 {[}hep-th{]}.

\bibitem{DoyonLowerBounds}B. Doyon, \textit{Nucl. Phys.} \textbf{B
892} (2015) 190-210, arXiv:1410.0292 {[}cond-mat.str-el{]}.

\bibitem{KastorMartinecTTBar}D. A. Kastor, E. J. Martinec and S.
H. Shenker, \textit{Nucl. Phys.} \textbf{B 316} (1989), 590.

\bibitem{ZamolodchikovTTBar}A. B. Zamolodchikov, \textit{Fractional
Spin Integrals of Motion in Perturbed Conformal Field Theory in the
Proceedings Conference Beijing 1989, \textquotedblright Fields, strings
and quantum gravity\textquotedblright }, 349.

\bibitem{Landauer}R. Landauer, \textit{IBM Journal of Research and
Development} \textbf{1} (1957) 223\textendash 231. 

\bibitem{Buttiker1}M. Büttiker,  Phys. Rev. Lett. \textbf{57} (1986)1761.

\bibitem{Buttiker2}M. Büttiker,  Phys. Rev. \textbf{B 38} (1988)
9375.

\bibitem{Landauer2}R. Landauer,  \textit{Phys. Scr.} \textbf{1992}
110.

\bibitem{ZamoNLIE}Al. B. Zamolodchikov,  \textit{Phys. Lett.} \textbf{B
335} (1994) 436\textendash 443.

\bibitem{DestriVegaNLIE}C. Destri and H. J. de Vega,  \textit{Phys.
Rev. Lett.} \textbf{69} (1992) 2313\textendash 2317.

\bibitem{DestriVegaNLIE2}C. Destri and H. J. de Vega,\textit{ Nucl.
Phys.} \textbf{B 438} (1995) 413\textendash 454, arXiv:hep-th/9407117.

\bibitem{MasslessNLIE}P. Dorey, C. Dunning and R. Tateo, \textit{Nucl.
Phys.} \textbf{B 578} (2000) 699-727, arXiv:hep-th/0001185.

\bibitem{TakiNLIE1}G. Feverati, F. Ravanini and G. Takács, \textit{Phys.
Lett.} \textbf{B 430} (1998) 264-273, hep-th/9803104.

\bibitem{TakiNLIE2}G. Feverati, F. Ravanini and G Takács, \textit{Phys.
Lett.} \textbf{B 444} (1998) 442-450, hep-th/9807160.

\bibitem{TakiNLIE3}G. Feverati, F. Ravanini and G. Takács,  \textit{Nucl.
Phys.} \textbf{B 540} (1999) 543-586, hep-th/9805117.

\bibitem{TakiNLIE4}G. Feverati, F. Ravanini and G. Takács, \textit{Nucl.
Phys.} \textbf{B 570} (2000) 615-643, hep-th/9909031.

\bibitem{W3p}V. A. Fateev and A. B. Zamolodchikov, \textit{Nucl.
Phys.} \textbf{B 280} (1987) 644-660.

\bibitem{W3pBetheA}M. J. Martins, \textit{Phys. Lett.} \textbf{B
277} (1992) 301-305, arXiv:hep-th/9201032.
\end{thebibliography}
\end{document}